\documentclass[a4paper,aps,pre,twocolumn,showpacs,nofootinbib,superscriptaddress,fleqn]{revtex4-1}

\usepackage{mathtools}
\usepackage{subcaption}
\usepackage{float}
\usepackage{fontawesome}
\usepackage{lipsum}
\usepackage{amsmath}
\usepackage{amssymb}
\usepackage{graphicx}
\usepackage{etoolbox} 
\usepackage{amsmath}
\usepackage{dsfont}
\usepackage{amsthm}
\usepackage{bold-extra}
\usepackage[title]{appendix}
\usepackage[usenames,dvipsnames,svgnames,table]{xcolor}
\usepackage[margin=5pt,font=small,labelfont=bf,labelsep=endash]{caption}
\usepackage{amsthm}
\usepackage{rotating}
\usepackage{hyperref}
\usepackage{slashed} 
\usepackage{array,multirow}
\usepackage{shuffle}
\usepackage{empheq}
\usepackage{marvosym}
\usepackage{yfonts}
\usepackage[title]{appendix}
\usepackage[bottom]{footmisc}
\usepackage{tikz}
\usetikzlibrary{calc,matrix}
\usepackage{stackrel}
\usepackage{textcomp}
\usepackage{fancyvrb}
\usepackage{tcolorbox}
\usepackage{scalefnt}
\usepackage{setspace}
\usepackage{enumitem}
\usepackage{csquotes}
\usepackage{adjustbox}
\usepackage{url}
\usepackage{hhline}
\usepackage[final]{pdfpages}
\usepackage{CJKutf8}

\newcommand{\cA}{{\cal A}}
\newcommand{\cB}{{\cal B}}

\newcommand{\cF}{{\cal F}}

\newcommand{\cH}{{\cal H}}

\newcommand{\cP}{{\cal P}}

\newcommand{\cS}{{\cal S}}

\newcommand{\rd}{\text{d}}
\newcommand{\re}{\text{e}}

\newcommand{\dRRP}{d_{\scriptscriptstyle\rm RRP}}
\newcommand{\dPPS}{d_{\scriptscriptstyle\rm PPS}}
\newcommand{\dSSR}{d_{\scriptscriptstyle\rm SSR}}
\newcommand{\dRRS}{d_{\scriptscriptstyle\rm RRS}}
\newcommand{\dPPR}{d_{\scriptscriptstyle\rm PPR}}
\newcommand{\dSSP}{d_{\scriptscriptstyle\rm SSP}}
\newcommand{\deq}{d_{\rm e}}
\newcommand{\dpol}{d_{\rm p}}
\newcommand{\bx}{\mathbf{x}}
\newcommand{\by}{\mathbf{y}}

\newcommand{\redR}{{{\color{red}\tt R}}}
\newcommand{\greenP}{{{\color{green!70!blue}\tt P}}}
\newcommand{\blueS}{{{\color{blue!70!gray}\tt S}}}
\newcommand{\grayEmptySet}{{{\color{black}$\emptyset$}}}
\newcommand{\lambdar}{\lambda_{\text{r}}}
\newcommand{\lambdai}{\lambda_{\text{i}}}

 \numberwithin{equation}{section}
\definecolor{shadecolor}{rgb}{0.97,0.97,0.97}
\definecolor{myblue}{rgb}{0.97,0.97,0.97}
\definecolor{mygreen}{rgb}{0.15,0.7,0.15}

\makeatletter
\AtBeginDocument{\let\LS@rot\@undefined}
\makeatother

\begin{document}

\title{\bf\textcolor{black}{Coevolutionary dynamics of a variant of the \\ cyclic Lotka-Volterra model with three-agent interactions}}

\author{Filippo Palombi}
\email{Corresponding author: filippo.palombi@enea.it}
\affiliation{ENEA---Italian National Agency for New Technologies, Energy and Sustainable Economic Development\\ Via Enrico Fermi 45, 00044 Frascati -- Italy\\[1.0ex]}

\author{Stefano Ferriani}
\affiliation{ENEA---Italian National Agency for New Technologies, Energy and Sustainable Economic Development\\ Via Martiri di Monte Sole 4, 40129 Bologna -- Italy\\[1.0ex]}

\author{Simona Toti}
\affiliation{ISTAT---Italian National Institute of Statistics\\ Via Cesare Balbo 16, 00184 Rome -- Italy}

\date{\today}

\begin{abstract}
  We study a variant of the cyclic Lotka-Volterra model with three-agent interactions. Inspired by a multiplayer variation of the Rock-Paper-Scissors game, the model describes an ideal ecosystem in which cyclic competition among three species develops through cooperative predation. Its rate equations in a well-mixed environment display a degenerate Hopf bifurcation, occurring as reactions involving two predators plus one prey have the same rate as reactions involving two prey  plus one predator. We estimate the magnitude of the stochastic noise at the bifurcation point, where finite size effects turn neutrally stable orbits into erratically diverging trajectories. In particular, we compare analytic predictions for the extinction probability, derived in the Fokker-Planck approximation, with numerical simulations based on the Gillespie stochastic algorithm. We then extend the analysis of the phase portrait to heterogeneous rates. In a well-mixed environment, we observe a continuum of degenerate Hopf bifurcations, generalizing the above one. Neutral stability ensues from a complex equilibrium between different reactions. Remarkably, on a two-dimensional lattice, all bifurcations disappear as a consequence of the spatial locality of the interactions. In the second part of the paper, we investigate the effects of mobility in a lattice metapopulation model with patches hosting several agents. We find that strategies propagate along the arms of rotating spirals, as they usually do in models of cyclic dominance. We observe propagation instabilities in the regime of large wavelengths. We also examine three-agent interactions inducing nonlinear diffusion. 
\\[-4.0ex]
\begin{center}
  \small\emph{``Three at play. That'll be the day!''}\\ (a child in \emph{Wings of desire} [W. Wenders, 1987])
\end{center}
  
\end{abstract}

\maketitle

\section{Introduction}\label{sect:1}

\begin{figure}[t!]
  \centering
  \includegraphics[width=0.09\textwidth]{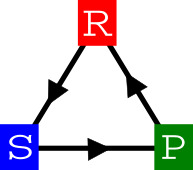}
  \vskip 0.0cm
  \caption{\footnotesize {\color{red} [color online]} {\tt RPS} cyclic chain.\label{fig:RPS}}
  \vskip -0.3cm
\end{figure}

Cyclic competition is distinctively associated with closed relational chains, describing aspects of the struggle for life such as feeding, hunting, mating. The prototype is a system made of three different species, interacting with one another like children playing Rock-Paper-Scissors ({\tt RPS}), the famous game where paper (\greenP) wraps rock, scissors (\blueS) cut paper and rock (\redR) crushes scissors, see Fig.~\ref{fig:RPS}.  Such schemes are at heart of biological systems spanning a wide range of length scales and complexity. Examples include the repressilator~\cite{Elowitz1}, the E. coli colicin E2 system~\cite{Kerr1}, the common side-blotched lizard~\cite{Sinervo1}, several plant systems~\cite{Taylor1,Cameron1,Lankau1} and so forth.

In principle, the absence of apex predators and bottom prey  in cyclic chains allows species to dominate in turn. As soon as one of them outperforms the others, it becomes itself a source of nourishment for the next one along the chain. This feature suggests that cyclic competition may serve as a fundamental mechanism facilitating coexistence and biodiversity. Experiments performed in Ref.~\cite{Kerr1} on three cyclically interacting strains of E. coli confirmed this thesis and also made it more precise: species can coexist provided ecological processes (interaction and dispersal) develop locally. In practice, to ensure coexistence in the experiments, colonies of different strains had to first grow in separate spatial domains and then be left free to invade neighboring colonies.

Theoretical insight into the role of locality and individual mobility for coexistence was achieved in a ground-breaking paper~\cite{Reichenbach1} thanks to an evolutionary model based on a three-state {\tt RPS} game. Results were then generalized to a four-state variant of the model~\cite{Peltomaki1}. Models unifying pairwise cyclic dominance with removal (as in Ref.~\cite{Reichenbach1}) and replacement (as in Ref.~\cite{Peltomaki1}) were later studied in Refs.~\cite{Szczesny1,Szczesny2,Rulands1}.
These studies revealed that:
\begin{itemize}[leftmargin=4mm,itemsep=-1mm]
\item[$\circ$]{individual mobility promotes coexistence by inducing self-organization of the strategies into spiral waves, traveling across the environment;}
\item[$\circ$]{stochastic noise, arising in finite populations, produces local inhomogeneities; nevertheless, it cannot prevent the creation of spiral waves;}
\item[$\circ$]{beyond a critical threshold, individual mobility leads to species extinction.}
\end{itemize}
  
Subsequent research dealt with a plethora of either induced or independent issues. They include the effects of competition on pattern formation~\cite{Jiang1}, the observation of multi-armed spirals~\cite{Jiang2}, the emergence of convective instability~\cite{Reichenbach2}, the role of stochastic noise~\cite{Reichenbach3} and mutations~\cite{Szczesny1,Szczesny2}, the analysis of coexistence and extinction basins~\cite{Rulands1,Shi1,Ni1}, the effects of uniform and nonuniform intra-specific competition~\cite{Yang1,Park1}, the influence of directional mobility on coexistence~\cite{Avelino1} and so forth. We omit to mention other relevant developments only for the sake of conciseness, while we refer the reader to Ref.~\cite{Szolnoki1} for a systematic review of concepts and results.

A common assumption in all the above-mentioned studies is that predation is a pairwise interaction, involving a single predator and a single prey. While this is the case for many living organisms, either animals or microbes, it is not so for others. An alternative strategy, favored by natural selection, is cooperative predation. Mammals such as  wolves~\cite{Mech1}, chimpanzees~\cite{Stanford1}, dolphins~\cite{Samuels1} and lions~\cite{Schaller1} cooperate in hunting. Some insects, including ants~\cite{Holldobler1}, behave analogously. Even bacteria can practice group hunting. Among them we mention Saprospira~\cite{Perez1}, Myxococcus Xanthus~\cite{Berleman1,Munoz1} and Lysobacter~\cite{Jurkevitch1}. They essentially \emph{``require a quorum of predators to degrade the prey, using excreted hydrolytic enzymes''}~\cite{Jurkevitch1}.

Another recently esablished mechanism for species coexistence involving multiagent interactions in microbial communities is the \emph{protection spillover}. Kelsic \emph{et al} have shown in Ref.~\cite{Kelsic} that bacteria protecting themselves from potential predators by excreting inhibitory enzyms, can protect bacteria of other species in the nearby area from the same predators. If we rephrase the the RPS chain as \emph{``P survives an attack by~R, which survives an attack by~S, which in turn survives an attack by P''}, then protection spillover means that~R helps~P survive an attack by~S. This protection mechanism stabilizes the species in a well-mixed environment \cite{Bergstrom} and gives rise, in structured populations, to spatial patterns which cannot be observed in the classical RPS game \cite{Szolnoki0}.

Other spatially structured models of cyclic competition featuring group predation were considered in Refs.~\cite{Szolnoki2,Cheng1,Cazaubiel1,Lutz1}. Specifically, the authors of Ref.~\cite{Szolnoki2} studied two variants of a three-state lattice model in which predation entails simultaneous pairwise interactions. One of them assumes that agents interact with their four von-Neumann neighbors, the other with their eight Moore neighbors~\cite{Toffoli1}. They found that increasing the interaction range can decelerate the propagation of predators and even revert the direction of species invasion contrary to its natural definition. The authors of Ref.~\cite{Cheng1} observed the emergence of mesoscopic subgroups of coexisting species in a five-state lattice model based on the Rock-Paper-Scissors-Lizard-Spock game. They developed a mean-field theory to show that group interactions at the mesoscopic scale must be taken into account to justify the observed states of coexistence. The authors of Ref.~\cite{Cazaubiel1} studied a stochastic lattice version of a model introduced by Lett, Auger, Gaillard~\cite{Lett1}. In this model, the abundances of prey  and predators are constant, while the fractions of each population using either an individual or a collective strategy coevolve. Their results include a complex phase diagram in which four different strategies (corresponding to prey/predators behaving collectively/individually in all possible combinations) dominate or coexist. Remarkably, in the pure coexistence phase, they found cyclic dominance of the four strategies. Finally, the same authors further explored the spatial version of the model to quantify some geometrical and percolative properties of the clusters formed by the four strategies~\cite{Lutz1}.

The reader will recognize that none of the abovementioned approaches considers ab initio multiagent microscopic reactions with independent rates, such as
\begin{equation}
  \label{eq:multiagent}
  \left\{\begin{array}{l}
    X_1X_2\dots X_k \to Y_1Y_2\dots Y_\ell, \text{ for } k,\ell\ge 3\,,\\[1.0ex]
    X_i,Y_j\in\{{\redR,\greenP,\blueS,\text{\grayEmptySet}}\}\,
  \end{array}
  \right.
\end{equation}
(in the literature \grayEmptySet\ conventionally denotes an empty site)\footnote{Reactions considered in Ref.~\cite{Szolnoki2} can be in fact recast in the form of Eq.~(\ref{eq:multiagent}) for $k=\ell=5,9$. Yet, their rates are somewhat intertwined in that they are functions of the pairwise rates $\delta_0,\delta_1$.}. In this paper, we study a simple variant of the cyclic Lotka-Volterra model~\cite{Reichenbach5,Frachebourg1,Frachebourg2,Frachebourg3,Provata1,Tsekouras1,Szabo1,Szabo2} with three-agent interactions, like Eq.~(\ref{eq:multiagent}) for $k=\ell=3$.  To explore the model, we use mathematical techniques developed in similar contexts, including nonlinear analysis of bifurcations, stochastic partial differential equations and numerical simulations.

We find that the underlying rate equations in a well-mixed environment exhibit a reactive fixed point falling in the universality class of the Hopf bifurcations. As such, they induce a macroscopic phenomenology qualitatively similar to models already studied in the literature. Nonetheless, the internal dynamics of the model is original and interesting. Two opposing forces drive the system. Interactions involving two prey plus one predator pull it towards the reactive fixed point, thus playing an equilibrating role. By contrast, interactions involving two predators plus one prey push the system away from the reactive fixed point, thus producing a polarizing effect. The relative strength of the two forces controls the evolution of the system in a spatially structured version of the model, with patches hosting several motile agents (metapopulation model). When polarizing interactions dominate over equilibrating ones, coexistence is achieved through the development of spatiotemporal patterns, taking the usual form of rotating spiral waves. Similar to Refs.~\cite{Szolnoki10,Szolnoki11,Szolnoki12,Szolnoki13}, the dynamics of three-agent interactions on a spatially structured population shows a largely different behavior from predictions based on well-mixed calculations.

We also find that spatial topology is critical for the evolution of species. Indeed, the phase portrait of the model changes drastically on a two-dimensional lattice with single agent per site and nearest-neighbor interactions. Locality makes the reactive fixed point stable for every choice of reaction rates. The disappearance of Hopf bifurcations indicates that local patches hosting several agents are essential for the development of patterns. 

The paper is organized as follows. First of all, in section~II, we define the model and study its equations in a well-mixed environment for homogeneous rates. In section~III, we estimate the magnitude of the stochastic noise affecting finite agent populations. As known~\cite{Reichenbach5}, fluctuations turn neutrally stable orbits into diverging trajectories, eventually resulting in the extinction of two species. We compare the extinction probability, derived in the Fokker-Planck approximation, with numerical simulations. In section~IV, we discuss the phase portrait of the model for heterogeneous rates. We also compare results in a well-mixed environment with those on a two-dimensional lattice.  In section~V, we introduce mobility reactions in a lattice metapopulation model with patches hosting several agents. In section VI, we examine three-agent interactions inducing nonlinear diffusion. Finally, in section~VII, we draw conclusions.   

\section{Rate equations}

Some time ago we happened to observe three children playing a three-player variant of {\tt RPS} in the hall of their school. Intrigued by the game, we asked them about it. In order not to leave anyone out---they proudly explained---they were playing all at once. The rules of the game were as follows. On each round, children had to deliver simultaneously one of the usual hand signals, representing \redR,\ \greenP\ and \blueS. Round by round they received payoffs, based on cyclic dominance and depending on the combination of delivered signals,  as we report in Table~\ref{tab:payoffs}\footnote{While writing the paper we realized that essentially the same three-player variant of {\tt RPS} is described in Ref.~\cite{Walker1}. Similar variants can be found on various websites. In a scene of \emph{Sonatine} (\begin{CJK}{UTF8}{min}ソナチネ\end{CJK}), a 1993 film by T.~Kitano, three yakuza gangsters play a three-player variant of RPS on the beach.}.

To translate the game into the language of evolutionary game theory, we assume a population of $N\gg 1$ agents, each adopting one of the competing strategies. We let $r$, $p$, $s$ denote the relative abundances of \redR, \greenP, \blueS\ respectively. As such they fulfill $r+p+s=1$. In our model, densities evolve in time as a consequence of microscopic interactions inspired by Table~\ref{tab:payoffs}. More precisely, in place of payoffs we consider stochastic reactions mediated by a dominance-replacement mechanism, namely
\begin{equation}
  \begin{array}{llcl}
    \text{\redR\,\redR\,\greenP}\ & \to\ \ \text{\redR\,\greenP\,\greenP} & \text{\ with rate} &  \dRRP\,, \\[0.7ex]
    \text{\greenP\,\greenP\,\blueS}\ & \to\ \ \text{\greenP\,\blueS\,\blueS} & \text{\ \, "} &  \dPPS\,,  \\[0.7ex]
    \text{\blueS\,\blueS\,\redR}\ & \to\ \ \text{\blueS\,\redR\,\redR} & \text{\ \, "} &  \dSSR\,,  \\[0.7ex]
    \text{\redR\,\redR\,\blueS}\ & \to\ \ \text{\redR\,\redR\,\redR} & \text{\ \, "} &  \dRRS\,,  \\[0.7ex]
    \text{\greenP\,\greenP\,\redR}\ & \to\ \ \text{\greenP\,\greenP\,\greenP} & \text{\ \, "} &  \dPPR\,,  \\[0.7ex]
    \text{\blueS\,\blueS\,\greenP}\ & \to\ \ \text{\blueS\,\blueS\,\blueS} & \text{\ \, "} &  \dSSP\,,
    \label{eq:transitions}
  \end{array}
\end{equation}
where rates (transition probabilities per unit time) are independent of one another. For the time being, we leave out transitions in which agents carry three different strategies before interacting. We also leave out transitions in which they all carry the same strategy, since such configurations yield no payoff in the classic formulation of the game. Rate equations (RE) including all contributions listed in Eqs.~(\ref{eq:transitions}) read as
\begin{align}
  \hskip -0.69cm
  \begin{array}{l}
  \dot r = \cF_{\rm r}(r,p,s)\,, \ \dot p = \cF_{\rm p}(r,p,s)\,, \ \dot s = \cF_{\rm s}(r,p,s)\,;\\[2.0ex]
  \cF_{\rm r} = r^2\left(\dRRS s - \dRRP p\right) + r\left(\dSSR s^2 - \dPPR p^2\right),\\[1.0ex]
  \cF_{\rm p} = p^2\left(\dPPR r - \dPPS s\right) + p\left(\dRRP r^2 - \dSSP s^2\right),\\[1.0ex]
  \cF_{\rm s} = s^2\left(\dSSP p - \dSSR r\right) + s\left(\dPPS p^2 - \dRRS r^2\right).
    \label{eq:rateeqs}
\end{array}
\end{align}
\begin{center}
  \begin{table}[t!]
    \small
    \begin{tabular}{c | c}
       outcome \  & payoffs \\
      \hline\hline
      \redR\ \redR\ \redR, \greenP\ \greenP\ \greenP, \blueS\ \blueS\ \blueS\raisebox{11pt} & \ all players receive 0 points\\[0.5ex]
      \hline
      \multirow{2}{*}{\redR\ \redR\ \greenP, \greenP\ \greenP\ \blueS, \blueS\ \blueS\ \redR} \raisebox{11pt} & \ the dominant player\\
       &  receives 1 point \\
      \hline
      \multirow{2}{*}{\redR\ \redR\ \blueS, \greenP\ \greenP\ \redR, \blueS\ \blueS\ \greenP} \raisebox{11pt} & \ each dominant player \\
       & receives 1/2 point \\
      \hline
      \redR\,\greenP\,\blueS \raisebox{11pt} & \ all players receive 0 points\\[0.5ex]
      \hline\hline
    \end{tabular}
    \vskip 0.1cm
    \caption{\footnotesize A three-player variant of {\tt RPS}.\label{tab:payoffs}}
  \vskip -0.4cm
  \end{table}
\end{center}
\vskip -0.5cm
Altogether, these cubic equations depend on six parameters. We can absorb one of them into a redefinition of time, thus obtaining five effective parameters. Moreover, we can drop the equation for $\dot s$ provided we insert $s=1-r-p$ into those for $\dot r$ and $\dot p$.  Eqs.~(\ref{eq:rateeqs}) describe correctly the evolution of strategies, induced by Eqs.~(\ref{eq:transitions}), in a well-mixed environment as $N\to\infty$. They represent a variant of the cyclic Lotka-Volterra equations \cite{Frachebourg1,Frachebourg2,Frachebourg3,Provata1,Tsekouras1,Szabo1,Szabo2,Reichenbach1} encompassing three-agent interactions.

To study the model, we first consider a simplified version in which rates are homogeneous. More precisely, we let
\begin{equation}
  \left.\begin{array}{l}
  \dRRP = \dPPS = \dSSR \equiv \deq\,,\\[1.0ex]
  \dRRS = \dPPR = \dSSP \equiv \dpol\,.\end{array}\right.
  \label{eq:homorates}
\end{equation}
As can be seen, homogeneity is not complete.  Eq.~(\ref{eq:homorates}) follows from separating reactions into two disjoint sets. The first three reactions in Eq.~(\ref{eq:transitions}) correspond to the first line of Eq.~(\ref{eq:homorates}). They are functionally homogeneous in that they involve two prey and one predator in the initial state. Their occurrence produces a local change of majority. The last three reactions in Eq.~(\ref{eq:transitions}) correspond to the second line of Eq.~(\ref{eq:homorates}). They start with two predators and one prey. Their occurrence results in a further increase of the local majority. The two groups of reactions play antagonistic roles. When the system is in a macroscopic state in which a strategy dominates, the former contributes to equilibrating it, whereas the latter contributes to further polarizing it. 

Resulting RE have three absorbing fixed points, $(r_1,p_1) = (1,0)$, $(r_2,p_2) = (0,1)$, $(r_3,p_3) = (0,0)$ and a reactive one, $(r_*,p_*) = (1/3,1/3)$. Since we are interested in the behavior of the system near the latter point, we let $r = 1/3 + x_\text{r}$ and $p = 1/3 + x_\text{p}$. Upon expanding the RE in Taylor series to first order around $x_\text{r} = x_\text{p} = 0$, we obtain
\begin{align}
  \left.\begin{array}{l}
    \displaystyle{\dot x_\text{r} = -\frac{\deq}{3}x_\text{r} - \frac{\deq+\dpol}{3}x_\text{p}}\,,\\[2.0ex]
    \displaystyle{\dot x_\text{p} = \ \,\frac{\dpol}{3}x_\text{p}+\frac{\deq+\dpol}{3}x_\text{r}}\,.
    \end{array}\right.
\end{align}
The eigenvalues of the Jacobian matrix are 
\begin{equation}
  \left.\begin{array}{l}
    \displaystyle{\lambda = \frac{1}{6}(\dpol - \deq) +\frac{\text{i}}{2\sqrt{3}}(\dpol+\deq)}\,,\\[2.0ex]
    \displaystyle{\bar\lambda = \frac{1}{6}(\dpol - \deq) -\frac{\text{i}}{2\sqrt{3}}(\dpol+\deq)}\,.
    \label{eq:linearev}
  \end{array}\right.
\end{equation}
The real part vanishes for $\deq = \dpol$, i.e. when equilibrating and polarizing forces compensate exactly. For this choice of rates we have a Hopf bifurcation. Starting near $(r_*,p_*)$, the system spirals inwards for $\deq>\dpol$, whereas it spirals outwards for $\deq<\dpol$. For $\deq=\dpol(=1)$ it travels on neutrally stable orbits enclosing the fixed point. They have angular frequency $\omega_0 = 1/\sqrt{3}$ and first integral $rps=\text{constant}$.

\begin{figure*}[t!]
  \centering
  \includegraphics[width=0.36\textwidth]{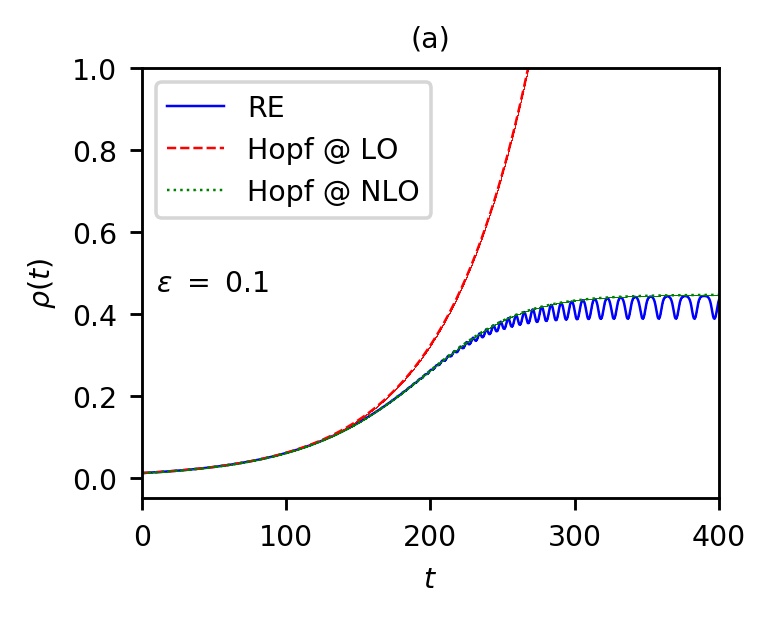}\hskip 2.0cm
  \includegraphics[width=0.36\textwidth]{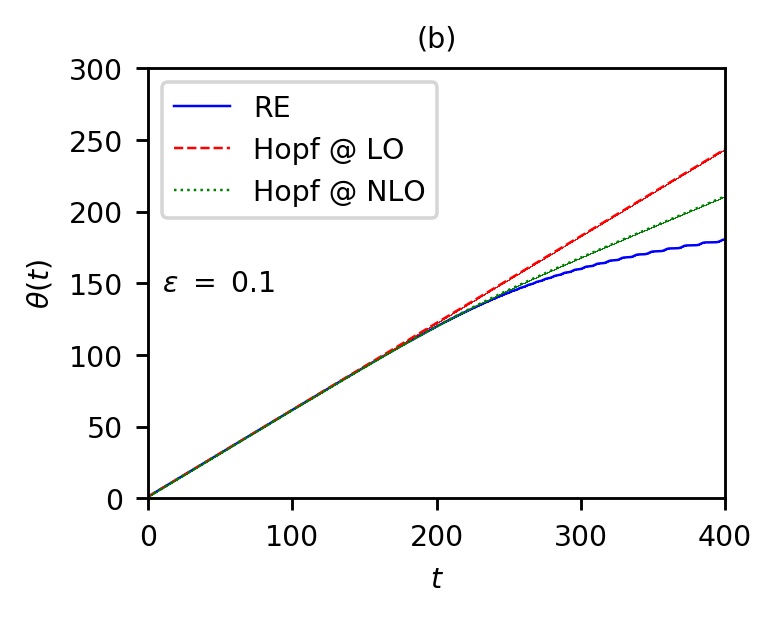}
  \vskip -0.3cm
  \caption{\footnotesize {\color{red} [color online]} Comparison between the RE and their Hopf normal form.\label{fig:hopf}}
  \vskip -0.2cm
\end{figure*}

To determine the type of the Hopf bifurcation, we need to bring the RE to normal form  according to a standard procedure, detailed, e.g., in Refs.~\cite{Wiggins1,Strogatz1}. To this aim, we let $\deq = 1$ and $\dpol = 1 + \epsilon$. We perform a linear change of variables, namely we let $y = Sx$ with
\begin{equation}
  S = \left(\begin{array}{cc}
    1/\sqrt{3} & - 1/\sqrt{3} \\[1.0ex] 1 & 1 \end{array}\right)\,.
\end{equation}
It yields equations
\begin{equation}
  \left.\begin{array}{l}
    \dot y_\text{r} = \text{Re}(\lambda)y_\text{r} -\text{Im}(\lambda)y_\text{p} + f_\text{r}(y_\text{r},y_\text{p},\epsilon)\,,\\[2.0ex]
    \dot y_\text{p} = \text{Im}(\lambda)y_\text{r} +\text{Re}(\lambda)y_\text{p} + f_\text{p}(y_\text{r},y_\text{p},\epsilon)\,,
    \end{array}\right.
\end{equation}
where the nonlinear functions $f_\text{r}$ and $f_\text{p}$ read as
\begin{align}
  f_\text{r}(y_\text{r},y_\text{p},\epsilon) & = \frac{\sqrt{3}}{4}(2+\epsilon)y_\text{r}^2 + \frac{\epsilon}{2}y_\text{r}y_\text{p} \nonumber\\[0.0ex]
  & \hskip -0.3cm - \frac{\sqrt{3}}{4}(1+\epsilon)y_\text{p}^2-\frac{3\epsilon}{4}y_\text{r}^3-\frac{3\epsilon}{4}y_\text{r}y_\text{p}^2\,,\\[1.0ex]
  f_\text{p}(y_\text{r},y_\text{p},\epsilon) & = \frac{\epsilon}{4}y_\text{r}^2-\frac{\sqrt{3}}{2}(2+\epsilon)y_\text{r}y_\text{p} \nonumber\\[0.0ex]
  & \hskip -0.3cm  -\frac{\epsilon}{4}y_\text{p}^2-\frac{3\epsilon}{4}y_\text{p}^3 -\frac{3\epsilon}{4}y_\text{p}y_\text{r}^2\,.
\end{align}
Next, we perform a change to complex variables $z = y_\text{r} + \text{i}y_\text{p}$ and ${\bar z} = y_\text{r} - \text{i} y_\text{p}$. Inverse relations are given by $y_r(z,\bar z) = (z +\bar z)/2$ and $y_p(z,\bar z) = (z-\bar z)/2\text{i}$. Inserting them into the above formulas yields
\begin{equation}
  \dot z = \lambda z + f(z,\bar z)\,,
  \label{eq:cplxhopf}
\end{equation}
with
\begin{align}
  f(z,\bar z) & = \left[\frac{\sqrt{3}}{2}\left(1+\frac{\epsilon}{2}\right)+\text{i}\frac{\epsilon}{4}\right]\bar z^2 - \frac{3\epsilon}{4} \nonumber\\[0.0ex]
  & \equiv R\bar z^2 -S \bar zz^2\,.
\end{align}
As can be seen, $f$ has quadratic and cubic terms. We can remove the former by performing an additional change of variable, viz. $z\to z + A\bar z^2$. We choose $A$ such that $(\lambda-2\bar\lambda)A=R$. After some algebra\footnote{Recall that $y=x + Ax^2$ cannot be inverted exactly. By iteration we have $x = y - Ax^2 = y - A(y - Ax^2)^2 = y - Ay^2 + 2A^2y^3 + \text{O}(y^4)\,$.
} we arrive at
\begin{equation}
  \dot z = \lambda z - c(\epsilon)z|z|^2\,,
  \label{eq:hopf}
\end{equation}
with $c(\epsilon) = a(\epsilon) + \text{i}b(\epsilon)$ and 
\begin{align}
  & a(\epsilon) = \frac{3}{2}\frac{4\epsilon^2 + 15\epsilon + 15}{7\epsilon^2+27\epsilon + 27}\epsilon\,,\\[1.0ex]
  & b(\epsilon) = \frac{9\sqrt{3}}{4}\frac{(\epsilon + 2)(\epsilon^2+3\epsilon+3)}{7\epsilon^2+27\epsilon+27}\,.
  \label{eq:hopfb}
\end{align}
Since $a(\epsilon)\approx (5/6)\,\epsilon$ as $\epsilon\to 0$, we conclude that the Hopf bifurcation is degenerate. In polar coordinates $z = \rho\re^{\text{i}\theta}$ Eq.~(\ref{eq:hopf}) turns into
\begin{equation}
  \dot\rho = \frac{\epsilon}{6}\rho - a(\epsilon)\rho^3\,, \qquad \dot\theta = \frac{2+\epsilon}{2\sqrt{3}}-b(\epsilon)\rho^2\,.
  \label{eq:polarhopf}
\end{equation}
In Fig.~\ref{fig:hopf}, we compare by numerical integration Eq.~(\ref{eq:polarhopf}) with the full RE for $\epsilon=0.1$. In both plots, dashed and dotted lines correspond respectively to the leading order approximation (LO) and the full next-to-leading order (NLO) version  of Eq.~(\ref{eq:polarhopf}). We always choose initial conditions with densities lying near the reactive fixed point. We observe no limit cycle. The asymptotic saturation of $\rho$ in Fig.~\ref{fig:hopf} (a) corresponds to heteroclinic cycles approaching the absorbing fixed points. 

A Hopf bifurcation of the same type as Eqs.~(\ref{eq:hopf})-(\ref{eq:hopfb}) characterizes the dynamics of the May-Leonard model~\cite{May1}. In that model, the bifurcation point corresponds to the vanishing of dominance-removal reactions. Therefore, the May-Leonard model is well-defined only on one side of the bifurcation. Ours makes sense on both sides. More importantly, on the unstable side, both models feature heteroclinic cycles. Owing to this similarity, their spatially structured versions, discussed respectively in Ref.~\cite{Reichenbach2} and section V, are phenomenologically equivalent. 

\section{Stochastic noise}

Intuition suggests that multiagent interactions should produce larger stochastic fluctuations than pairwise ones since they involve more fluctuating degrees of freedom. A non-trivial and surprising consequence is that strategies have to fight longer before one of them prevails on the others. As a result, the probability of coexistence increases. Using the Fokker-Planck equation, we derive an accurate estimate of the magnitude of the stochastic noise in our model. We follow Ref.~\cite{Reichenbach5}, where calculations are fully detailed. The idea is to consider a specific setting in which the RE predict neutrally stable orbits, namely $\deq = \dpol = 1$ in our case. For $N<\infty$ the conservation law $rps=\text{const}.$ is broken by $\text{O}(N^{-1})$ terms. Hence, the system follows an erratic trajectory, interpolating between different neutrally stable orbits. Eventually, it ends up on one of the absorbing fixed points. We can derive the intrinsic magnitude of the stochastic noise from the Master Equation
\begin{align}
  \partial_t \cP(\phi,t) & = \sum_{\delta\phi}\left\{w(\phi+\delta\phi\to\phi)\cP(\phi+\delta\phi,t)\right.\nonumber\\
  & - \left. w(\phi\to\phi+\delta\phi)\cP(\phi,t)\right\}\,,
  \label{eq:master}
\end{align}
where $\phi = (r,p)$ and $w(\phi\to\phi')$ denotes the transition probability per unit time (rate) from $\phi$ to $\phi'$. The sum over $\delta\phi$ includes all possible microscopic transitions characterizing the model. We choose conventionally the macroscopic time unit as the interval including $N$  reactions. In Table~\ref{tab:transprobs}, we list rates and density variations corresponding to this choice (one has to replace $s=1-r-p$ in all expressions).

\begin{table}[!h]
  \begin{center}
    \small
    \begin{tabular}{|c|c|c|c|}
      \hline
      reaction \raisebox{10pt} & $w$ & $N\delta r$ & $N\delta p$ \\[0.2ex]
      \hline\hline
      \ \ \redR\,\redR\,\greenP\ $\to$ \redR\,\greenP\,\greenP \raisebox{10pt} \ \ & \ \ $\deq Nr^2p$ \ \ & \ \ $-1$ \ \ & \ \ $\phantom{-}1$ \ \ \\
      \ \ \greenP\,\greenP\,\blueS\ $\to$ \greenP\,\blueS\,\blueS \raisebox{5pt} \ \ & $\deq Np^2s$ & $\phantom{-}0$ & $-1$ \\
      \ \ \blueS\,\blueS\,\redR\ $\to$ \blueS\,\redR\,\redR \raisebox{5pt} \ \ & $\deq Ns^2r$ & $\phantom{-}1$ & $\phantom{-}0$ \\
      \ \ \redR\,\redR\,\blueS\ $\to$ \redR\,\redR\,\redR \raisebox{5pt} \ \ & $\dpol Nr^2s$ & $\phantom{-}1$ & $\phantom{-}0$ \\
      \ \ \greenP\,\greenP\,\redR\ $\to$ \greenP\,\greenP\,\greenP \raisebox{5pt} \ \ & $\dpol Np^2r$ & $-1$ & $\phantom{-}1$ \\
      \ \ \blueS\,\blueS\,\greenP\ $\to$ \blueS\,\blueS\,\blueS \raisebox{5pt} \ \ & $\dpol Ns^2p$ & $\phantom{-}0$ & $-1$ \\
      \hline
    \end{tabular}
    \caption{\footnotesize Transition rates and density variations.\label{tab:transprobs}}
  \end{center}
\end{table}

Eq.~(\ref{eq:master}) yields an exact description for $N<\infty$. Unfortunately, we cannot solve it analytically. We can obtain an effective approximation by means of the Kramers–Moyal expansion, which, upon truncation to second order, yields the Fokker-Planck equation (FPE)
\begin{align}
  \partial_t\cP(\phi,t) & = -\sum_{i=\text{r},\text{p}}\partial_i\left[\alpha_i(\phi)\cP(\phi,t)\right] \nonumber\\[-1.0ex]
  & + \frac{1}{2}\sum_{i,j=\text{r},\text{p}}\partial_i\partial_j\left[B_{ij}(\phi)\cP(\phi,t)\right]\,.
\end{align}
The functions $\alpha_i$ and $B_{ij}$, respectively known as \emph{drift} and \emph{diffusion} functions, are given by
\begin{align}
  \alpha_i(\phi) & = \sum_{\delta\phi}\delta v_i w(\phi\to\phi+\delta\phi)\,,\\
  B_{ij}(\phi) & = \sum_{\delta\phi}\delta \phi_i\delta \phi_j w(\phi\to\phi+\delta\phi)\,.
\end{align}
From Table~\ref{tab:transprobs} it follows that
\begin{align}
  & \alpha_\text{r}(\phi) = r(s-p)\,,\\[1.0ex]
  & \alpha_\text{p}(\phi) = p(r-s)\,,
\end{align}
\begin{align}
  & B_{\text{rr}}(\phi) = \frac{1}{N}r(p+s-2ps)\,,\\[0.0ex]
  & B_{\text{rp}}(\phi) = B_{\text{pr}}(\phi) = -\frac{1}{N}rp(r+p)\,,\\[0.0ex]
  & B_{\text{pp}}(\phi) = \frac{1}{N}p(r+s-2rs)\,.
\end{align}
Just like in section II, we let $\phi = (1/3,1/3) + x$, then we expand the whole FPE around $x = 0$ (Van Kampen's linear noise approximation~\cite{VanKampen1}). Accordingly, we obtain
\begin{align}
  \partial_t\cP(x,t) & = -\sum_{i,j=\text{r},\text{p}}\partial_i[\cA_{ij}x_j\cP(x,t)] \nonumber\\[0.0ex]
  & + \frac{1}{2}\cB_{ij}\partial_i\partial_j\cP(x,t)\,,
\end{align}
with
\begin{equation}
  \cA = -\frac{1}{3}\left(\begin{array}{rr} \!\!1 & 2 \\ \!\!-2 & 1\end{array}\right)\,,\quad
    \cB = \frac{2}{27N}\left(\begin{array}{rr} \!\! 2 & -1 \\ -1 & 2 \end{array}\right)\,.
\end{equation}
As a final step, we perform another change of variables, namely we let $x\to y = \cS x$, with $\cS = \frac{\sqrt{3}}{2}\left(\begin{smallmatrix}2\omega_0 & \omega_0\\ 0 & 1\end{smallmatrix}\right)$. Accordingly, $\cA$ and $\cB$ turn into
\begin{align}
  &  \cA\to\tilde\cA = \cS\cA\cS^{-1} = \omega_0\left(\begin{array}{rr}0 & -1 \\ -1 & 0\end{array}\right)\,,\\[1.0ex]
&    \cB\to\tilde\cB = \cS\cB\cS^{\text{T}} = \frac{1}{9N}\left(\begin{array}{rr}1 & 0 \\ 0 & 1 \end{array}\right)\,.
\end{align}
The diffusion matrix $\tilde \cB$ is now diagonal. Hence, the FPE takes the simplified form
\begin{align}
  \partial_t\cP(y,t) & = -\omega_0[y_\text{p}\partial_\text{r} - y_\text{r}\partial_\text{p}]\cP(y,t)\nonumber\\[0.0ex]
  & + \frac{1}{18N}[\partial^2_\text{r} + \partial^2_\text{p}]\cP(y,t)\,.
  \label{eq:FPfinal}
\end{align}
The diffusion constant $D = 1/(18N)$  is smaller than the analogous constant in the original cyclic Lotka-Volterra model by a factor of~2/3, whereas $\omega_0$ is the same, see Eq.~(25) of Ref.~\cite{Reichenbach5}\footnote{The drift term in Eq.~(\ref{eq:FPfinal}) has an overall minus sign with respect to Eq.~(25) of Ref.~\cite{Reichenbach5}. It is due to the opposite invasion order: Ref.~\cite{Reichenbach5} assumes $a\to b\to c \to a$, while here we have $\redR\leftarrow \greenP\leftarrow \blueS\leftarrow \redR$.}. The reduction factor is rather suggestive, as it equals the ratio of simultaneous players in the two models. The value of 2/3 can be easily explained. The original model has three microscopic reactions, ours has twice this number. Each reaction contributes positively to the stochastic noise. This yields a multiplicative factor of 2 in the diffusion matrix. Moreover, the contribution from each reaction is quadratic in the original model while cubic in ours. In the linear noise approximation, we calculate the diffusion matrix at the reactive fixed point. This yields an additional multiplicative factor of 1/3 and that is all. In general, the more agents partake in the interactions, the larger the number of possible microscopic reactions. The magnitude of the induced stochastic noise depends eventually on combinatorial factors, including the number and degree of reactions. Of course, it depends as well on the value of the strategy densities at equilibrium. 

\begin{figure}[t!]
  \centering
  \includegraphics[width=0.38\textwidth]{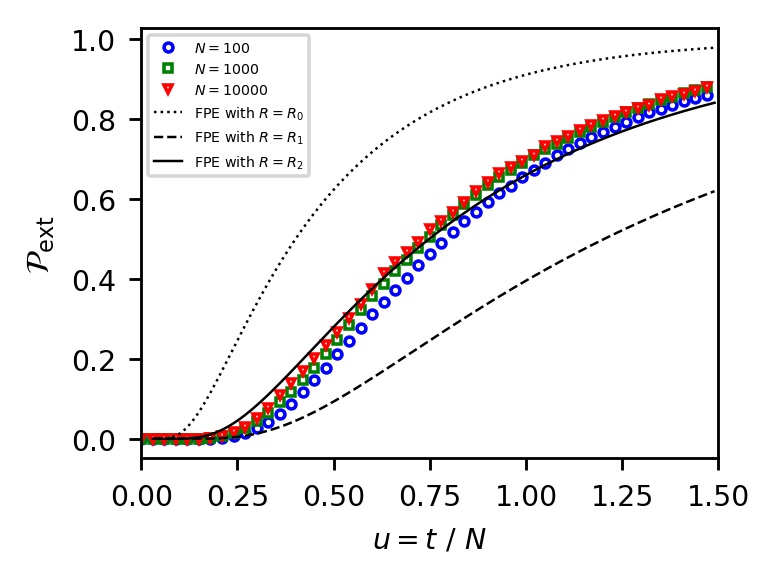}
  \caption{\footnotesize {\color{red} [color online]} Extinction probability.\label{fig:pext}}
\end{figure}

We can calculate the probability $\cP_{\rm ext}$ that two species go extinct after a certain time in the Fokker-Planck approximation as the authors of Ref.~\cite{Reichenbach5} do. There is no need to repeat the derivation here, since it applies identically to our model. The final formula is
\begin{equation}
  \text{LT}\{\cP_{\rm ext}(u)\} = \frac{1}{sI_0(R\sqrt{Ds})}\,,
  \label{eq:ltpext}
\end{equation}
where LT stands for Laplace Transform, $u=t/N$ is a scaling variable measuring time in units of $N$ and $I_0$ denotes the Bessel function of first kind and  0th order. $R$~represents the distance traveled by the system on its random walk from the reactive fixed point to one of the absorbing fixed points. We can regard it as the radius of an \emph{absorbing sphere}. The authors of Ref.~\cite{Reichenbach5} adopt three possible definitions of $R$, namely $R_0 = 1/3$, $R_1 = 1/\sqrt{3}$ and $R_2 = (R_0 + R_1)/2$. They yield three different probability functions. Fig.~\ref{fig:pext} shows them together with the results of numerical simulations based on Gillespie's algorithm~\cite{Gillespie1,Gillespie2}. To compute the extinction probability from Eq.~(\ref{eq:ltpext}) we use a numerical implementation of the inverse LT. Moreover, we expand $I_0$ asymptotically to the 10th order, as also the authors of Ref.~\cite{Reichenbach5} do. To simulate extinction times correctly, we need to rescale all rates by appropriate volume factors. Notice that all reactions in our model involve two reactants of the same species and one of another. For such reactions, the right definition of the reaction parameters $c_\text{XXY}$ in Gillespie's algorithm is $c_\text{XXY} = 2d_\text{XXY}/N^2$, for $\text{X,Y}\in\{\redR,\greenP,\blueS\}$~\cite{Gillespie1,Gillespie2}. Similar to the original Lotka-Volterra model, the analytic prediction that best fits numerical data is the one corresponding to~$R_2$.   

As anticipated, the extinction probability is uniformly lower in the cyclic Lotka-Volterra model with three-agent interactions than with two-agent ones, although the former are intrinsically noisier than the latter. This result is only apparently counterintuitive. In fact, it has a simple interpretation. When three agents interact, strategies fluctuate longer around the reactive fixed point before one of them prevails on the others. Group interactions help the system stay in equilibrium. Hence, we conclude, they promote species coexistence.

\vbox{\vskip -0.0cm}

\section{Heterogeneous rates}

The reactive fixed point is symmetric for homogeneous rates, independently of whether $\deq>\dpol$ or $\deq<\dpol$. Things become more interesting as soon as we break homogeneity. Unfortunately, studying the model in full generality is not simple, due to the high dimensionality of the parameter space. A reasonable compromise is to let $\dRRP = \dPPS = \dSSR \equiv \deq = 1$ and to leave all other rates unconstrained. Under this assumption, the RE still exhibit four fixed points, three absorbing plus one reactive. The former coincide with the vertices of the ternary diagram, as can be easily checked from Eqs.~(\ref{eq:rateeqs}). The latter has a complex algebraic structure. Indeed, it reads  
\begin{align}
  \left.\begin{array}{l}
  r_* = \frac{(\dSSP-\dPPR)X^2 + (1+2\dPPR)X - \dPPR}{1-\dPPR+(\dPPR-\dRRS)X}\,,\\[3.0ex]
  p_* = \frac{(\dRRS-\dSSP)X^2 - (2+\dRRS)X + 1}{1-\dPPR+(\dPPR-\dRRS)X}\,,
  \end{array}\right.
  \label{eq:heterofixed}
\end{align}
\vskip 0.cm
with $X$ fulfilling the cubic equation
\begin{widetext}
  \small
  \begin{align}
    0 & = a_3X^3 + a_2X^2 + a_1X + a_0\,,\nonumber\\[2.0ex]
    a_0 & = 1-\dRRS\,\dPPR^2\,,\nonumber\\[1.0ex]
    a_1 & = -2\,\dRRS+\dPPR-3-\dSSP\,\dPPR+2\,\dRRS\,\dPPR+\dSSP+3\,\dRRS\,\dPPR^2-\dPPR^2\,,\\[1.0ex]
    a_2 & = 3\,\dRRS-3\,\dSSP-2\,\dRRS\,\dSSP-3\,\dRRS\,\dPPR-3\,\dRRS\,\dPPR^2+\dRRS^2 + 3\,\dRRS\,\dPPR\,\dSSP \nonumber\\[0.0ex]
    & \hskip 0.40cm +\,2\,\dSSP\,\dPPR +2\,\dPPR^2\,,\nonumber\\[0.0ex]
    a_3 & = \dSSP^2\dPPR+\dSSP\,\dPPR+\dRRS\,\dSSP-3\,\dRRS\,\dPPR\,\dSSP+\dRRS\,\dPPR^2 +\dRRS\,\dPPR-\dSSP^2-\dRRS^2\nonumber\\[0.0ex]
    & \hskip 0.40cm -\,\dPPR^2 +\dRRS^2\dSSP\,.\nonumber
  \end{align}
\end{widetext}
\begin{figure*}[t]
    \centering
    \begin{adjustbox}{varwidth=\textwidth,center}
      \centering
      \vbox{\vskip 0.3cm}
      \begin{subfigure}{0.23\textwidth}
        \includegraphics[width=35mm]{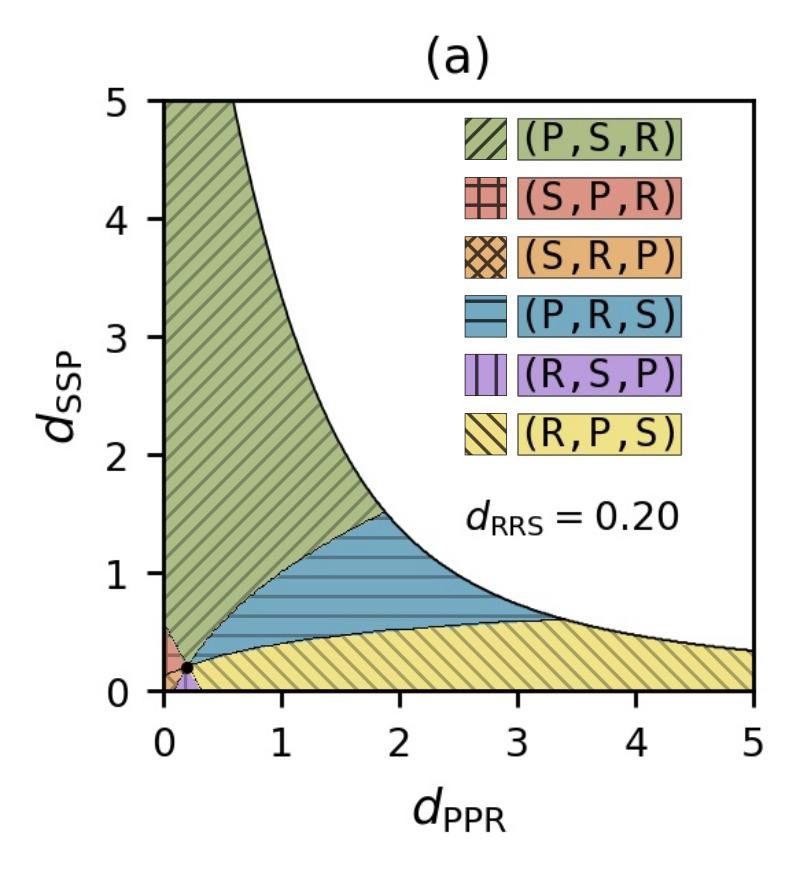}
        \vskip -0.3cm
        \caption*{}
      \end{subfigure}
      \begin{subfigure}{0.23\textwidth}
        \includegraphics[width=35mm]{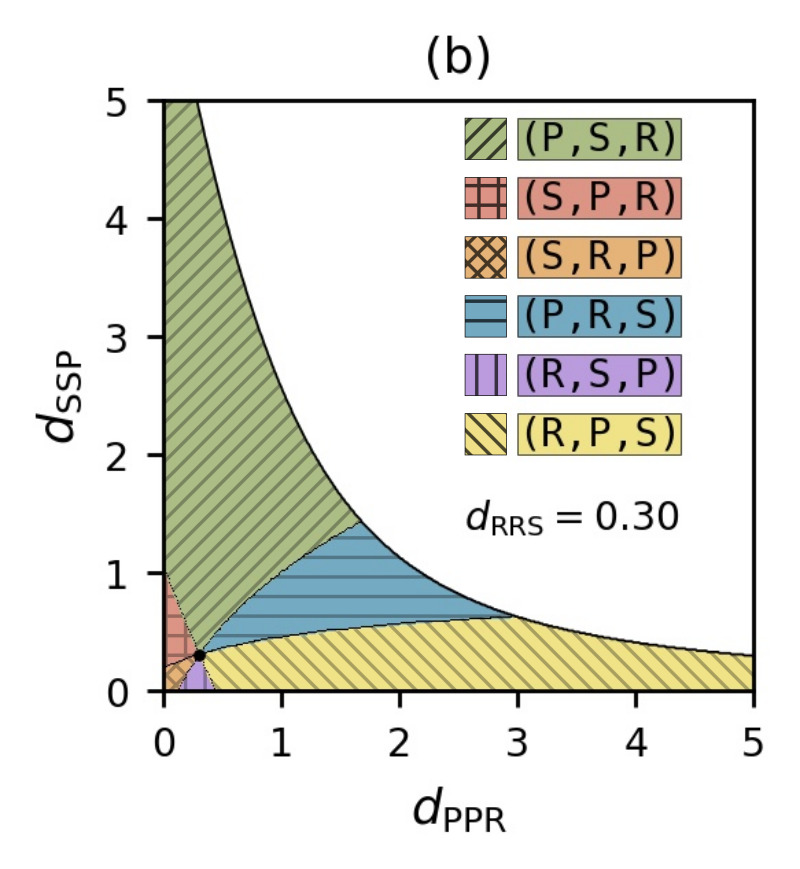}
        \vskip -0.3cm
        \caption*{}
      \end{subfigure}
      \begin{subfigure}{0.23\textwidth}
        \includegraphics[width=35mm]{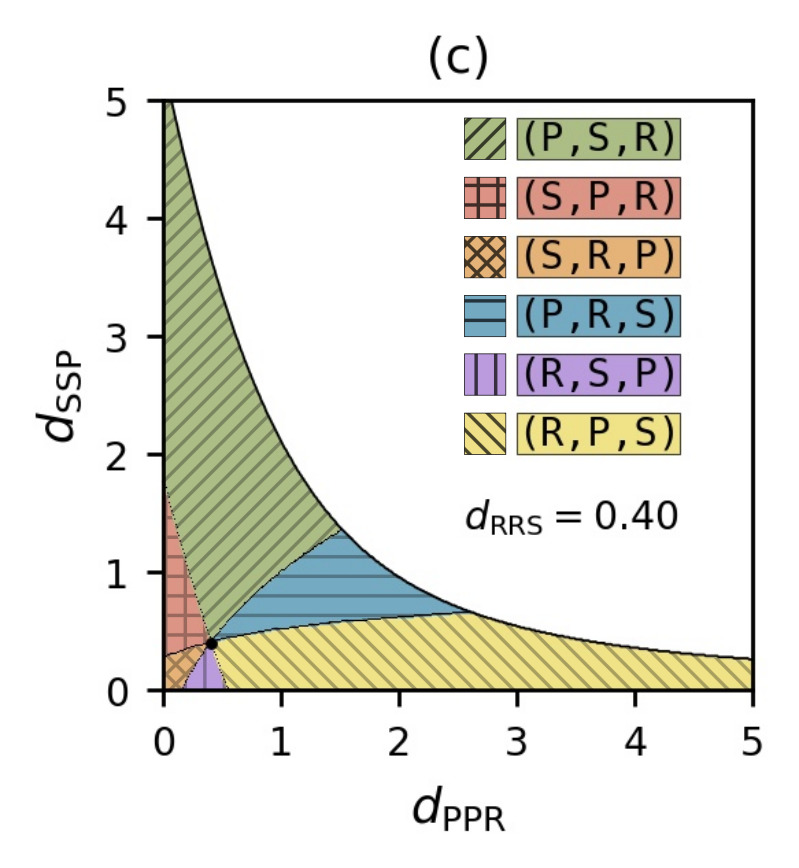}
        \vskip -0.3cm
        \caption*{}
      \end{subfigure}
      \begin{subfigure}{0.23\textwidth}
        \includegraphics[width=35mm]{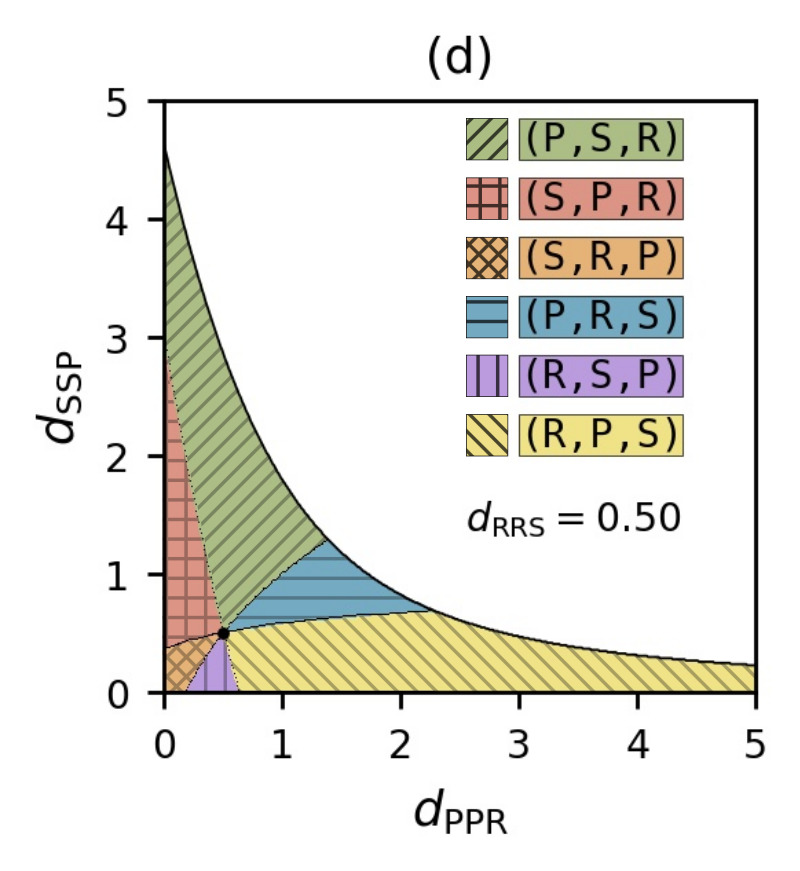}
        \vskip -0.3cm
        \caption*{}
      \end{subfigure}
      \\[-4.0ex]
      \begin{subfigure}{0.23\textwidth}
        \includegraphics[width=35mm]{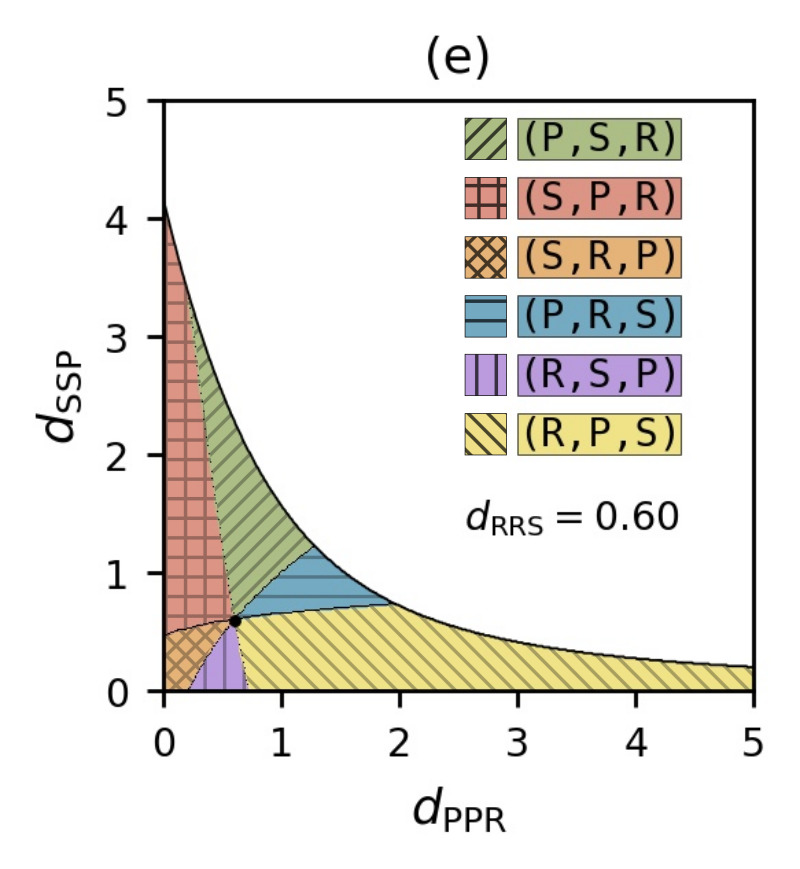}
        \vskip -0.3cm
        \caption*{}
      \end{subfigure}
      \begin{subfigure}{0.23\textwidth}
        \includegraphics[width=35mm]{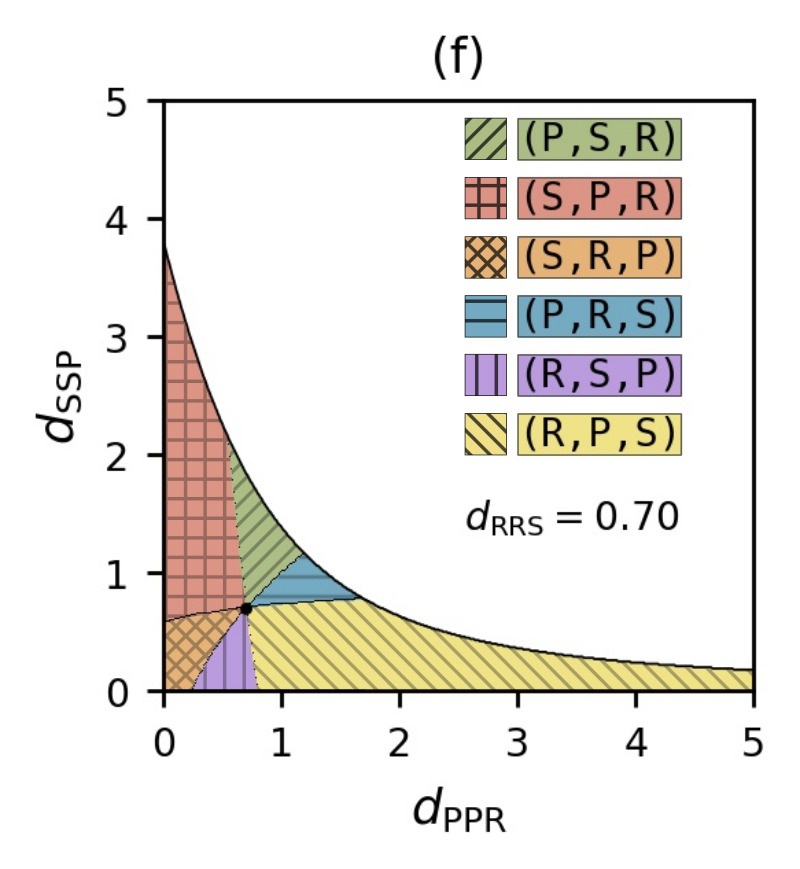}
        \vskip -0.3cm
        \caption*{}
      \end{subfigure}
      \begin{subfigure}{0.23\textwidth}
        \includegraphics[width=35mm]{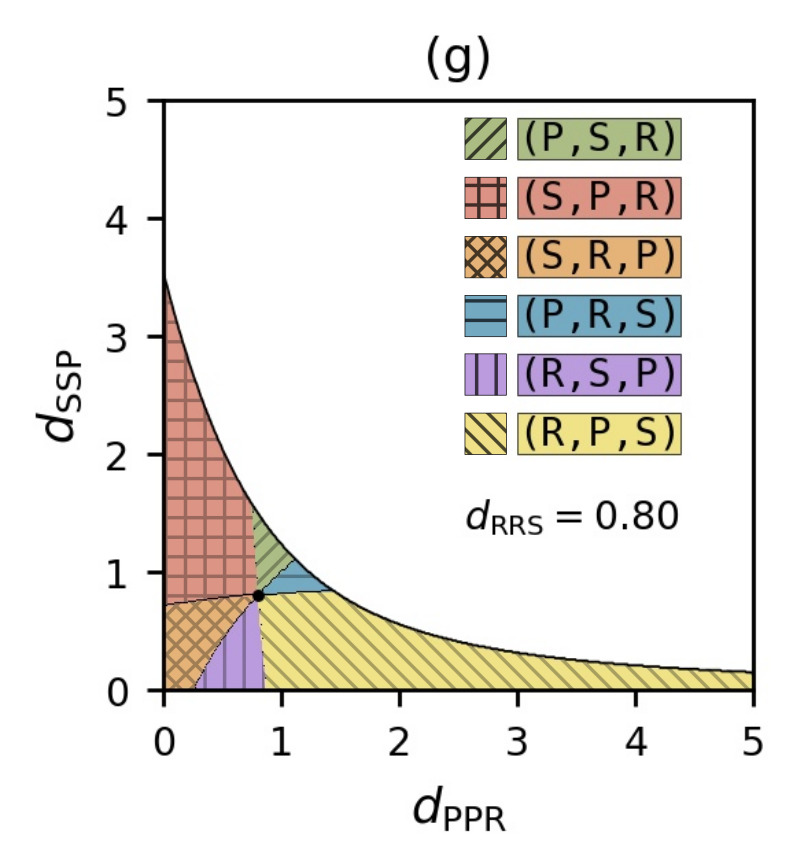}
        \vskip -0.3cm
        \caption*{}
      \end{subfigure}
      \begin{subfigure}{0.23\textwidth}
        \includegraphics[width=35mm]{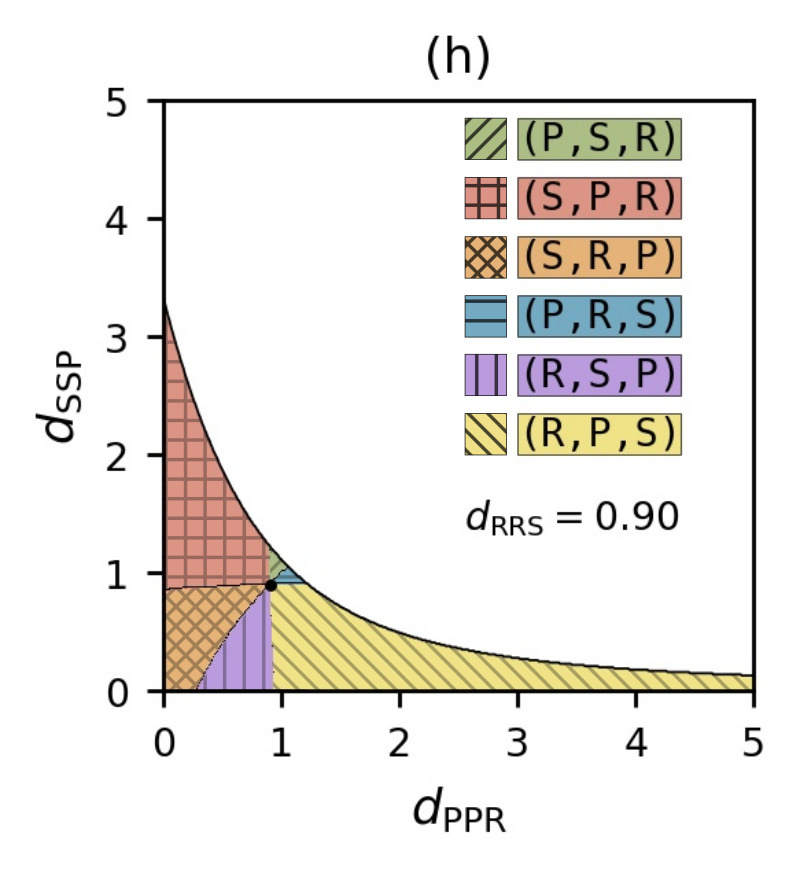}
        \vskip -0.3cm
        \caption*{}
      \end{subfigure}
      \\[-4.0ex]
      \begin{subfigure}{0.23\textwidth}
        \includegraphics[width=35mm]{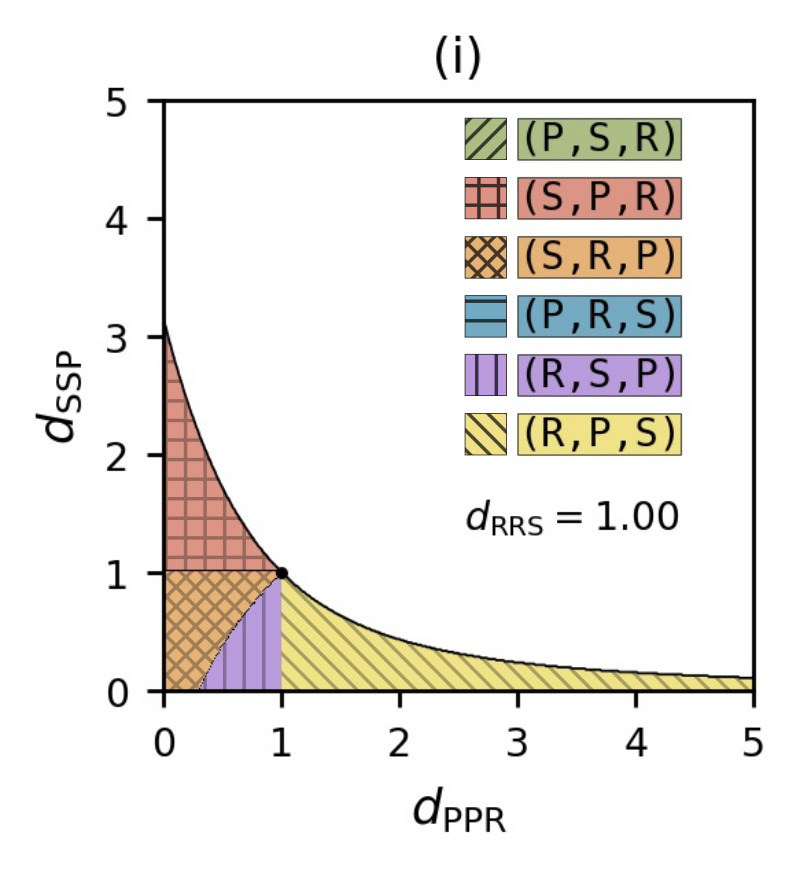}
        \vskip -0.3cm
        \caption*{}
      \end{subfigure}
      \begin{subfigure}{0.23\textwidth}
        \includegraphics[width=35mm]{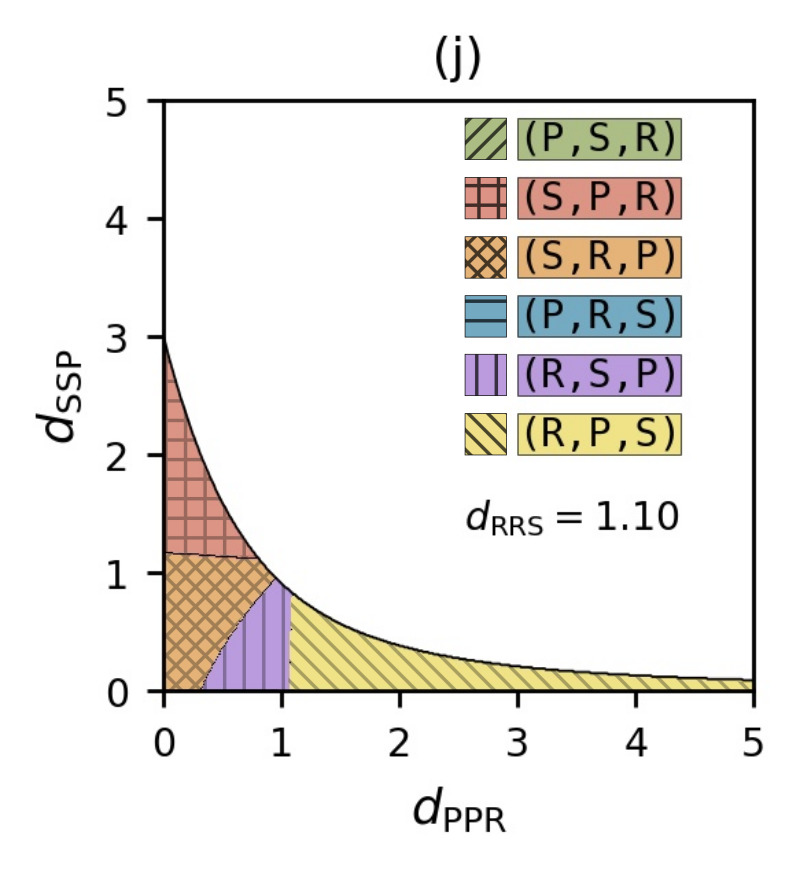}
        \vskip -0.3cm
        \caption*{}
      \end{subfigure}
      \begin{subfigure}{0.23\textwidth}
        \includegraphics[width=35mm]{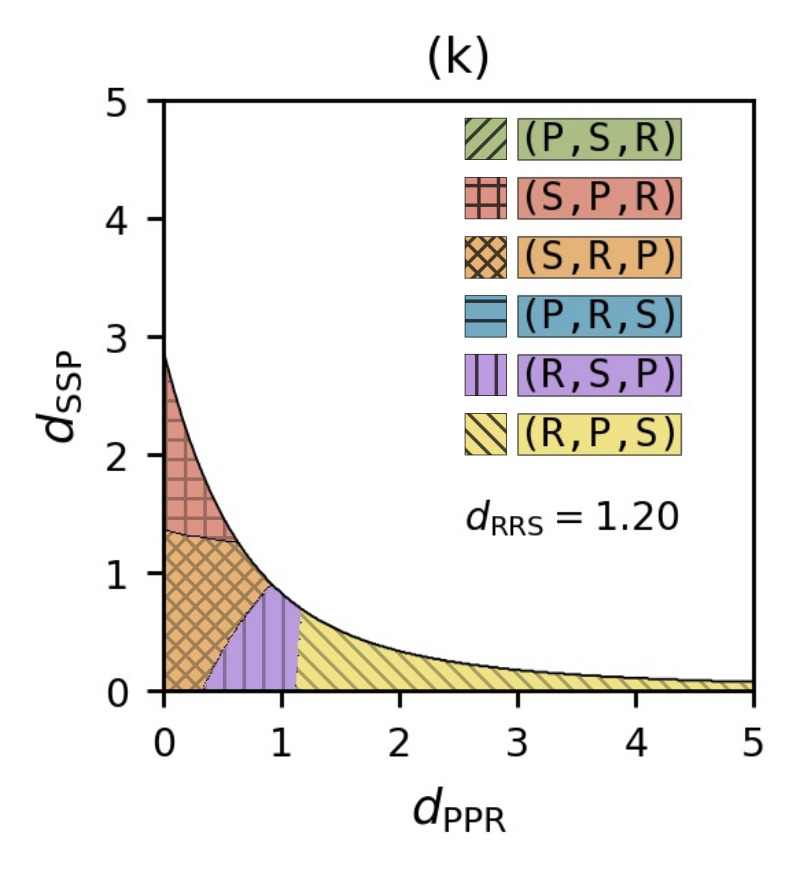}
        \vskip -0.3cm
        \caption*{}
      \end{subfigure}
      \begin{subfigure}{0.23\textwidth}
        \includegraphics[width=35mm]{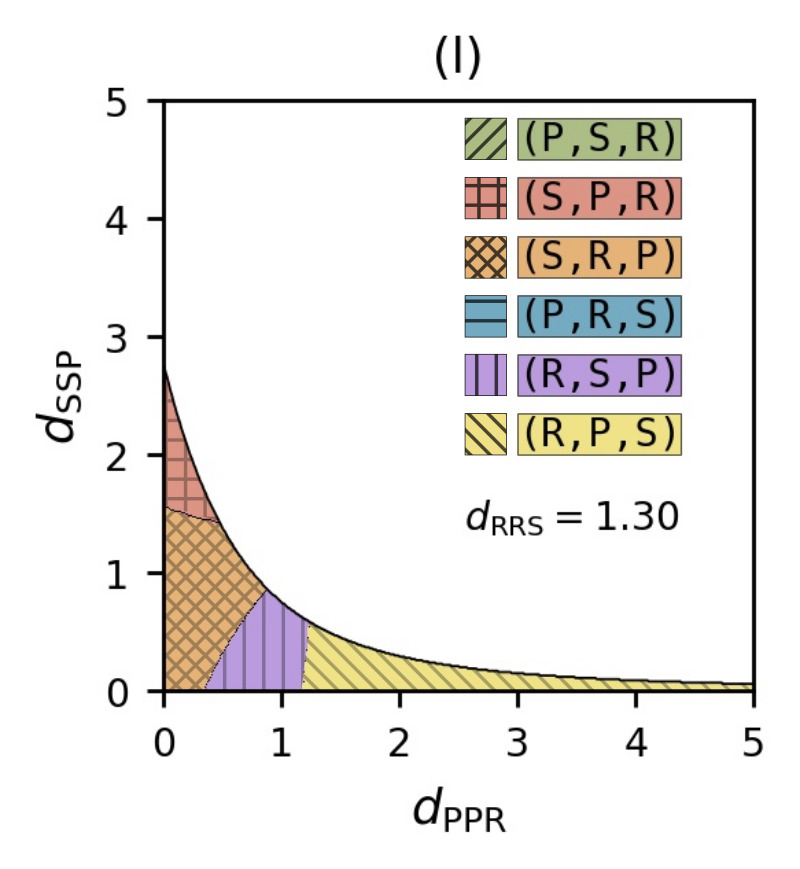}
        \vskip -0.3cm
        \caption*{}
      \end{subfigure}
      \\[-4.0ex]
      \begin{subfigure}{0.23\textwidth}
        \includegraphics[width=35mm]{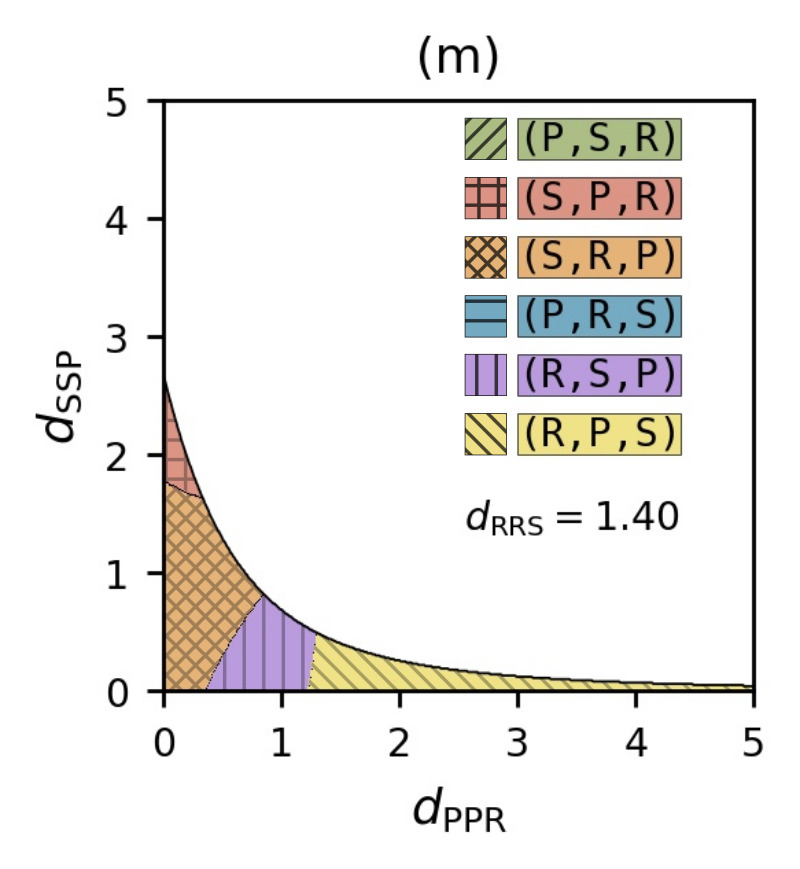}
        \vskip -0.3cm
        \caption*{}
      \end{subfigure}
      \begin{subfigure}{0.23\textwidth}
        \includegraphics[width=35mm]{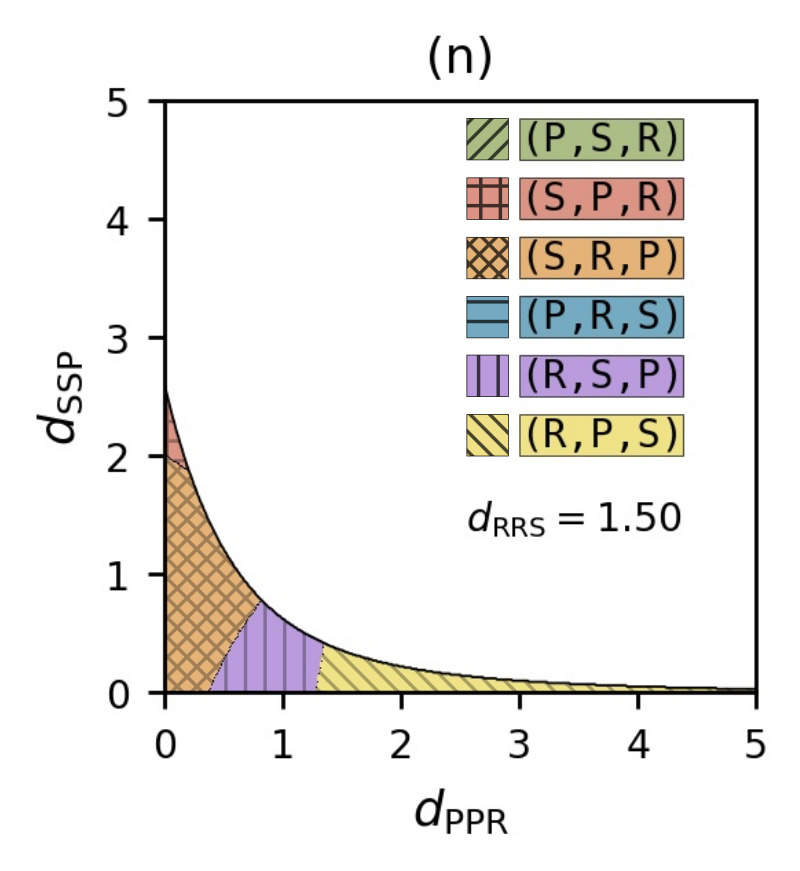}
        \vskip -0.3cm
        \caption*{}
      \end{subfigure}
      \begin{subfigure}{0.23\textwidth}
        \includegraphics[width=35mm]{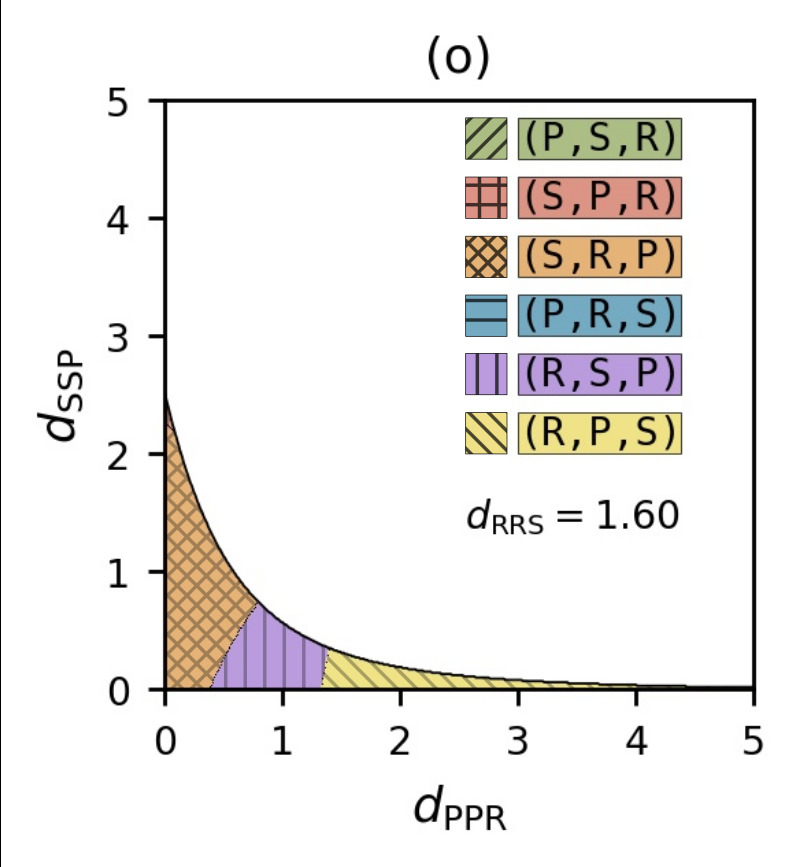}
        \vskip -0.3cm
        \caption*{}
      \end{subfigure}
      \begin{subfigure}{0.23\textwidth}
        \includegraphics[width=35mm]{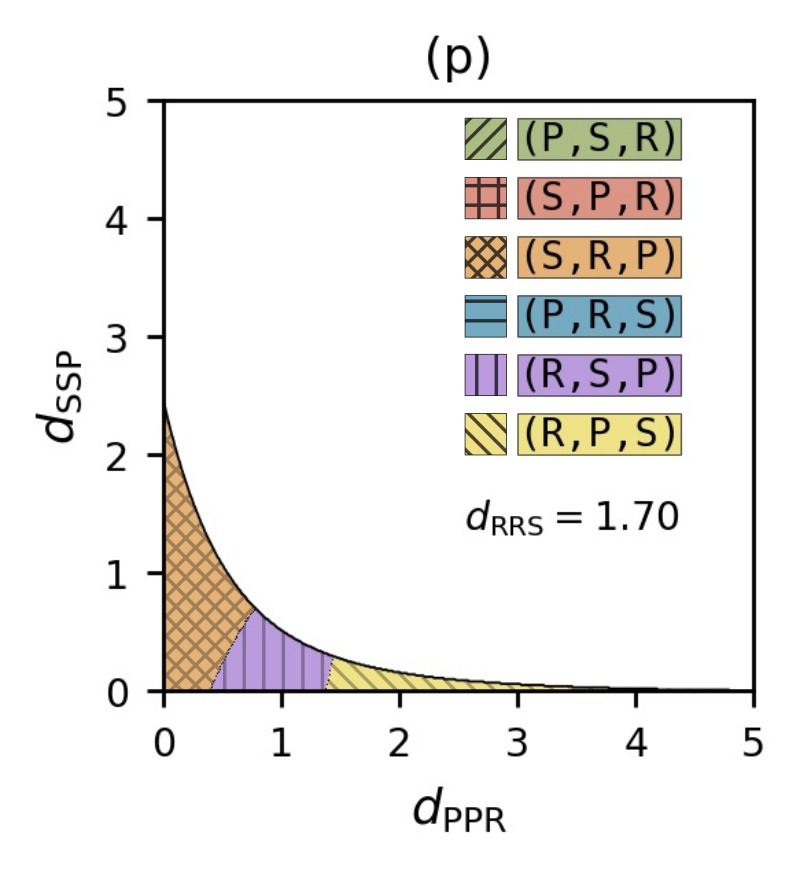}
        \vskip -0.3cm
        \caption*{}
      \end{subfigure}
      \\[-4.0ex]
      \begin{subfigure}{0.23\textwidth}
        \includegraphics[width=35mm]{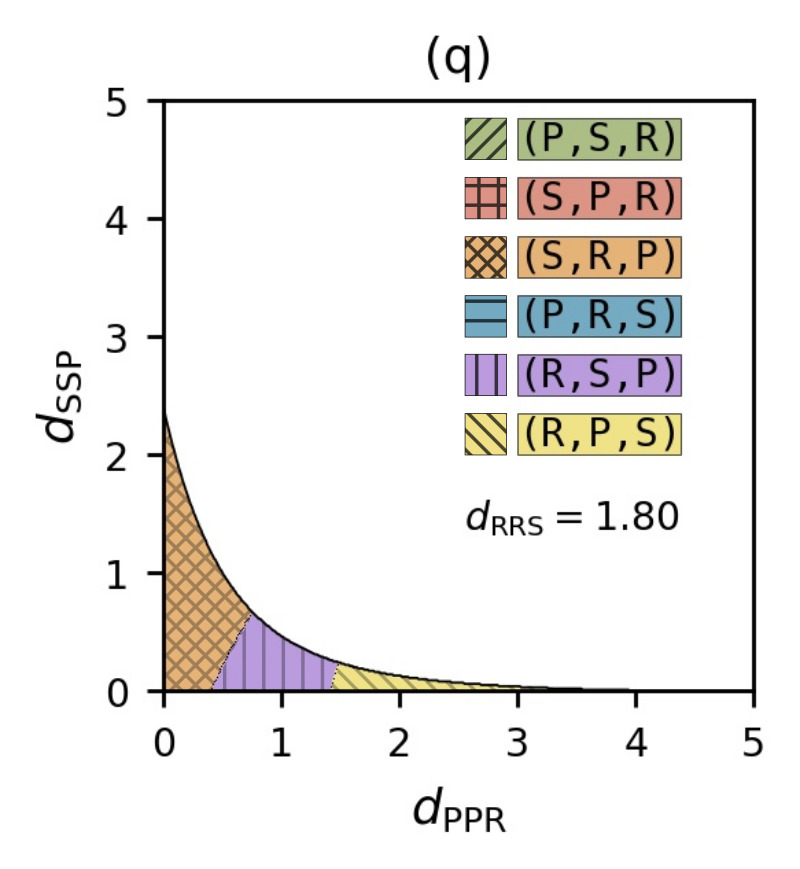}
        \vskip -0.3cm
        \caption*{}
      \end{subfigure}
      \begin{subfigure}{0.23\textwidth}
        \includegraphics[width=35mm]{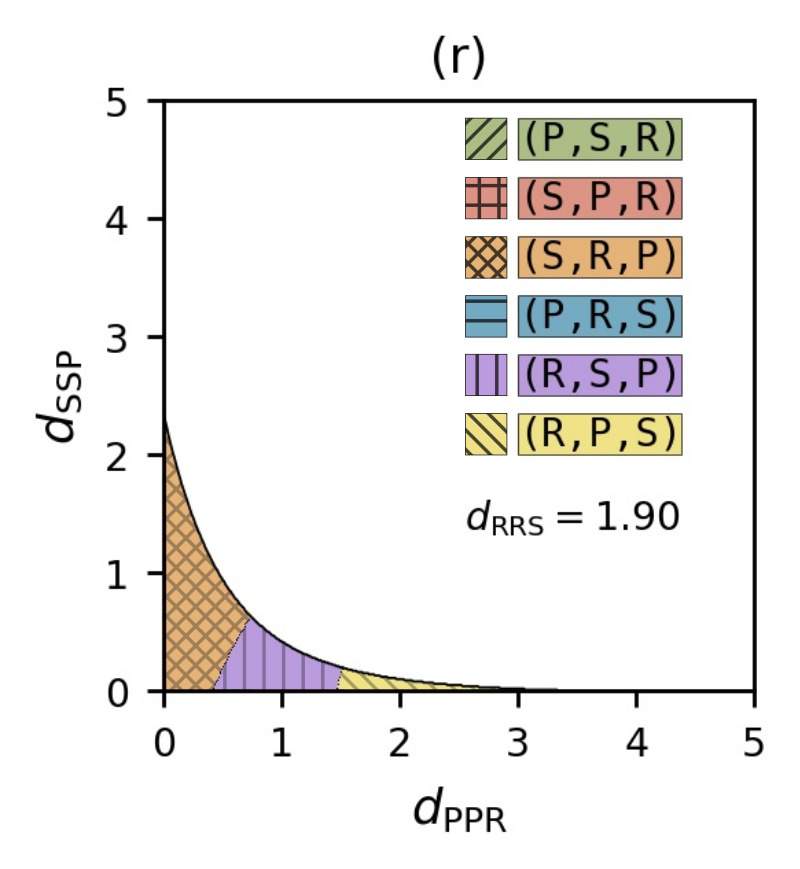}
        \vskip -0.3cm
        \caption*{}
      \end{subfigure}
      \begin{subfigure}{0.23\textwidth}
        \includegraphics[width=35mm]{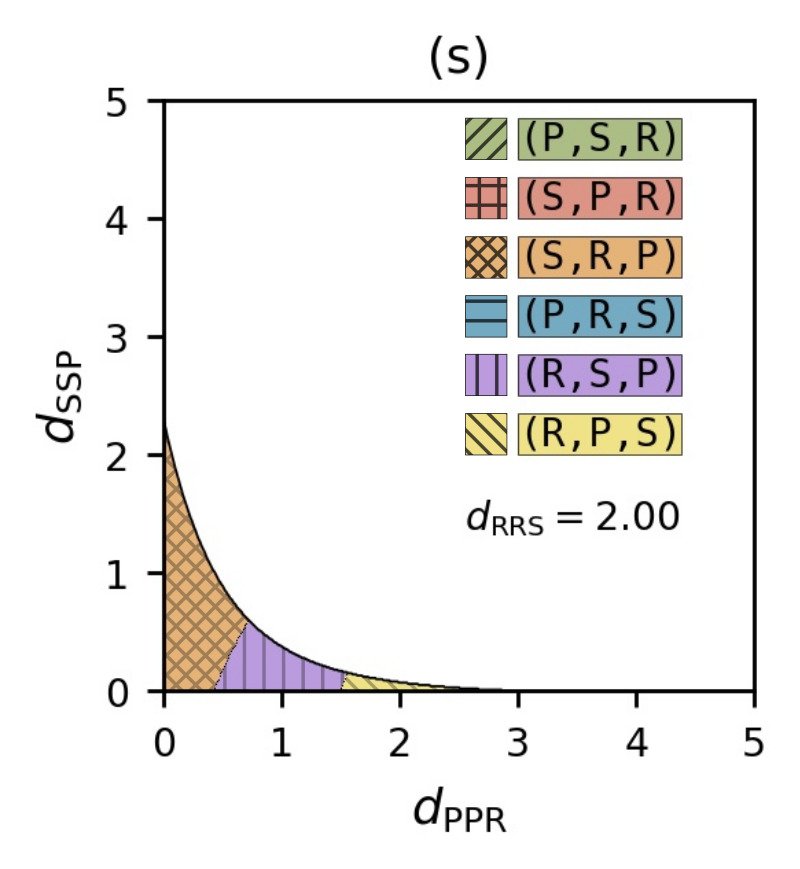}
        \vskip -0.3cm
        \caption*{}
      \end{subfigure}
      \begin{subfigure}{0.23\textwidth}
        \includegraphics[width=35mm]{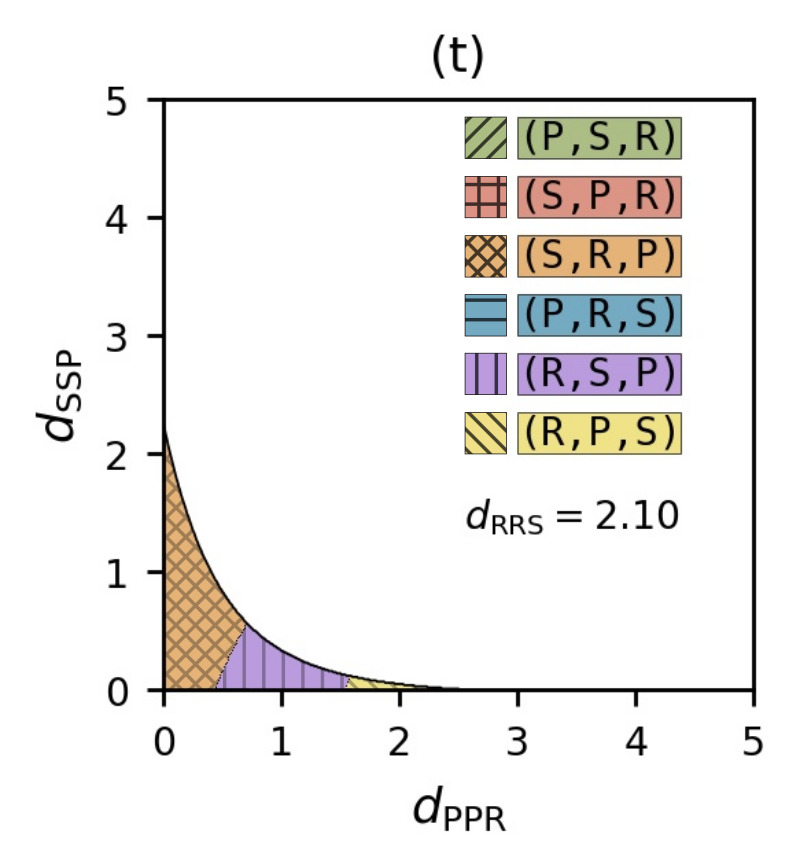}
        \vskip -0.3cm
        \caption*{}
      \end{subfigure}
    \end{adjustbox}
    \vskip -0.1cm
    \caption{\footnotesize {\color{red} [color online]} Phase structure of the model in a well-mixed environment for heterogeneous rates.\label{fig:phases}}
    \vskip 0.2cm
\end{figure*}

We wish to examine how strategies rank at equilibrium depending on the polarizing rates. To this aim, we proceed like the authors of Ref.~\cite{Szolnoki2}. Specifically, we denote by ($\tt A$,$\tt B$,$\tt C$) a domain in parameter space for which $a_*\le b_*\le c_*$, for $a_*,b_*,c_*$ a permutation of $r_*,p_*,s_*\equiv 1-r_*-p_*$ and $\tt A$, $\tt B$, $\tt C$ the corresponding permutation of \redR, \greenP, \blueS. In accordance with Ref.~\cite{Szolnoki2}, we adopt the name of \emph{phase} for all such domains, although this word is somewhat misleading (at least in our case). Indeed, both the equilibrium densities and their derivatives are smooth functions. In particular, they do not show discontinuities at phase transitions. In Fig.~\ref{fig:phases}, we report the phase structure for a sequence of values of $\dRRS$, ranging from 0.2 to 2.1 in steps of 0.1. In each plot, we notice a white region and a variously colored one. The former corresponds to unstable spirals, diverging from the reactive fixed point (they eventually turn into heteroclinic cycles), whereas the latter corresponds to stable spirals, converging to it. A continuum of Hopf bifurcations lies along the lines separating colored and white regions, as we discuss below. Colored regions split into six phases or less, depending on $\dRRS$. Each phase makes contact with all others for $\dPPR=\dSSP=\dRRS$. The contact point lies in the stable region only for $\dRRS\le 1$. The overall surface occupied by stable phases reduces as $\dRRS$ increases (the larger $\dRRS$ the stronger the polarizing force corresponding to fixed values of $\dPPR$ and $\dSSP$).

A glance to plot~(i), corresponding to $\dRRS = 1$, suggests a recipe for predicting the phase of the system for given values of $\dRRS,\dPPR,\dSSP$. It consists of three steps:
\begin{enumerate}[leftmargin=4mm]
  \setlength\itemsep{-1.0ex}
\item[$1.$]{rank the polarizing rates in ascending order (e.g. $\dPPR<\dRRS<\dSSP$);}
\item[$2.$]{extract the losing strategy from each rate label (it yields $(\redR,\blueS,\greenP)$ in the above example);}
\item[$3.$]{turn each strategy into the immediately inferior one (it finally yields $(\blueS,\greenP,\redR)$).}
\end{enumerate}
\vskip -0.1cm
The recipe has a straightforward interpretation: the more a species is preyed on by its predator, the more its prey has room to develop. Its main drawback is that it is only approximately exact due to nonlinear effects. For instance, the transition line separating $(\blueS,\redR,\greenP)$ from $(\redR,\blueS,\greenP)$ should bisect the phase plane for $\dRRS=1$, while it does not. Moreover, all other transition lines develop a slope for $\dRRS\ne 1$. We notice that by no reason the phases of Fig.~\ref{fig:phases} should be symmetric for $\dSSP\leftrightarrow\dPPR$. The only symmetry of the system is the cyclic one, namely $\redR\leftarrow\greenP\leftarrow\blueS\leftarrow\redR$. If we fix $\dRRS$, we lose it.

Altogether, the phase structure looks more complex than observed in Ref.~\cite{Szolnoki2}, although our model features simpler multiagent interactions. As far as we understand, the rationale behind this is that we consider fully independent rates, while the authors of Ref.~\cite{Szolnoki2} build multiagent rates as accumulated payoffs, depending on the pairwise rates. As a consequence, they explore a two-dimensional manifold embedded in a larger and more complex phase space.

We can provide a better characterization of the transition lines separating white and colored regions. The Jacobian of the linearized RE has two complex conjugate eigenvalues $\lambdar \pm\text{i} \lambdai$. We can express both the real and the imaginary part as functions of the fixed-point densities, namely
\begin{align}
  \lambdar & = \frac{1}{2}(\dRRP-\dRRS)r_*^2 + \frac{1}{2}(\dPPS-\dPPR)p_*^2 \nonumber\\[0.0ex]
  & + \frac{1}{2}(\dSSR-\dSSP)s_*^2  + (\dPPR-\dRRP)r_*p_* \nonumber\\[0.7ex]
  & + (\dSSP-\dPPS)p_*s_* + (\dRRS-\dSSR)r_*s_*\,,
\end{align}
\begin{align}
  \lambdai & = \frac{1}{2}\{c_\text{r4}r_*^4 + c_\text{p4}p_*^4 + c_\text{s4}s_*^4\nonumber\\[0.5ex]
  & + c_\text{r3p1}r_*^3p_* + c_\text{r3s1}r_*^3s_* + c_\text{p3r1}p_*^3r_*\nonumber\\[0.5ex]
  & + c_\text{p3s1}p_*^3s_* + c_\text{s3r1}s_*^3r_* + c_\text{s3p1}s_*^3p_*\nonumber
\end{align}
\begin{align}
  & + c_\text{r2p2}r_*^2p_*^2 + c_\text{r2s2}r_*^2s_*^2 + c_\text{p2s2}p_*^2s_*^2\nonumber\\[1.0ex]
  & + c_\text{r2ps}r_*^2p_*s_* + c_\text{p2sr}p_*^2s_*r_* + c_\text{s2rp}s_*^2r_*p_*\}^{1/2}\,,
\end{align}
with coefficients reading
\begin{widetext}
  \small
  \begin{align}
    c_\text{r4} & = (\dRRP + \dRRS)^2\,,\qquad c_\text{p4} = (\dPPS + \dPPR)^2\,,\quad c_\text{s4} = (\dSSR + \dSSP)^2\,, \nonumber\\[2.0ex]
    c_\text{r3p1} & = 4(\dRRS + \dRRP)(\dPPR-\dRRP)\,,\quad c_\text{r3s1} = 4(\dRRP + \dRRS)(\dSSR-\dRRS)\,, \nonumber\\[0.0ex]
    c_\text{p3r1} & = 4(\dPPS + \dPPR)(\dRRP-\dPPR)\,,\quad c_\text{p3s1} = 4(\dPPR + \dPPS)(\dSSP-\dPPS)\,, \nonumber\\[0.0ex]
    c_\text{s3r1} & = 4(\dSSR + \dSSP)(\dRRS-\dSSR)\,,\quad c_\text{s3p1} = 4(\dSSP + \dSSR)(\dPPS-\dSSP)\,, \nonumber\\[2.0ex]
    c_\text{r2p2} & = 4\dRRP^2 + 4\dPPR^2 - 2\dRRS\dPPR - 2\dRRS\dPPS - 2\dRRP\dPPS - 10\dRRP\dPPR\,, \\[0.0ex]
    c_\text{r2s2} & = 4\dRRS^2 + 4\dSSR^2 - 2\dRRS\dSSP - 2\dRRP\dSSP - 2\dRRP\dSSR - 10\dRRS\dSSR\,,  \nonumber\\[0.0ex]
    c_\text{p2s2} & = 4\dPPS^2 + 4\dSSP^2 - 2\dPPS\dSSR - 2\dPPR\dSSR - 2\dPPR\dSSP - 10\dPPS\dSSP\,,  \nonumber\\[2.0ex]
    c_\text{r2ps} & = -4 (\dRRS \dPPS - 2 \dPPR \dSSR + 2 \dPPR \dRRS + 2 \dSSR \dRRP  + \dSSP \dRRP + \dRRS \dSSP + 2 \dRRS \dRRP + \dPPS \dRRP)\,,\nonumber\\[0.0ex]
    c_\text{p2sr} & = -4 (2 \dPPR \dSSP + \dRRS \dPPS - 2 \dSSP \dRRP + 2 \dPPS \dRRP + \dPPR \dSSR + \dSSR \dPPS + 2 \dPPR \dPPS + \dPPR \dRRS)\,,\nonumber\\[0.0ex]
    c_\text{s2rp} & = -4 (2 \dRRS \dSSP + 2 \dSSR \dPPS + \dSSP \dRRP + \dPPR \dSSR - 2 \dRRS \dPPS + 2 \dSSR \dSSP + \dSSR \dRRP + \dPPR \dSSP)\,.\nonumber
  \end{align}
\end{widetext}
Both $(r_*,p_*)$ and $\lambdar$ depend nonlinearly upon $\dRRS\,,\dPPR\,,\dSSP$. Any point, belonging to a transition line separating white and colored regions in Fig.~\ref{fig:phases}, corresponds by definition to a reactive fixed point yielding $\lambdar=0$. Keeping $(r_*,p_*)$ fixed means imposing two constraints. Since we have three degrees of freedom, we remain with one. We conclude that there is a one-dimensional (non-planar) manifold, characterized by $(r_*,p_*)=\text{const.}$, crossing the transition line at the chosen point. All other points belonging to this manifold correspond to reactive fixed points yielding $\lambdar\ne 0$. Apart from possible exceptions (that we never observed in our numerical experiments), the manifold splits into two parts, one having $\lambdar>0$, the other $\lambdar<0$. Therefore, the system undergoes a Hopf bifurcation along the manifold for $\lambdar=0$. Hence, the RE exhibit a continuum of Hopf bifurcations at all transition lines of Fig.~\ref{fig:phases}. Unfortunately, we have been unable to clarify whether RE at the bifurcation points have first integrals, like $rps$ for homogeneous rates and, if affirmative, how they look analytically.

\begin{figure*}[!t]
  \centering
  \begin{adjustbox}{varwidth=\textwidth,center}
  \centering
    \vbox{\vskip 0.3cm}
    \hskip 0.5cm
    \begin{subfigure}{0.31\textwidth}
      \includegraphics[width=36mm]{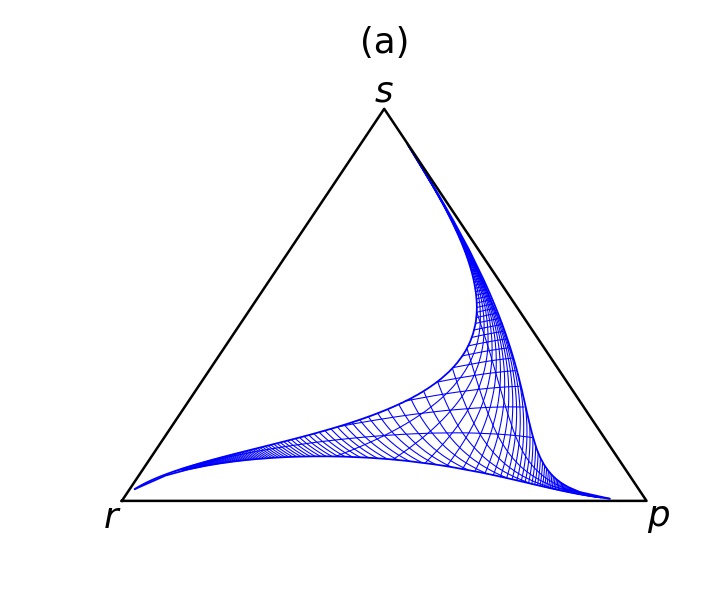}
      \vskip -0.3cm
      \caption*{\footnotesize \hskip 0.2cm $d_\text{RRP}={\color{red}{\mathbf{0.1}}},\ d_\text{PPS}={\color{red}\mathbf{0.1}},\ d_\text{SSR}={\color{red}\mathbf{1.0}}$}
    \end{subfigure}
    \begin{subfigure}{0.31\textwidth}
      \includegraphics[width=36mm]{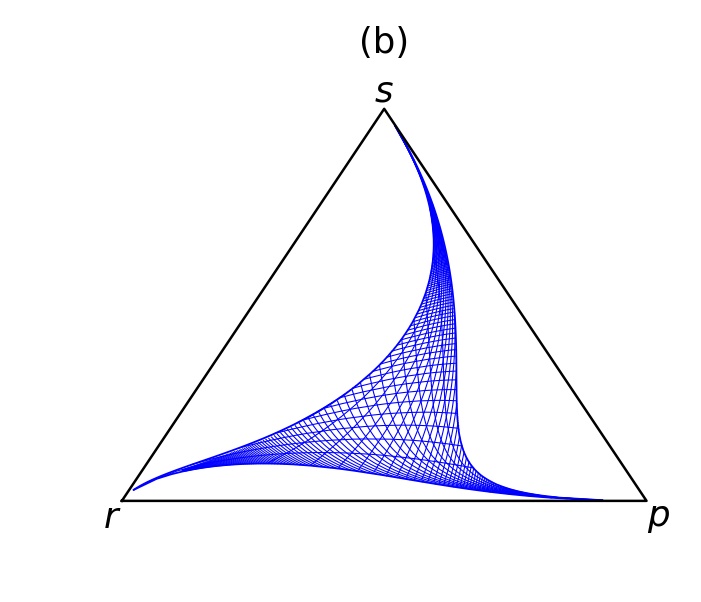}
      \vskip -0.3cm
      \caption*{\footnotesize \hskip 0.2cm $d_\text{RRP}={\color{red}{\mathbf{0.1}}},\ d_\text{PPS}={\color{red}\mathbf{0.5}},\ d_\text{SSR}={\color{red}\mathbf{1.0}}$}
    \end{subfigure}
    \begin{subfigure}{0.31\textwidth}
      \includegraphics[width=36mm]{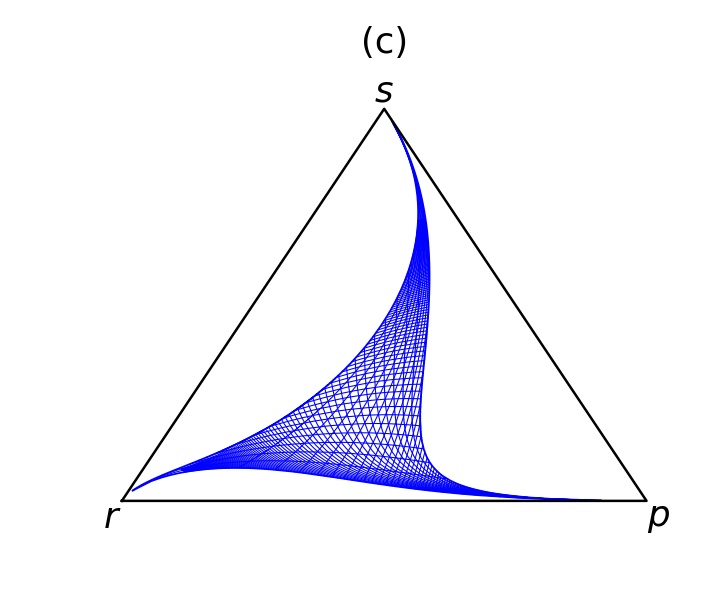}
      \vskip -0.3cm
      \caption*{\footnotesize \hskip 0.2cm $d_\text{RRP}={\color{red}{\mathbf{0.1}}},\ d_\text{PPS}={\color{red}\mathbf{1.0}},\ d_\text{SSR}={\color{red}\mathbf{1.0}}$}
    \end{subfigure}
    \\[0.0ex]
    \hskip 0.5cm
    \begin{subfigure}{0.31\textwidth}
      \includegraphics[width=36mm]{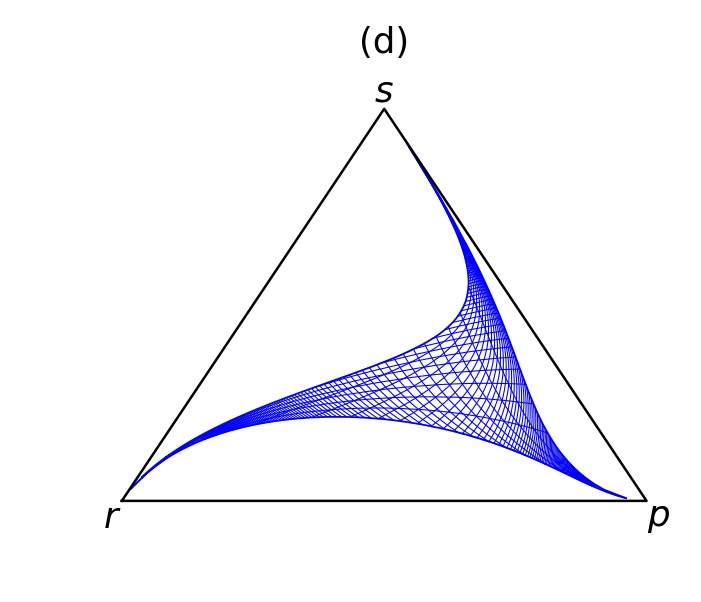}
      \vskip -0.3cm
      \caption*{\footnotesize \hskip 0.2cm $d_\text{RRP}={\color{red}{\mathbf{0.5}}},\ d_\text{PPS}={\color{red}\mathbf{0.1}},\ d_\text{SSR}={\color{red}\mathbf{1.0}}$}
    \end{subfigure}
    \begin{subfigure}{0.31\textwidth}
      \includegraphics[width=36mm]{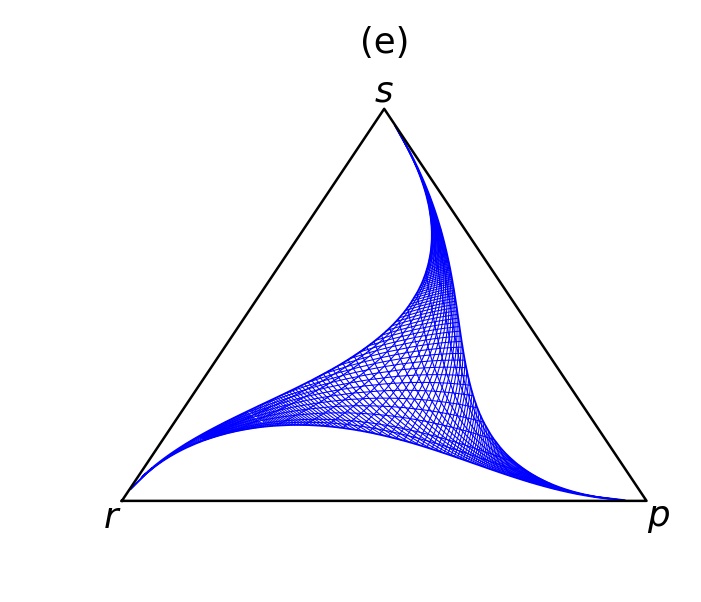}
      \vskip -0.3cm
      \caption*{\footnotesize \hskip 0.2cm $d_\text{RRP}={\color{red}{\mathbf{0.5}}},\ d_\text{PPS}={\color{red}\mathbf{0.5}},\ d_\text{SSR}={\color{red}\mathbf{1.0}}$}
    \end{subfigure}
    \begin{subfigure}{0.31\textwidth}
      \includegraphics[width=36mm]{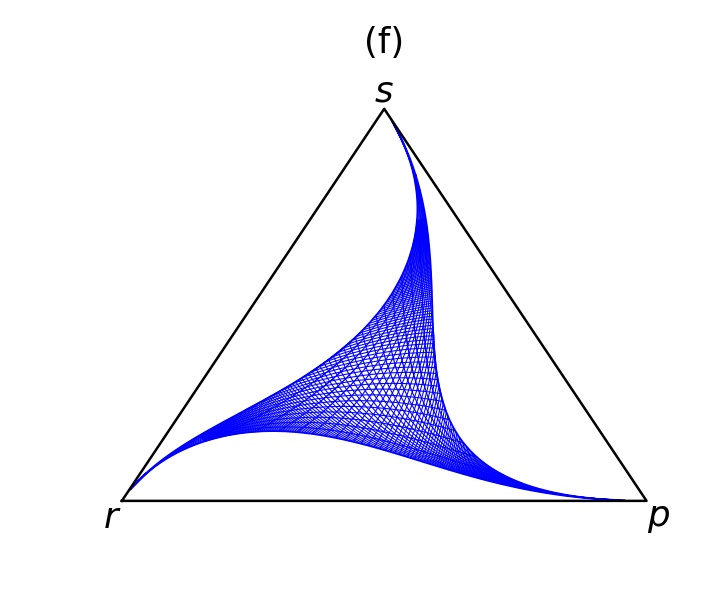}
      \vskip -0.3cm
      \caption*{\footnotesize \hskip 0.2cm $d_\text{RRP}={\color{red}{\mathbf{0.5}}},\ d_\text{PPS}={\color{red}\mathbf{1.0}},\ d_\text{SSR}={\color{red}\mathbf{1.0}}$}
    \end{subfigure}
    \\[0.0ex]
    \hskip 0.495cm
    \begin{subfigure}{0.31\textwidth}
      \includegraphics[width=36mm]{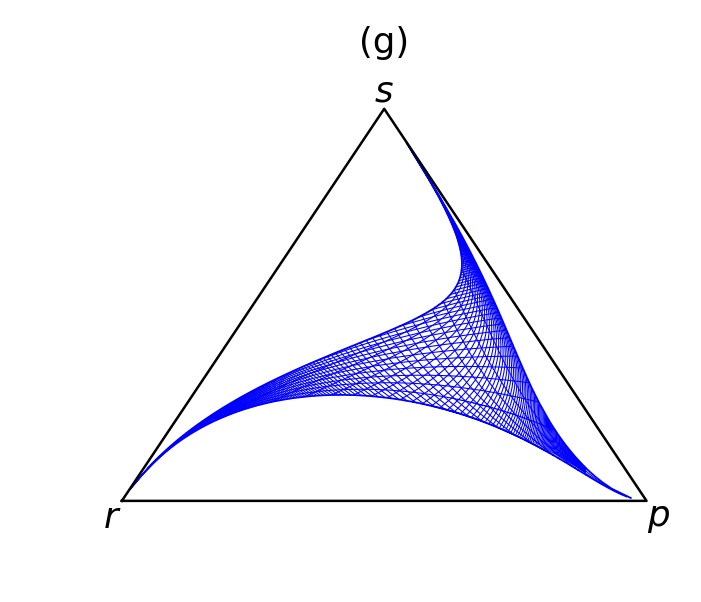}
      \vskip -0.3cm
      \caption*{\footnotesize \hskip 0.2cm $d_\text{RRP}={\color{red}{\mathbf{1.0}}},\ d_\text{PPS}={\color{red}\mathbf{0.1}},\ d_\text{SSR}={\color{red}\mathbf{1.0}}$}
    \end{subfigure}
    \begin{subfigure}{0.31\textwidth}
      \includegraphics[width=36mm]{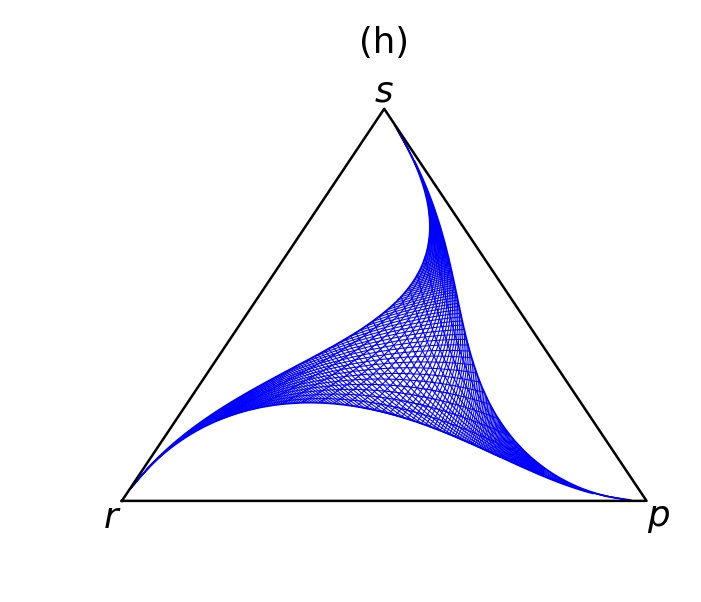}
      \vskip -0.3cm
      \caption*{\footnotesize \hskip 0.2cm $d_\text{RRP}={\color{red}{\mathbf{1.0}}},\ d_\text{PPS}={\color{red}\mathbf{0.5}},\ d_\text{SSR}={\color{red}\mathbf{1.0}}$}
    \end{subfigure}
    \begin{subfigure}{0.31\textwidth}
      \includegraphics[width=36mm]{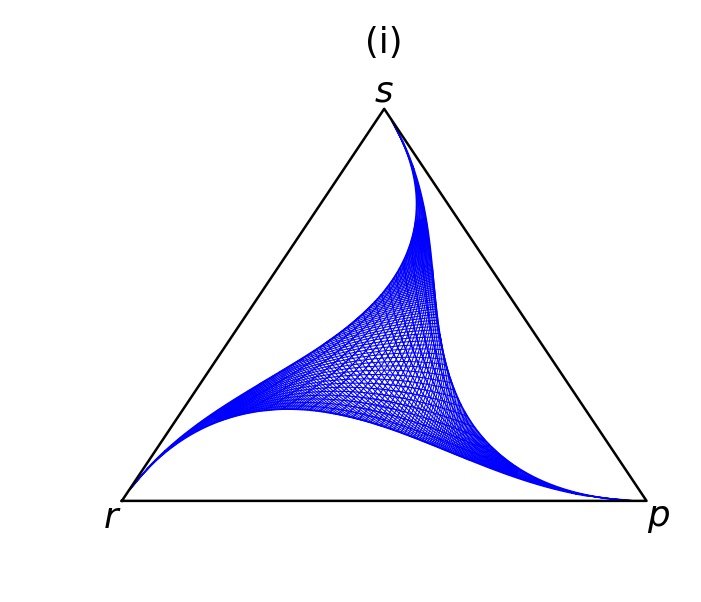}
      \vskip 0.1cm
      \caption*{\footnotesize \hskip 0.2cm $d_\text{RRP}={\color{red}{\mathbf{1.0}}},\ d_\text{PPS}={\color{red}\mathbf{1.0}},\ d_\text{SSR}={\color{red}\mathbf{1.0}}$}
    \end{subfigure}
  \end{adjustbox}
  \caption{\footnotesize {\color{red} [color online]} Ensemble $\cH$ of neutrally stable fixed points.\label{fig:hdom}}
  \vskip -0.4cm
\end{figure*}

So far, we have assumed that $\dRRP = \dPPS = \dSSR = 1$. We now relax this constraint to examine another feature of Eqs.~(\ref{eq:rateeqs}). We know that the cyclic Lotka-Volterra model with pairwise interactions has only neutrally stable orbits~\cite{Reichenbach5}. Moreover, the ensemble of fixed points fills the ternary diagram, as can be seen from Eq.~(4) of Ref.~\cite{Reichenbach5}: for each pair $(r_*,p_*)$, there exists a choice of rates for which $(r_*,p_*)$ is a reactive fixed point. The reader may ask whether this feature holds similarly in our model. The problem is non-trivial because $(r_*,p_*)$ has a complex dependency on three-agent rates, as Eq.~(\ref{eq:heterofixed}) shows. To answer the question, we let $\cH$ denote the ensemble of neutrally stable fixed points corresponding to a given choice of equilibrating rates, namely
\begin{align}
  & \hskip -0.5cm \cH(\dRRP,\dPPS,\dSSR) = \left\{(r,p):  \ \cF_\text{r} = \cF_\text{p} = 0\,, \right. \nonumber\\[1.0ex]
  & \hskip 0.7cm  \left. \lambda_\text{r}=0 \text{ and } \lambda_\text{i}\ne 0 \ \bigr|\ \dRRP,\dPPS,\dSSR\right\}\,,
  \label{eq:Hensemble}
\end{align}
where integration over polarizing rates is understood. In Fig.~\ref{fig:hdom}, we reconstruct $\cH$ numerically for several values of $\dRRP,\,\dPPS,\,\dSSR$. Continuous lines correspond to sequences of polarizing rates.  Points lying between neighboring lines belong to the ensembles as well. They just correspond to polarizing rates we did not consider numerically. Plot (i) shows that $\cH(1,1,1)$ is a proper subset of the ternary diagram, symmetric under cyclic permutations. Changing one or two rates distorts its shape: the smaller the rates, the closer $\cH$ shifts towards the boundaries. By extrapolation, Fig.~\ref{fig:hdom} suggests that $\bigcup_{\dRRP,\dPPS,\dSSR}\cH(\dRRP,\dPPS,\dSSR)$ covers the whole diagram. However, one-to-one correspondence between fixed points and reaction rates is lost: several distinct sets of rates yield the same reactive fixed point. Therefore, we conclude, our model has a more complex algebraic structure than the cyclic Lotka-Volterra model with pairwise interactions.

Besides, Fig.~\ref{fig:hdom} highlights that the \emph{the law of the weakest}, first described in Ref.~\cite{Berr1}, holds here as well. Take for instance plots (g)-(h)-(i): the lower $\dPPS$, the closer $\cH$ approaches the $(p,s)$ boundary. Recall that $\dPPS$ mediates the equilibrating transition $\greenP\,\greenP\,\blueS\to\greenP\,\blueS\,\blueS$. As such, it yields a relative measure of the strength of $\blueS$ versus $\greenP$. Any neutrally stable orbit surrounds the corresponding fixed point and flows counterclockwise on the ternary diagram. Therefore, the lower $\dPPS$ the higher the probability that the system leaves its orbit by stochastic noise and falls eventually on the $(p,s)$ boundary, where $r=0$. When this happens, the dynamics terminates with $s=1$. Hence, \blueS, the weakest strategy, survives, while \redR\ and \greenP\ go extinct.

\begin{figure}[H]
  \centering
  \includegraphics[width=0.19\textwidth]{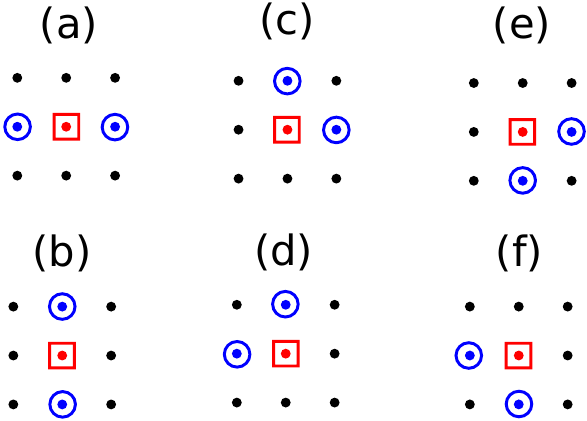}
  \caption{\footnotesize {\color{red} [color online]} An agent ({\color{red}$\boxdot$}) and two interacting neighbors ({\color{blue}$\odot$}) on a two-dimensional lattice.\label{fig:neighbors}}
\end{figure}

We finally investigate how the phase structure of the model changes off a well-mixed environment. To this aim, we make use of the Gillespie algorithm to simulate the dynamics on a two-dimensional lattice. We consider a square grid with $N=L\times L$  sites and periodic boundary conditions. We assume that each lattice site is occupied by a single agent carrying one of the strategies. Local interactions involve one agent and two of its nearest neighbors, placed in one of six possible configurations, as shown in Fig.~\ref{fig:neighbors}.

\begin{figure*}[t!]
  \centering
  \includegraphics[width=0.32\textwidth]{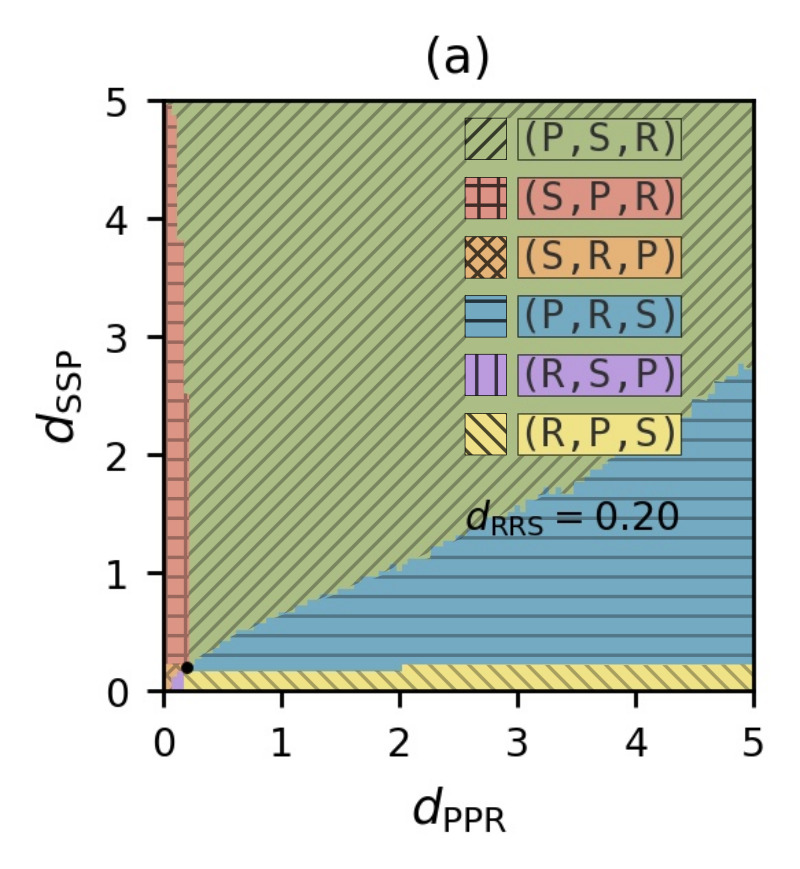}\hskip 2.0cm
  \includegraphics[width=0.32\textwidth]{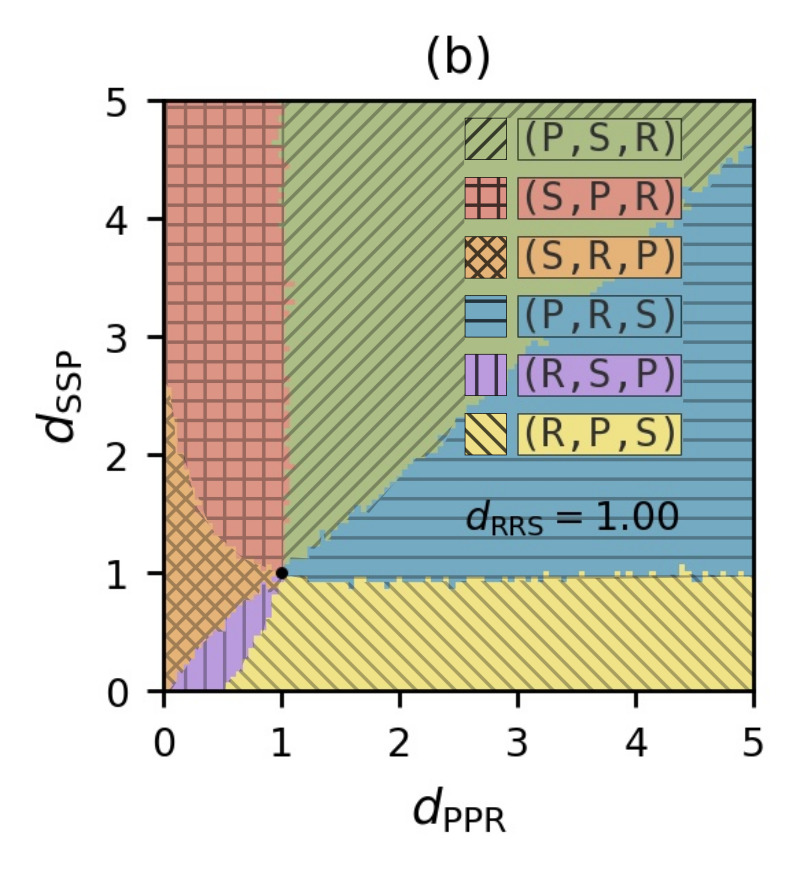}\\[0.0cm]
  \includegraphics[width=0.32\textwidth]{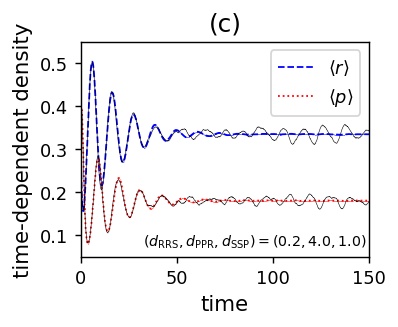}\hskip 2.0cm
  \includegraphics[width=0.32\textwidth]{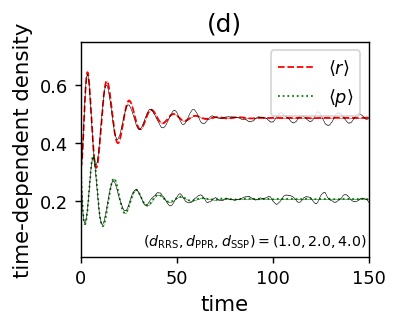}
  \vskip 0.2cm
  \caption{\footnotesize {\color{red} [color online]} (a)-(b) Phase structure of the model on a two-dimensional lattice. (c)-(d) Representative trajectories (continuous lines) and averages (dashed/dotted lines) of global densities over 100 sample trajectories. Both plots correspond to equilibrating rates $(\dRRP,\dPPS,\dRRS) = (1,1,1)$ \label{fig:2dlat}}. 
  \vskip 0.0cm
\end{figure*}

Fig.~\ref{fig:2dlat} (a)-(b) shows phases corresponding to $\dRRS = 0.2$ and $\dRRS=1.0$ for $N = 256\times 256$. The most remarkable difference with respect to the well-mixed environment is the absence of bifurcations. In both plots the phase plane is entirely filled by stable fixed points with global densities (here meaning \emph{strategy fractions over the whole lattice}) spiralling inwards. A comparison with plots (a) and (i) of Fig.~\ref{fig:phases} indicates that the relative position of phases is essentially the same in the region of stable equilibrium, even if their shapes are deformed. For instance, for $\dRRS=1.0$, $\dPPR<1$, $\dSSP\gtrsim 1$, $\greenP$ has a relative majority on the lattice, whereas $\redR$ has a relative majority in a well-mixed environment. We conclude that changing spatial topology induces distortive effects analogous to those observed in Ref.~\cite{Szolnoki2}.

As we discussed in section~III, stochastic noise perturbates the evolution of strategies for $N<\infty$. On a two-dimensional lattice, stable spirals do not converge exactly. Global densities end up fluctuating erratically around the fixed point. Fig.~\ref{fig:2dlat} (c)-(d) shows this effect for two sets of rates, both corresponding to unstable equilibria in a well-mixed environment. In both plots, continuous lines are representative trajectories of the global densities, while dashed/dotted lines represent averages over 100 sample trajectories. As can be seen, stochastic noise averages to zero. The amplitude of similar erratic oscillations was studied with full detail in a model featuring species mutations~\cite{Mobilia2}. In that context, a resonance amplification, occurring at a specific frequency, influenced fluctuations. The resonant frequency was estimated by the power spectrum method in the Fokker-Planck approximation. Here, we have less analytic information concerning the reactive fixed point. Hence, we do not attempt such an analysis. We only report that we observed a very light dependence of the fluctuation amplitude upon the polarizing rates in our numerical experiments. However, this result might depend on the excessive coarseness of the grid of values we chose.

\section{Agent mobility}

In the previous sections, we have allowed strategies to propagate as a result of predation. Now, we let them diffuse explicitly via additional pair exchange reactions, namely 
\begin{equation}
  {\tt X\,Y} \to {\tt Y\,X}\,, \quad {\tt X,Y}\in\{\redR,\greenP,\blueS\}\,,
\end{equation}
all occurring with rate $\gamma_2$. We set up a lattice metapopulation model along the lines of Refs.~\cite{McKane1,Lugo1,Szczesny1,Lamouroux1,Rulands1}. Specifically, we consider a two-dimensional square lattice with $N=L\times L$ sites and periodic boundary conditions. We interpret lattice sites as patches having a carrying capacity of $M\le \infty$ agents. We consider all reactions listed in Eq.~(\ref{eq:transitions}) as local processes, meaning that they always involve agents lying on the same patch. For simplicity, we assume homogeneous rates, as specified by Eqs.~(\ref{eq:homorates}). By contrast, we interpret exchange reactions as bilocal processes, involving agents lying on two  neighboring patches. Finally, we choose the lattice spacing to be $h = 1/L$ so that the whole lattice has length one. As $N\to\infty$, the density field $\phi \equiv \left(r,p\right)(\bx,t)$ is governed by stochastic partial differential equations (SPDE), reading as
\begin{align}
  \left.\begin{array}{l}
    \partial_t r = D_2\Delta r + \cF_{\rm r}(r,p,s) + \sum_{i={\rm r,p}}C_{{\rm r}i}\xi_i\,,\\[2.0ex]
    \partial_t p = D_2\Delta p + \cF_{\rm p}(r,p,s) + \sum_{i={\rm r,p}}C_{{\rm p}i}\xi_i\,,
    \end{array}\right.
  \label{eq:SPDE}
\end{align}
where $D_2 = \gamma_2/N$ is a scaling diffusion constant (in order to keep it finite we have to scale $\gamma_2\propto N$ as $N\to\infty$), $\Delta = \partial^2_x+\partial^2_y$ is the two-dimensional Laplace operator and  $\xi_i$ denotes uncorrelated Gaussian noise. The matrix $C$ fulfills $CC^{\rm\scriptscriptstyle T} = B$, where $B$ is the diffusion matrix of the FPE. In section~III, we obtained a simplified formula for $B$, holding for $\deq = \dpol = 1$. The most general expression to be used in Eqs.~(\ref{eq:SPDE}) reads
\begin{align}
  B_{\rm r,r} & = \frac{\deq}{M}(r^2p+s^2r) + \frac{\dpol}{M}(r^2s+p^2r)\,,\\[0.0ex]
  B_{\rm r,p} & = B_{\rm p,r} = -\frac{\deq}{M}r^2p - \frac{\dpol}{M}p^2r\,,\\[0.0ex]
  B_{\rm p,p} & = \frac{\deq}{M}(r^2p+p^2s) + \frac{\dpol}{M}(p^2r+s^2p)\,.
\end{align}
We compute $C$  from $B$ via Cholesky decomposition whenever $B$ is a positive definite matrix. If $\phi$ falls on a vertex of the ternary diagram, we let $C_{ij}=0$ by continuity. 

The dynamics of the system is trivial for $\deq>\dpol$. In this case, $\phi$ converges uniformly to the reactive fixed point $(1/3,1/3)$ on all patches, up to small fluctuations, independently of the initial conditions. A convenient way to studying the dynamics for $\dpol\ge\deq$ is to set up initial conditions such that each strategy occupies exclusively a finite portion of the lattice, as originally proposed and implemented in Refs.~\cite{Szczesny2,Szczesny4}. We proceed identically, namely for $\bx = (x,y)$ we let
\begin{equation}
  \phi(\bx,0) =
  \left\{\begin{array}{ll}
  (0,0) & \text{ for } 0\le x<L/2\\[0.0ex]
  & \text{ and } L/2\le y< L\,,\\[2.0ex]
   (0,1) & \text{ for } 0\le x<L/2\\[0.0ex]
  & \text{ and } 0\le y<L/2\,,\\[2.0ex]
   (1,0) & \text{ for } L/2\le x<L\,.
  \end{array}
  \right.
  \label{eq:initcond}
\end{equation}
The advantage of such initial conditions is related to the special role of the four lattice points $\bx = (0,0),(L/2,0),(0,L/2),(L/2,L/2)$. Here, all strategies meet and give rise to spiral waves for $t>0$. Just for the sake of completeness, we briefly review why this happens. To this aim, we introduce the topological current
\begin{equation}
  J_\mu(\bx,t) = \frac{1}{2}\epsilon_{\mu\nu\rho}\epsilon_{ab}\partial_\nu\phi_a(\bx,t)\partial_\rho\phi_b(\bx,t)\,.
  \label{eq:topcur}
\end{equation}
We assume that Greek indices take values $\{0,1,2\}$, while Latin indices take values $\{1,2\}$. Repeated indices are conventionally understood to be summed over their respective domains. The symbols $\epsilon_{\mu\nu\rho}$ and $\epsilon_{ab}$ denote totally antisymmetric tensors with three and two components respectively. We let $\partial_0 \equiv \partial_t$.  It takes no effort to show that $J_\mu$ fulfills the local conservation law $0 = \partial_\mu J_\mu = \partial_0 J_0 + \partial_k J_k$. We define the topological charge in the thermodynamic limit as 
\begin{equation}
  Q(t) = \int\rd\bx\,J_0(\bx,t) = \int\rd\bx\, (\partial_1 r\,\partial_2 p - \partial_1p\,\partial_2 r) \,.
\end{equation}
From the conservation of $J_\mu$ and the periodicity of the boundary conditions, it follows that $\rd Q/\rd t = \int\rd\bx\,\partial_0J_0 = -\int\rd\bx\,\partial_kJ_k = 0$, hence $Q$ is invariant in time. We then let $J_0^{\pm}(\bx,t) = \max\{\pm J_0(\bx,t),0\}$. We notice that $J_0^{\pm}$ is strictly positive only near points around which the three strategies follow cyclically in counterclockwise/clockwise order. In particular, $J_0^{+}(\bx,0)$ is positive (infinite) for $\bx = (0,L/2)$, $(L/2,0)$, while $J_0^{-}(\bx,0)$ is positive (infinite) for $\bx = (0,0)$, $(L/2,L/2)$ and $J_0^{\pm}(\bx,0)=0$ elsewhere. As a result, we have four topological charges, two positive and two negative, localized on the four mentioned points, yielding $0 = Q(0) = Q(t)$ for all $t>0$. While the topological density is sharply peaked for $t=0$, it becomes somewhat smooth for $t>0$. This corresponds to the appearance of two spirals plus two anti-spirals originating from the four points. Whether these objects last forever or disappear sooner or later is not a matter of topology; it depends on Eqs.~(\ref{eq:SPDE}).

\begin{figure*}[!t]
  \centering
  \begin{adjustbox}{varwidth=1.0\textwidth,center}
  \centering
  \vbox{\vskip 0.3cm}
    \begin{subfigure}{0.24\textwidth}
      \includegraphics[width=42mm]{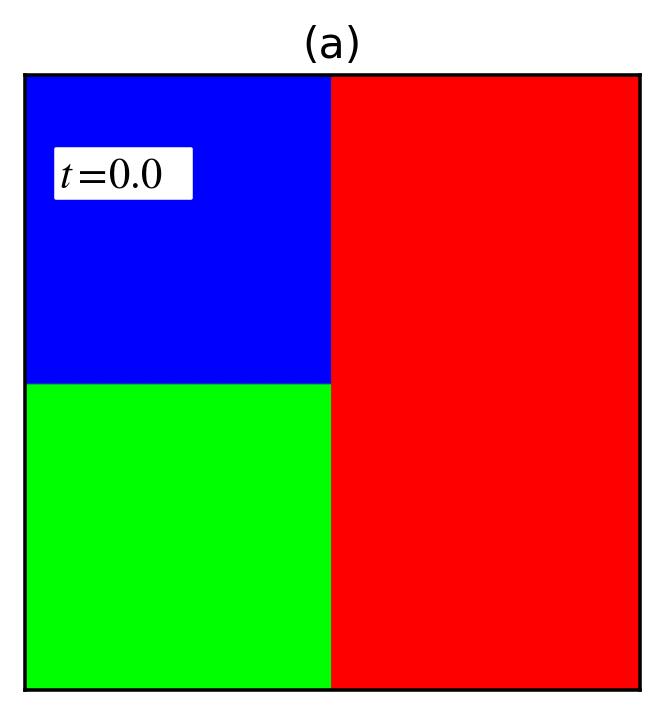}
      \vskip -0.3cm
      \caption*{}
    \end{subfigure}
    \begin{subfigure}{0.24\textwidth}
      \includegraphics[width=42mm]{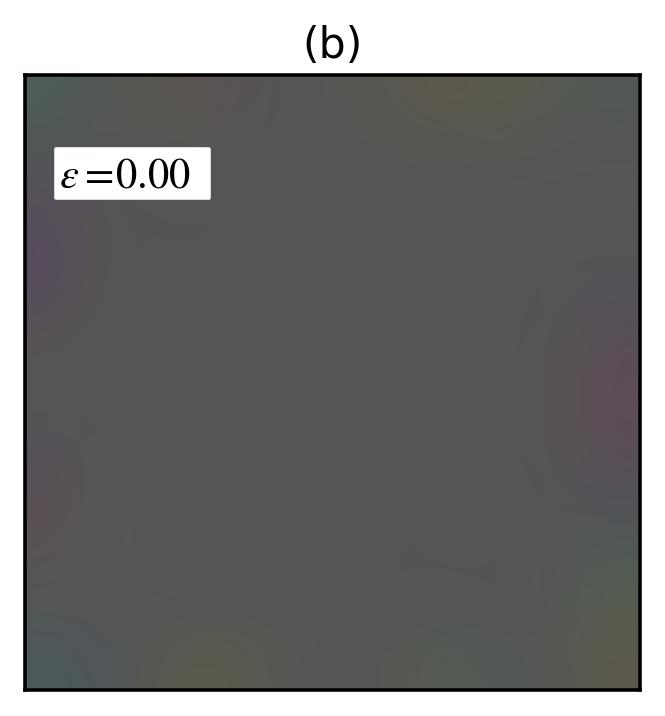}
      \vskip -0.3cm
      \caption*{}
    \end{subfigure}
    \begin{subfigure}{0.24\textwidth}
      \includegraphics[width=42mm]{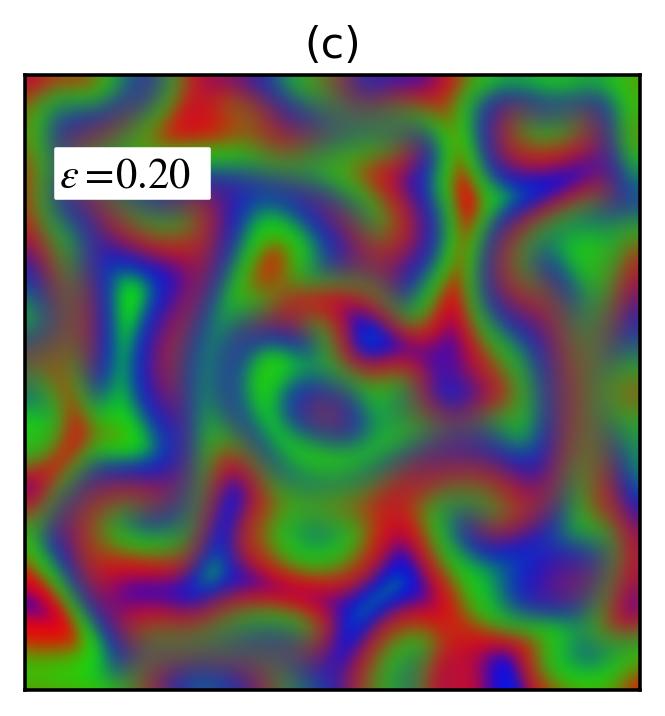}
      \vskip -0.3cm
      \caption*{}
    \end{subfigure}
    \begin{subfigure}{0.24\textwidth}
      \includegraphics[width=42mm]{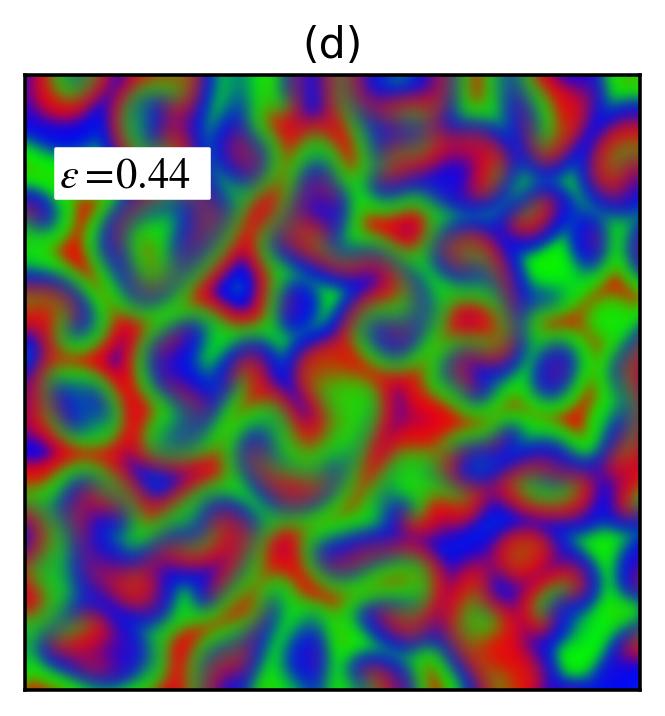}
      \vskip -0.3cm
      \caption*{}
    \end{subfigure}
    \\[-2ex]
    \begin{subfigure}{0.24\textwidth}
      \includegraphics[width=42mm]{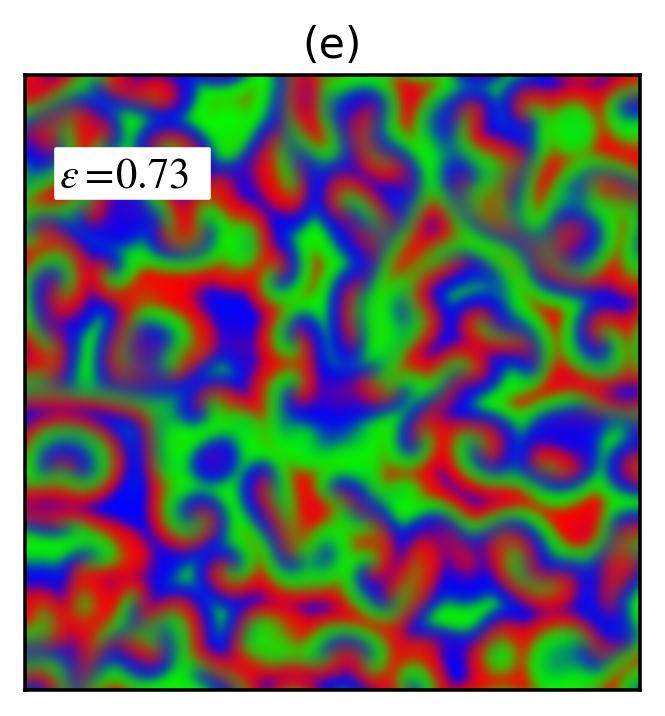}
      \vskip -0.3cm
      \caption*{}
    \end{subfigure}
    \begin{subfigure}{0.24\textwidth}
      \includegraphics[width=42mm]{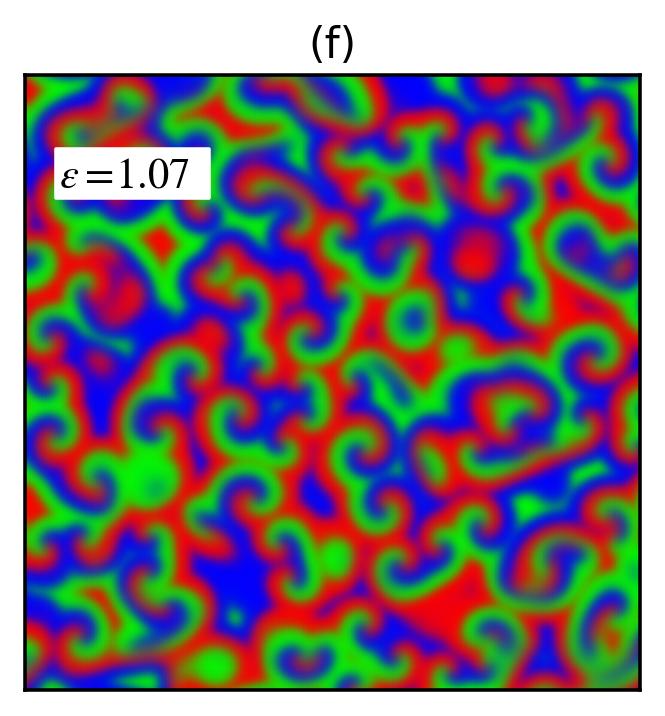}
      \vskip -0.3cm
      \caption*{}
    \end{subfigure}
    \begin{subfigure}{0.24\textwidth}
      \includegraphics[width=42mm]{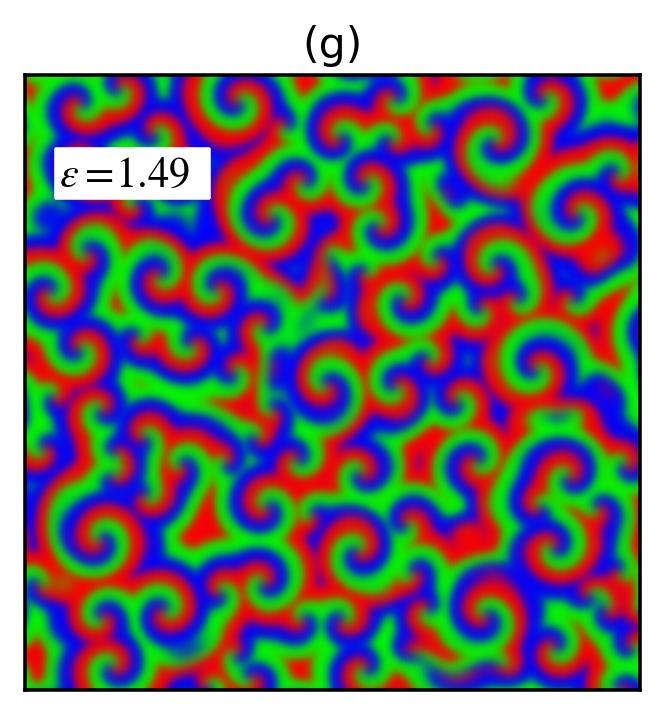}
      \vskip -0.3cm
      \caption*{}
    \end{subfigure}
    \begin{subfigure}{0.24\textwidth}
      \includegraphics[width=42mm]{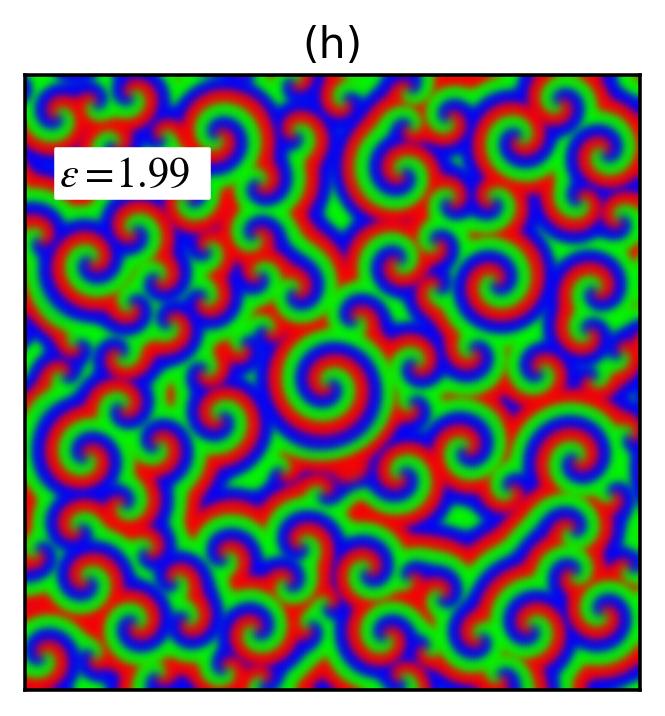}
      \vskip -0.3cm
      \caption*{}
    \end{subfigure}
    \\[-2ex]
    \begin{subfigure}{0.24\textwidth}
      \includegraphics[width=42mm]{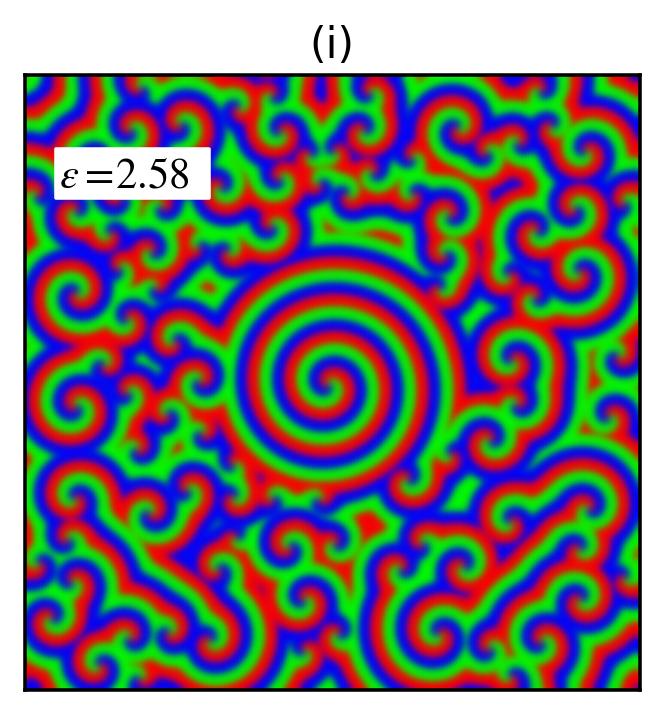}
      \vskip -0.3cm
      \caption*{}
    \end{subfigure}
    \begin{subfigure}{0.24\textwidth}
      \includegraphics[width=42mm]{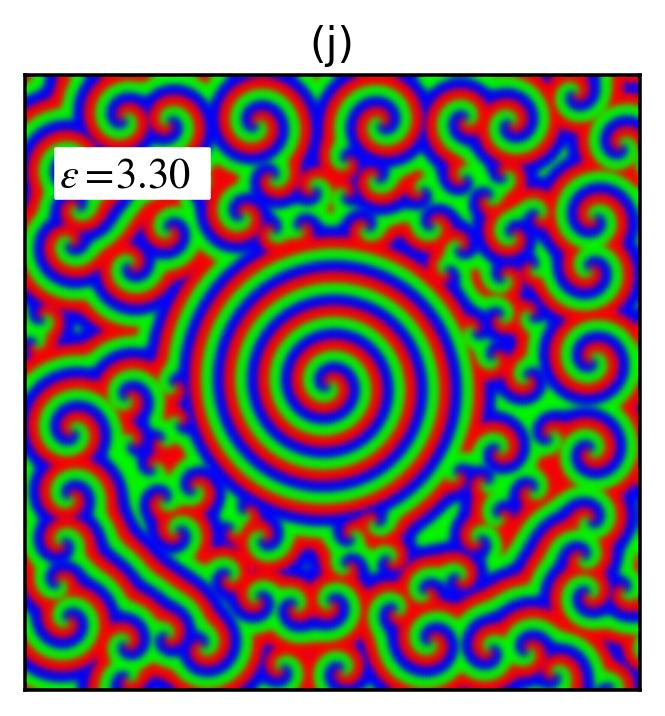}
      \vskip -0.3cm
      \caption*{}
    \end{subfigure}
    \begin{subfigure}{0.24\textwidth}
      \includegraphics[width=42mm]{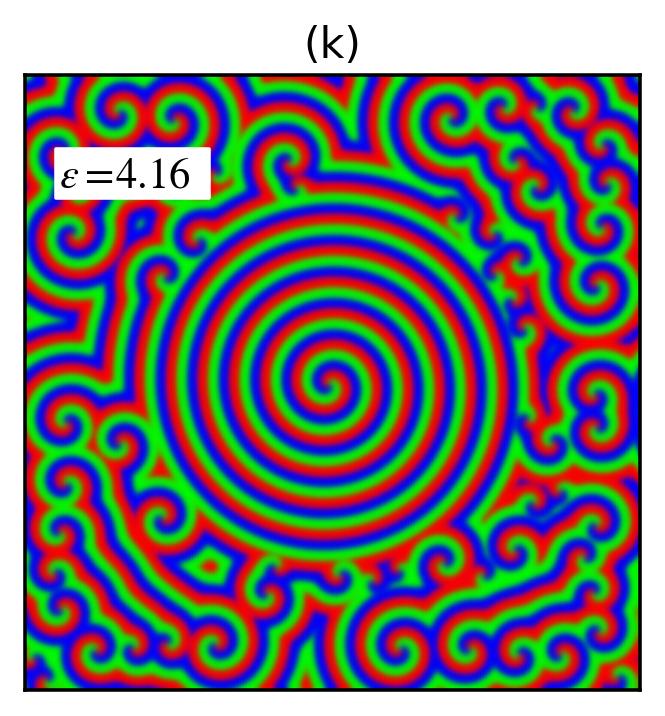}
      \vskip -0.3cm
      \caption*{}
    \end{subfigure}
    \begin{subfigure}{0.24\textwidth}
      \includegraphics[width=42mm]{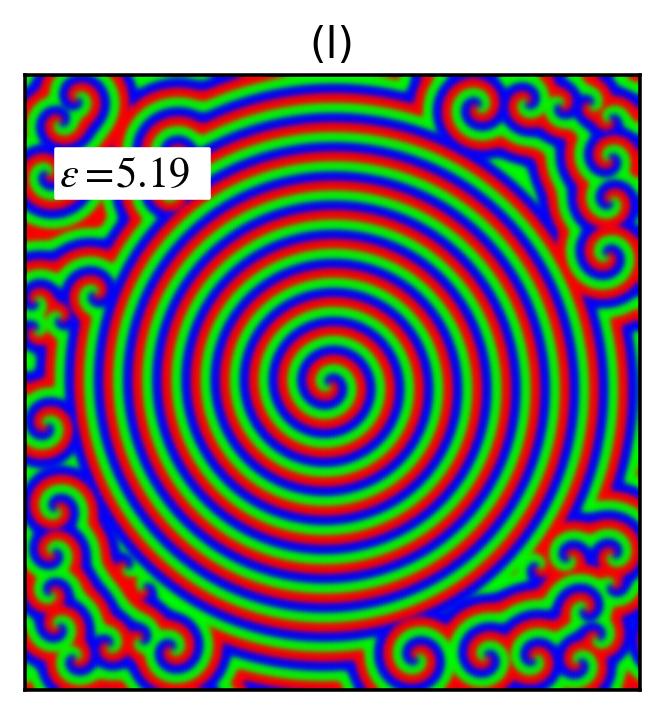}
      \vskip -0.3cm
      \caption*{}
    \end{subfigure}
    \\[-2ex]
    \begin{subfigure}{0.24\textwidth}
      \includegraphics[width=42mm]{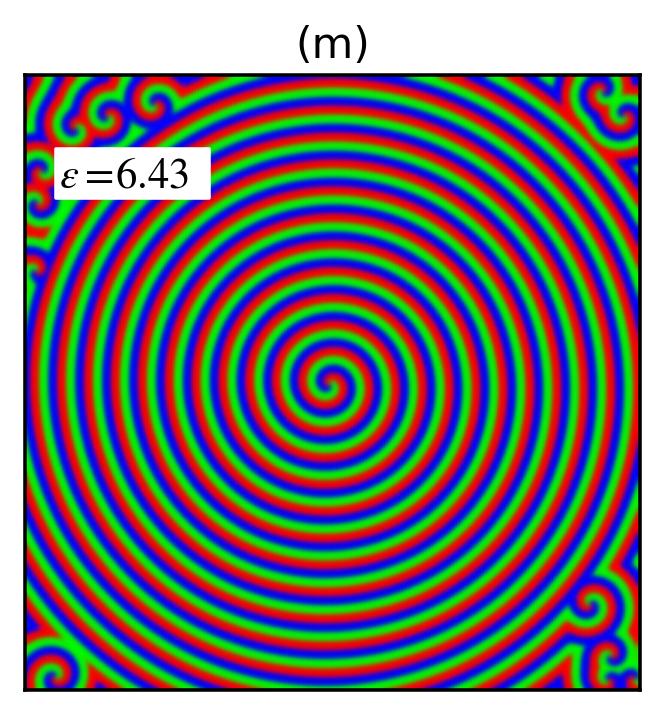}
      \vskip -0.3cm
      \caption*{}
    \end{subfigure}
    \begin{subfigure}{0.24\textwidth}
      \includegraphics[width=42mm]{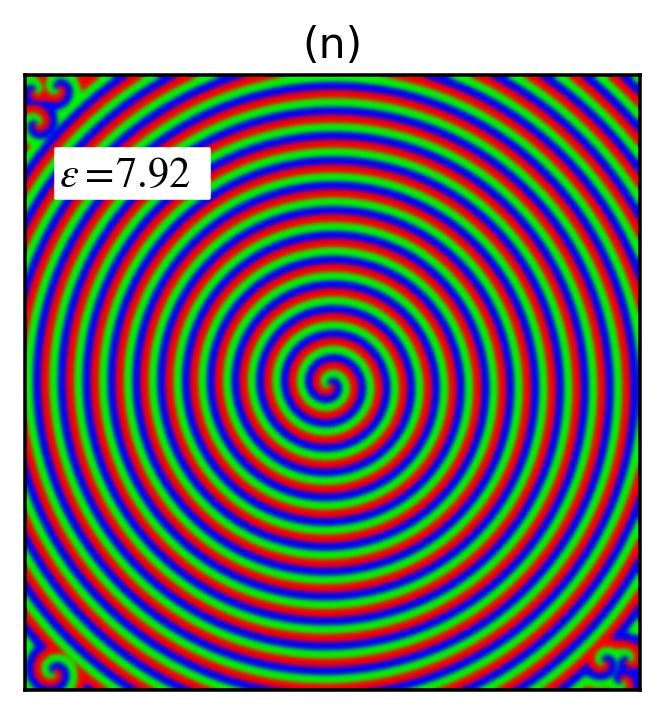}
      \vskip -0.3cm
      \caption*{}
    \end{subfigure}
    \begin{subfigure}{0.24\textwidth}
      \includegraphics[width=42mm]{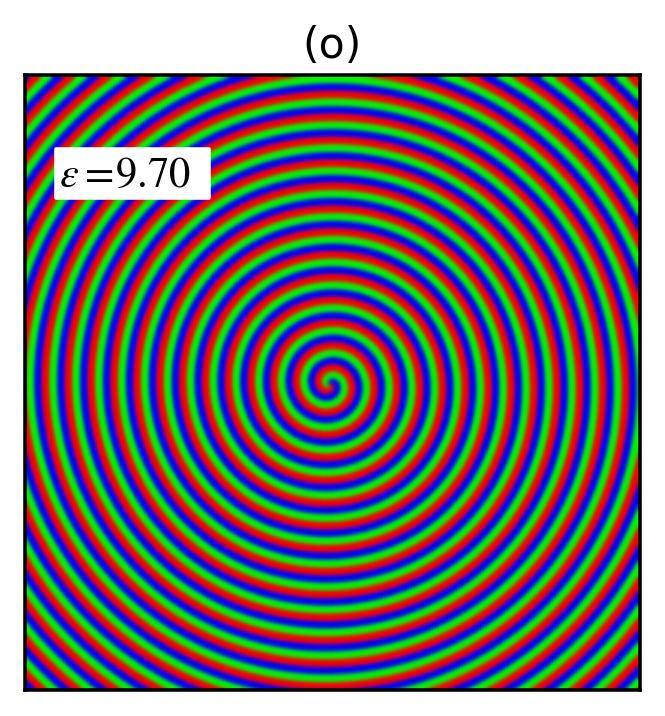}
      \vskip -0.3cm
      \caption*{}
    \end{subfigure}
    \begin{subfigure}{0.24\textwidth}
      \includegraphics[width=42mm]{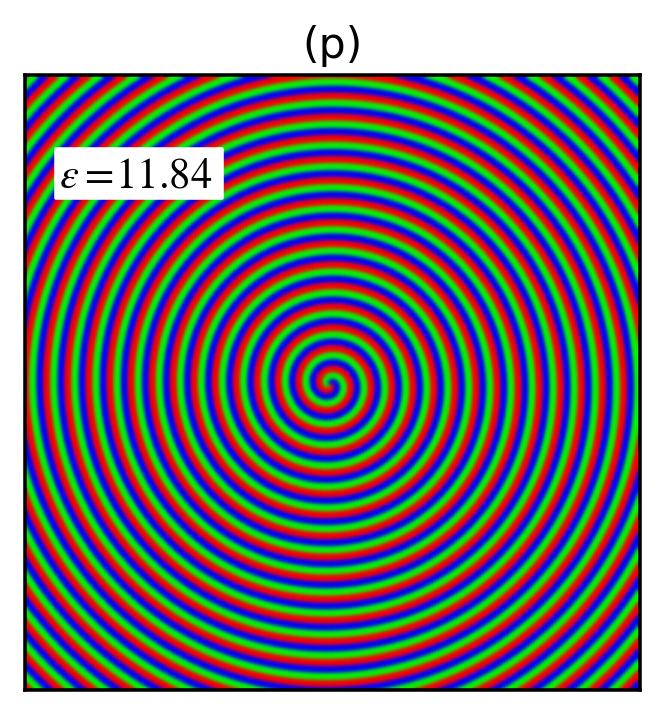}
      \vskip -0.3cm
      \caption*{}
    \end{subfigure}
  \end{adjustbox}
  \vskip -0.45cm
  \caption{\footnotesize {\color{red} [color online]} Snapshots of the strategy density field in RGB representation for \\ $N=2048\times 2048$, $\gamma_2 = 16$, $M=\infty$, $t=400$.\label{fig:PDE}}
  \vskip -0.3cm
\end{figure*}

In Fig.~\ref{fig:PDE} we report a sequence of snapshots of $\phi$ in RGB representation. The sequence refers to the following choice of parameters:
\begin{itemize}[itemsep=0.0pt]
\item[$\circ$]{$N = 2048\times 2048$;}
\item[$\circ$]{$\gamma_2 = 16$ ($D_2 \simeq 3.81\times 10^{-6}$);}
\item[$\circ$]{$\deq=1\,,\ \dpol = 1.2^k$, for $k=0,1,\ldots,14$;}
\item[$\circ$]{$M=\infty$.}
\end{itemize}
All pictures represent $\phi$ in the central area of the lattice, namely for $L/4<x_1,x_2<3L/4$. Plot (a) corresponds to $t=0$, all others to $t=400$. The latter time is sufficiently large to ensure that transient effects have disappeared and the system has evolved to a steady state. To obtain Fig.~\ref{fig:PDE}, we integrated Eqs.~(\ref{eq:SPDE}) numerically via the ETD2RK scheme, introduced in Ref.~\cite{Cox1}. We summarize below the main features of the plots:
\begin{itemize}[itemsep=0.0pt]
\item[$\circ$]{chaotic patterns with blurred and stretched shapes are predominant for $\epsilon<1$;}
\item[$\circ$]{blurring reduces as $\epsilon$ increases;}
\item[$\circ$]{small spirals emerge from chaos for $\epsilon\gtrsim 1$;}
\item[$\circ$]{a central spiral arises in a background of smaller spirals for $\epsilon\gtrsim 2$;}
\item[$\circ$]{the propagation radius of the central spiral increases progressively, until it becomes largest for $\epsilon\approx 10$;}
\item[$\circ$]{for even larger $\epsilon$ the central spiral keeps definitively stable;}
\item[$\circ$]{the wavelength of spatial patterns decreases monotonically as $\epsilon$ increases.}
\end{itemize}

\begin{figure*}[!t]
  \centering
  \includegraphics[width=0.35\textwidth]{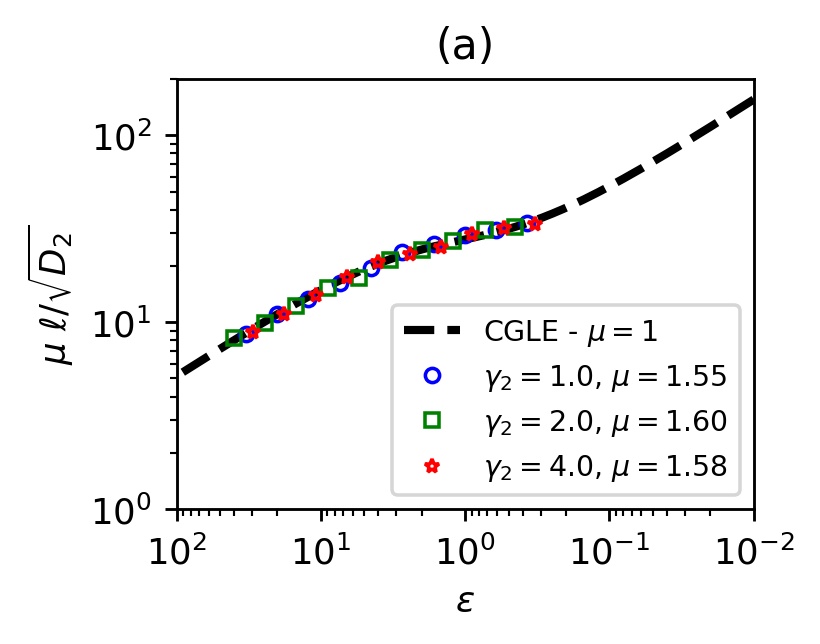} \hskip 1.0cm
  \includegraphics[width=0.35\textwidth]{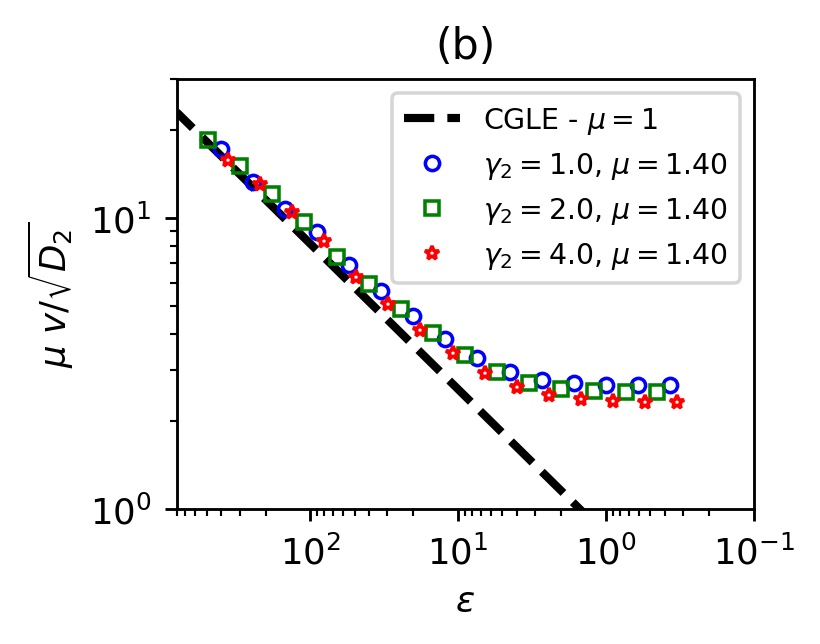}
  \vskip -0.3cm
  \caption{\footnotesize {\color{red} [color online]} {Comparison of wavelength and propagation speed of spiral waves from the CGLE with data obtained from numerical integration of Eqs.~(\ref{eq:SPDE}) for $N=512\times 512$ and $M=\infty$.\label{fig:fignine}}}
  \vskip -0.1cm
\end{figure*}

Patterns characterized by very similar behavior emerge in the spatial version of the May-Leonard model, featuring different particle interactions but analogous Hopf bifurcation~\cite{Reichenbach2}. This similarity provides a confirmation that macroscopic phenomena induced by cyclic competition are robust. They depend only on the type of the Hopf bifurcation, whereas the details of the interactions have no qualitative (and little quantitative) influence~\cite{Reichenbach1,Frey1}. We follow Ref.~\cite{Reichenbach2} for the analysis of spiral waves in terms of the complex Ginzburg-Landau equation (CGLE). In our case, the CGLE reads (after rescaling the complex amplitude and shifting its phase)
\begin{equation}
  \partial_t z = D_2\Delta z + \lambda_\text{r}\,z - [1+\text{i}\alpha(\epsilon)]\,z|z|^2\,,
\end{equation}
with $\lambda_\text{r} = \epsilon/6$ being the real part of the eigenvalues of the linearized RE, see Eq.~(\ref{eq:linearev}), and with $\alpha(\epsilon)$ a function parameter given by
\begin{equation}
  \alpha(\epsilon) = \frac{b(\epsilon)}{a(\epsilon)} = \frac{3\sqrt{3}}{2}\frac{(2+\epsilon)(\epsilon^2 + 3\epsilon + 3)}{\epsilon(4\epsilon^2 + 15\epsilon + 15)}\,.
\end{equation}
It is important to recall that the CGLE is accurate only in the vicinity of a supercritical Hopf bifurcation. In that case, the role of the Laplacian is to synchronize limit cycles at neighboring lattice sites, thus giving rise to coherent spatiotemporal patterns. In our model, we have heteroclinic cycles, just like in Ref.~\cite{Reichenbach2}. Hence, the CGLE is not guaranteed to describe correctly the dynamics of the system. Nonetheless, we can compare predictions of the CGLE with numerical observations. In particular, in the approach of Ref.~\cite{VanSaarloos} the wavelength $\ell(\epsilon)$ of spiral waves and their propagation speed $v(\epsilon)$ read
\begin{equation}
  \ell(\epsilon) = \frac{2\pi\alpha(\epsilon)\sqrt{D_2/\lambda_\text{r}}}{\sqrt{1+\alpha(\epsilon)^2}-1}\,,\qquad v(\epsilon) = 2\sqrt{D_2\lambda_\text{r}}\,,\quad
  \label{eq:wavelength}
\end{equation}
Fig.~\ref{fig:fignine} shows a comparison of  Eq.~(\ref{eq:wavelength}) with numerical data corresponding to $N=512\times 512$, $M=\infty$ and $\gamma_2=1,2,4$. With regard to the wavelength, the agreement is perfect at all tested scales up to a multiplicative constant $\mu = 1.55\div 1.60$, depending on $\gamma_2$ but not on $\epsilon$. This constant includes nonlinear effects which are not properly captured by the CGLE. Surprisingly, $\mu$ is very similar in size to the analogous constant in Ref.~\cite{Reichenbach2}. As for the propagation speed, the agreement becomes very good only in the asymptotic regime $\epsilon\to\infty$. Moreover, the multiplicative constant $\mu$ is slightly smaller. This yields evidence that nonlinear effects may change from one observable to another\footnote{A different theoretical approach can be found in Ref.~\cite{Szczesny2}. Here, $\ell$ and $v$ are expressed in the plane wave approximation as functions of $|z|^2$. The latter quantity is computed from numerical solutions of the CGLE via global averaging over the whole space.}.

An intuitive explanation of why the RGB representation of $\phi$ blurs as $\epsilon\to 0$ follows from the observation that $\dot\rho$ and $\dot\theta$ in Eq.~(\ref{eq:polarhopf}) are monotonic functions of $\epsilon$ at LO. This feature has straightforward consequences for the dynamics in the metapopulation model. Indeed, the smaller $\epsilon$, the closer $\phi$ keeps wandering chaotically around the reactive fixed point before walking over heteroclinic cycles. As a result, RGB colors representing $\phi$ shift progressively to gray (the color corresponding to equal densities), while spatial patterns fade out.

Closer inspection suggests that two different physical mechanisms interfere with the spatiotemporal coherence of rotating spirals in regimes of intermediate and small $\epsilon$, as discussed in Refs.~\cite{Reichenbach2,Szczesny2,Szczesny4}. For $2\lesssim\epsilon\lesssim 5$ small disturbances propagate radially together with the wavefronts. They intensify while traveling away from the core of spirals until they result in a far-field break-up of spatial coherence (convective instability). As $\epsilon\to 0$, the blurring of $\phi$ becomes very strong. In this limit, coherent propagation is fully compromised. Small disturbances grow locally. Their overall effect is to twist and stretch spatial patterns (absolute instability). 
\begin{figure}[!h]
  \centering
  \includegraphics[width=0.156\textwidth]{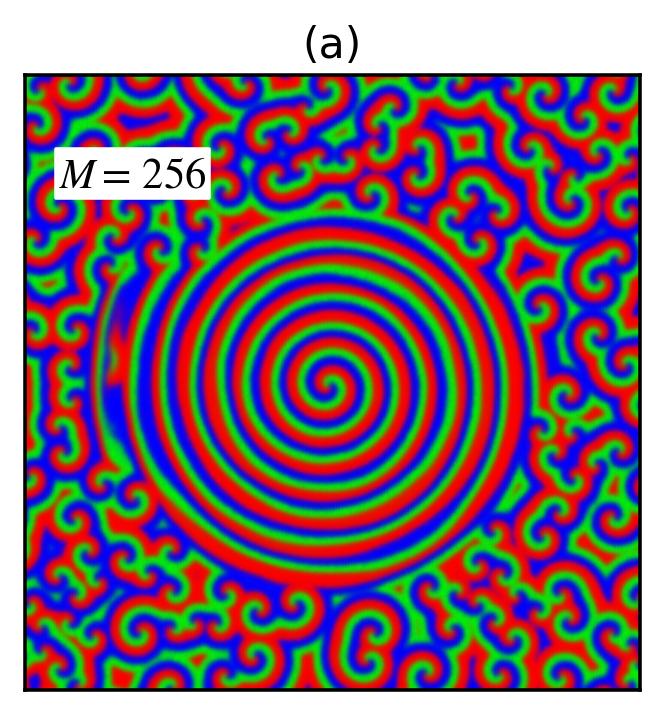}
  \includegraphics[width=0.156\textwidth]{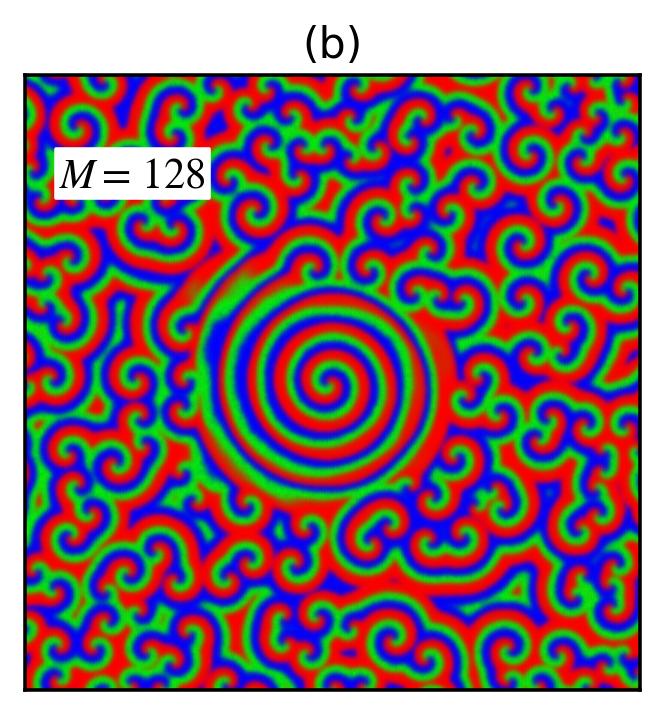}
  \includegraphics[width=0.156\textwidth]{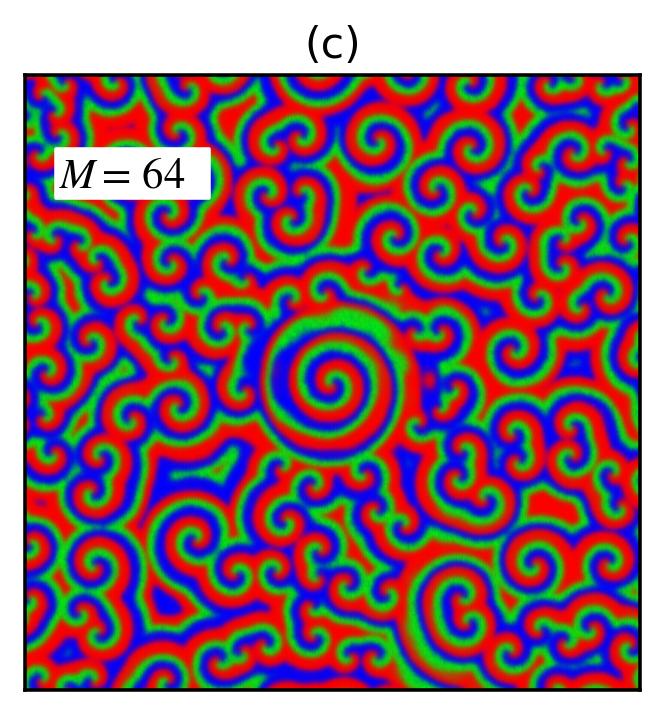}
  \caption{\footnotesize {\color{red} [color online]} Effects of stochastic noise on the propagation of spiral waves.{\label{fig:finiteM}}}
\end{figure}


In Fig.~\ref{fig:finiteM}, three snapshots of $\phi$, corresponding respectively to $M=256$~(a), $128$~(b) and $64$~(c), illustrate how stochastic noise induced by finite $M$ perturbates the stability of the central spiral wave. We obtained all plots from the numerical integration of Eqs.~(\ref{eq:SPDE}) for $N=2048 \times 2048$, $\gamma_2 = 16$, $\epsilon = 6.43$, $t=400$ and initial conditions given by Eq.~(\ref{eq:initcond}). The increase rate of the spiral radius with $M$ indicates relatively slow convergence to the asymptotic limit, represented by Fig.~\ref{fig:PDE}~(m).

\section{Three-agent chase reactions}

Spatially structured games, admitting empty sites, usually assume that agents move via site hopping and pair exchange. These two processes correspond to distinct mobility reactions, namely
\begin{alignat}{3}
    {\tt X}\,\emptyset\,\to\, \emptyset\,{\tt X}\,, & \quad \text{occurring with rate } \gamma_\text{h}\,, \label{eq:hopping} \\[1.0ex] 
    {\tt X}\,{\tt Y}\,\to\, {\tt Y}\,{\tt X}\,, & \quad \text{occurring with rate } \gamma_\text{e}\,, \label{eq:exchange}
\end{alignat}
with ${\tt X},{\tt Y}\in\{\redR,\greenP,\blueS\}$. As $N\to\infty$, they induce differential variations of the strategy densities, quantified by
\vskip -0.5cm
\begin{equation}
  \delta a = \frac{\gamma_\text{h}}{N}\Delta a +  \frac{(\gamma_\text{h}-\gamma_\text{e})}{N}[a\Delta (b+c)-(b+c)\Delta a]\,,
  \label{eq:gendiff}
\end{equation}
for $(a,b,c)$ a permutation of $(r,p,s)$. Eq.~(\ref{eq:gendiff}) reduces to Gaussian diffusion for $\gamma_\text{h}=\gamma_\text{e}$.  Nonlinear terms, arising for $\gamma_\text{h}\ne \gamma_\text{e}$, can disrupt the stability of spiral waves by producing perturbations that result in the far-field break-up of spatial coherence~\cite{Szczesny1}. Hence, they could be important to clarify in which circumstances growing bacterial colonies develop coherent patterns and in which they do not~\cite{Szczesny1}. Anyway, hopping and pair exchange yield homogeneous and isotropic diffusion, regardless of how $\gamma_\text{h}$ and $\gamma_\text{e}$ are chosen.

\begin{figure}[H]
  \centering
  \includegraphics[width=0.34\textwidth]{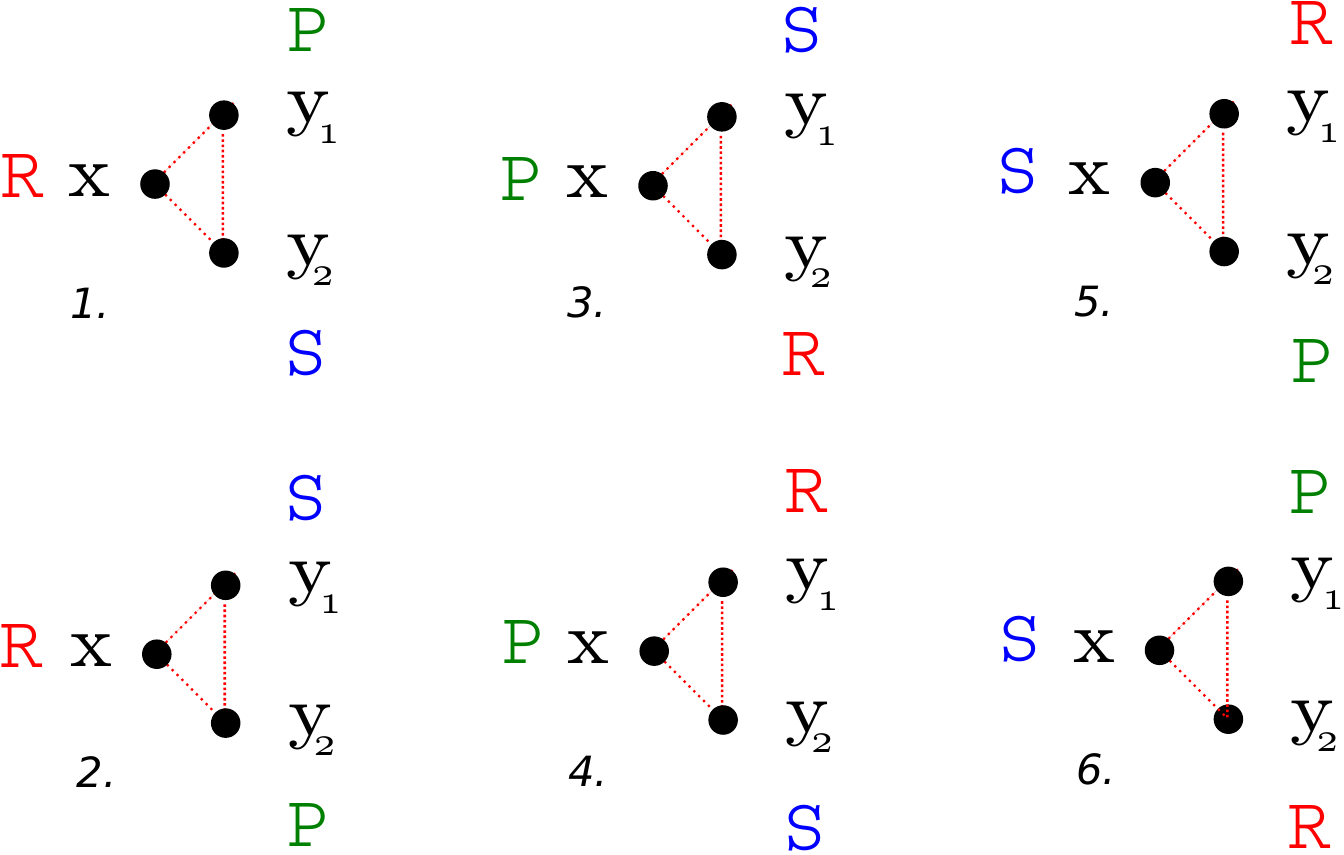}
  \caption{\footnotesize {\color{red} [color online]} Configurations of three neighboring chasing agents.\label{fig:ringofroses}}
\end{figure}

In this section, we address the issue of whether we can introduce nonlinear diffusion in our model. Since we admit no empty sites, Eq.~(\ref{eq:hopping}) is ruled out. Hence, Eq.~(\ref{eq:gendiff}) makes sense no more. The mobility operator induced by pair exchange is the Laplacian, $\delta a_\text{ex} = (\gamma_2/N)\Delta a$ for $a=r,p,s$. We have three possibilities: either we give up species homogeneity in pair exchange, or we break isotropy, or, less trivially,  we introduce more complex mobility reactions. Going back to Table~\ref{tab:payoffs} and Eq.~(\ref{eq:transitions}), we realize that we can tentatively use for our purpose transitions where agents carry initially three different strategies. Given a lattice site $\bx$ and two nearest neighbors~$\by_1,\,\by_2$, we introduce the \emph{chase} reactions
\begin{equation}
  {\tt X}(\bx)\,{\tt Y}(\by_1)\,{\tt Z}(\by_2)\,\to\,{\tt Y}(\bx)\,{\tt Z}(\by_1)\,{\tt X}(\by_2)\,,
  \label{eq:ringofroses1}
\end{equation}
for $({\tt X},{\tt Y},{\tt Z})$ an even permutation of $(\redR,\greenP,\blueS)$ and
\begin{equation}
  {\tt X}(\bx)\,{\tt Y}(\by_1)\,{\tt Z}(\by_2)\,\to\,{\tt Z}(\bx)\,{\tt X}(\by_1)\,{\tt Y}(\by_2)\,,
  \label{eq:ringofroses2}
\end{equation}
for $({\tt X},{\tt Y},{\tt Z})$ an odd permutation of $(\redR,\greenP,\blueS)$. Incidentally, the interpretation of Eqs.~(\ref{eq:ringofroses1})-(\ref{eq:ringofroses2}) as reactions where species chase one another cyclically is not the only possible one. We could, equivalently and perhaps more imaginatively, regard them as an evolutionary version of the playground singing game \emph{Ring a ring of roses}. Anyway, we assume that all such reactions occur with rate $\gamma_3$. Fig.~\ref{fig:ringofroses} shows the initial state corresponding to all possible permutations.

We wish to derive the mobility operator arising from Eqs.~(\ref{eq:ringofroses1})-(\ref{eq:ringofroses2}), as $N\to\infty$. To this aim, it is sufficient to focus on one of the strategies, e.g. \redR. The average variation of $r(\bx)$, due to the above reactions, receives two negative and two positive contributions, corresponding respectively to initial configurations 1.-2. and 5.-6., namely
\begin{align}
  & \hskip -0.5cm \delta r_\text{chase}(\bx\,|\,\by_1,\by_2) = \sum_{k=1,2,5,6}\delta r_k(\bx\,|\,\by_1,\by_2)\,,
\end{align}
with
\begin{align}
  & \delta r_1(\bx\,|\,\by_1,\by_2) = -\gamma_3\,r(\bx)p(\by_1)s(\by_2)\,,\\[1.0ex]
  & \delta r_2(\bx\,|\,\by_1,\by_2) = -\gamma_3\,r(\bx)s(\by_1)p(\by_2)\,,\\[1.0ex]
  & \delta r_5(\bx\,|\,\by_1,\by_2) = \phantom{-}\gamma_3\,s(\bx)r(\by_1)p(\by_2)\,,\\[1.0ex]
  & \delta r_6(\bx\,|\,\by_1,\by_2) = \phantom{-}\gamma_3\,s(\bx)p(\by_1)r(\by_2)\,.
\end{align}
In principle, there are six possible ways of choosing $\by_1,\,\by_2$ given $\bx$. An elegant expression for the mobility operator follows provided we discard configurations (a)-(b) of Fig.~\ref{fig:neighbors} and keep all others. The variation of $r(\bx)$ corresponding to configuration (c) of Fig.~\ref{fig:neighbors} reads as
\begin{align}
  \delta r_\text{c}(\bx) = & -\gamma_3\, r(\bx)\left[p(\bx+h\hat\bx)s(\bx+h\hat\by) \right. \nonumber\\[1.0ex]
    & \hskip 1.23cm  \left. + s(\bx+h\hat\bx)p(\bx+h\hat\by)\right]\nonumber\\[2.0ex]
  & + \gamma_3\, s(\bx)\left[p(\bx+h\hat\bx)r(\bx+h\hat\by)\right. \nonumber\\[1.0ex]
    & \hskip 1.23cm  \left. + r(\bx+h\hat\bx)p(\bx+h\hat\by)\right]\,.
\end{align}
Expanding it in Taylor series around $h=0$ yields
\begin{align}
  \delta r_\text{c}(\bx) & = \delta r_\text{c}^{(0)}(\bx) + h\,\delta r_\text{c}^{(1)}(\bx) \nonumber\\[2.0ex]
  & + h^2\,\delta r_\text{c}^{(2)}(\bx) + \text{O}(h^3)\,,
\end{align}
with
\begin{align}
  \hskip -0.8cm \delta r_\text{c}^{(0)}(\bx) & = 0\,,\\[3.0ex]
  \hskip -0.8cm \delta r_\text{c}^{(1)}(\bx) & =  -\gamma_3\, r(\bx)[p(\bx)\partial_y s(\bx) + p(\bx)\partial_x s(\bx)]\nonumber\\[1.0ex]
  & \hskip 0.0cm \phantom{=\,} +\gamma_3\, s(\bx)[p(\bx)\partial_y r(\bx) + p(\bx)\partial_x r(\bx)]\,,\\[3.0ex]
  \hskip -0.8cm \delta r_\text{c}^{(2)}(\bx) & = \nonumber\\[1.0ex]
  & \hskip -1.0cm -\frac{1}{2}\gamma_3\, r(\bx)\bigl\{p(\bx)\Delta s(\bx) \nonumber\\[1.0ex]
  & \hskip -0.4cm + 2[\partial_x p(\bx)][\partial_y s(\bx)] + 2[\partial_y p(\bx)][\partial_x s(\bx)]\bigr\}\nonumber\\[2.0ex]
  & \hskip -1.0cm +\frac{1}{2}\gamma_3\, s(\bx)\bigl\{p(\bx)\Delta r(\bx) \nonumber\\[1.0ex]
  & \hskip -0.4cm + 2[\partial_x p(\bx)][\partial_y r(\bx)] + 2[\partial_y p(\bx)][\partial_x r(\bx)]\bigr\}\,.
\end{align}
We find analogous expressions $\delta r_\text{k}(\bx)$ for k = d,\,e,\,f, corresponding respectively to initial configurations (d), (e), (f) of Fig.~\ref{fig:neighbors}.  For the sake of conciseness, we leave the reader with the exercise of deriving them. Upon adding $\delta r_\text{c}$ and $\delta r_\text{d}$, the mixed-derivative terms at $\text{O}(h^2)$ cancel, hence we obtain
\begin{align}
  & \hskip -0.5cm \delta r_\text{c}^{(0)} + \delta r_\text{d}^{(0)} = 0\,,\\[2.0ex]
  & \hskip -0.5cm \delta r_\text{c}^{(1)}(\bx) + \delta r_\text{d}^{(1)}(\bx) = \nonumber\\[1.0ex]
  & \label{eq:h1terms1}
 \hskip -0.55cm  -2\gamma_3\, r(\bx)p(\bx)\,\partial_y s(\bx) + 2\gamma_3\, s(\bx)p(\bx)\,\partial_y r(\bx)\,,\\[2.0ex]
  & \hskip -0.5cm \delta r_\text{c}^{(2)}(\bx) + \delta r_\text{d}^{(2)}(\bx) = \nonumber\\[1.0ex]
 & \gamma_3\, \left[s(\bx)p(\bx)\Delta r(\bx) - r(\bx)p(\bx)\Delta s(\bx)\right]\,.
   \label{eq:h2terms1}  
\end{align}
Analogously, adding $\delta r_\text{e}$ and $\delta r_\text{f}$ yields
\begin{align}
  & \hskip -0.2cm \delta r_\text{e}^{(0)} + \delta r_\text{f}^{(0)} = 0\,,\\[3.0ex]
  & \hskip -0.2cm \delta r_\text{e}^{(1)}(\bx) + \delta r_\text{f}^{(1)}(\bx) = \nonumber\\[1.0ex]
  \label{eq:h1terms2}
  & \hskip -0.1cm  2\gamma_3\, r(\bx)p(\bx)\,\partial_y s(\bx) - 2\gamma_3\, s(\bx)p(\bx)\,\partial_y r(\bx)\,,\\[3.0ex]
  & \hskip -0.2cm \delta r_\text{c}^{(2)}(\bx) + \delta r_\text{d}^{(2)}(\bx) = \nonumber\\[1.0ex]
  & \gamma_3\,\left[ s(\bx)p(\bx)\Delta r(\bx) - r(\bx)p(\bx)\Delta s(\bx)\right]\,.
  \label{eq:h2terms2}  
\end{align}
As can be seen, the $\text{O}(h)$ contributions, Eqs.~(\ref{eq:h1terms1}), (\ref{eq:h1terms2}), are equal and opposite, while the $\text{O}(h^2)$ ones, Eqs.~(\ref{eq:h2terms1}), (\ref{eq:h2terms2}), are just equal. Consequently, the former cancel whereas the latter add up. At the end, we get
\begin{align}
  & \hskip -0.4cm \delta r_\text{chase}(\bx)  = D_3\, \left[ s(\bx)p(\bx)\Delta r(\bx) - r(\bx)p(\bx)\Delta s(\bx)\right] \nonumber\\[1.0ex]
  & \hskip 1.1cm + \text{O}(N^{-3/2})\,,
  \label{eq:3agentmob}
\end{align}
with scaling diffusion constant $D_3 = 2\gamma_3/N$.

Eq.~(\ref{eq:3agentmob}) yields a continuous chase operator. It is just one of several possible definitions. We could produce others, for instance, by differently choosing configurations from Fig.~\ref{fig:neighbors}. Unsurprisingly, $\delta r_\text{chase}$ is nonlinear (cubic) in the strategy densities. At first sight, the expression in square brackets looks smaller than $\Delta r(\bx)$ because two densities multiply the Laplace operators. In view of this, we should reasonably expect that Eq.~(\ref{eq:3agentmob}) contributes as a small perturbation to Eq.~(\ref{eq:SPDE}) for $\gamma_3 \approx \gamma_2$. To make $\delta r_\text{chase}$ comparable in strength to $\delta r_\text{ex}$, we should let $\gamma_3$ larger that $\gamma_2$ by a factor of about $3\div 5$. Generally, reaction-diffusion equations with nonlinear diffusion operators cannot be mapped onto a CGLE. Hence, the formation of stable spirals is ruled out. 

\begin{figure}[H]
  \centering
  \includegraphics[width=0.156\textwidth]{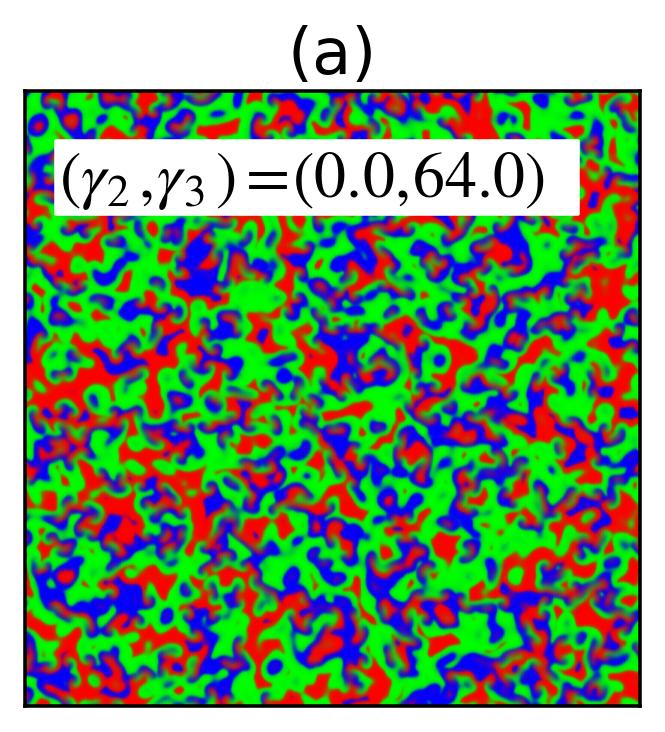}
  \includegraphics[width=0.156\textwidth]{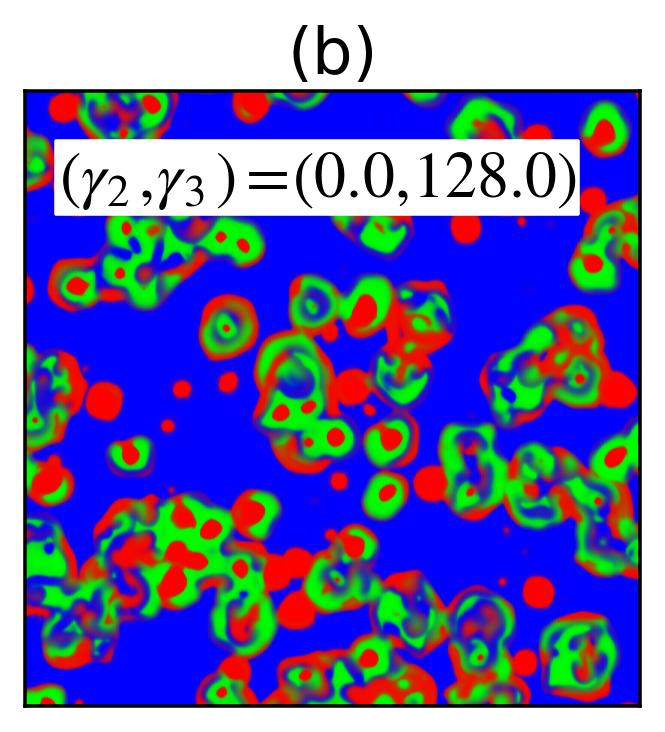}
  \includegraphics[width=0.156\textwidth]{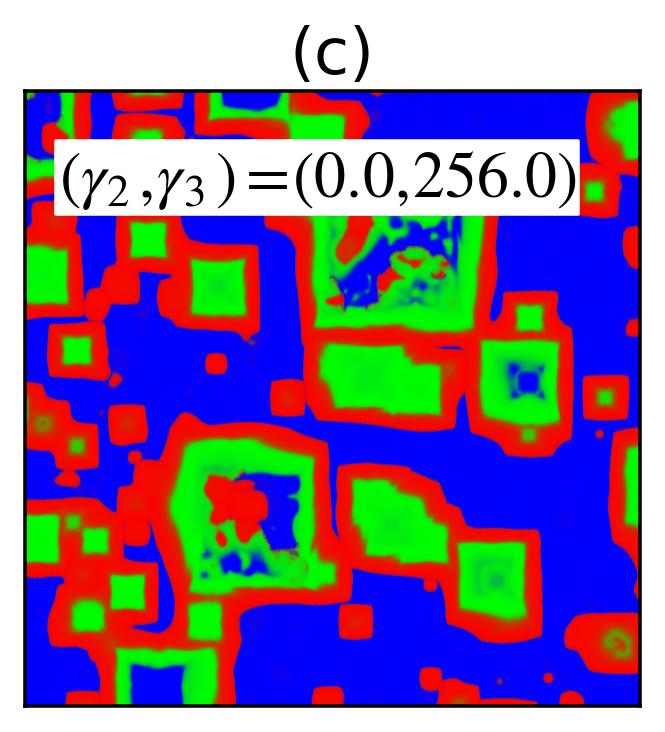}
  \\[2.0ex]
  \includegraphics[width=0.156\textwidth]{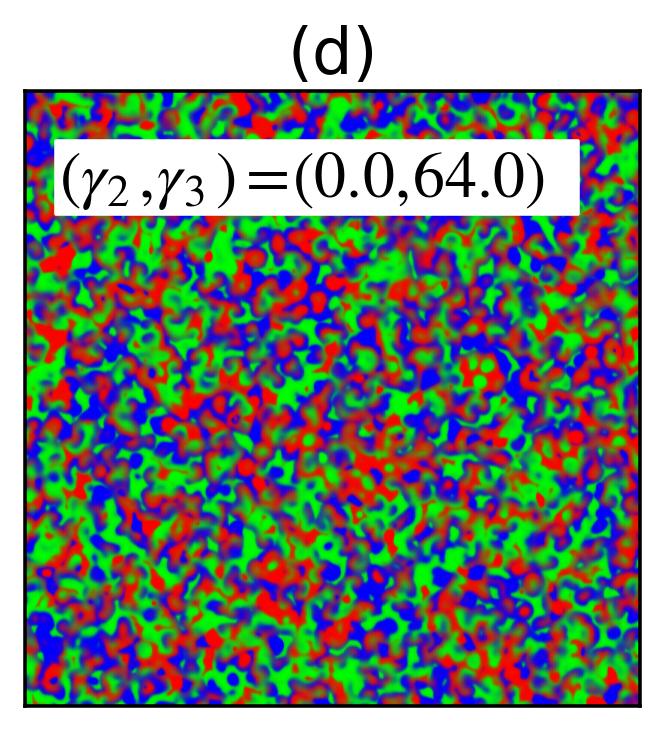}
  \includegraphics[width=0.156\textwidth]{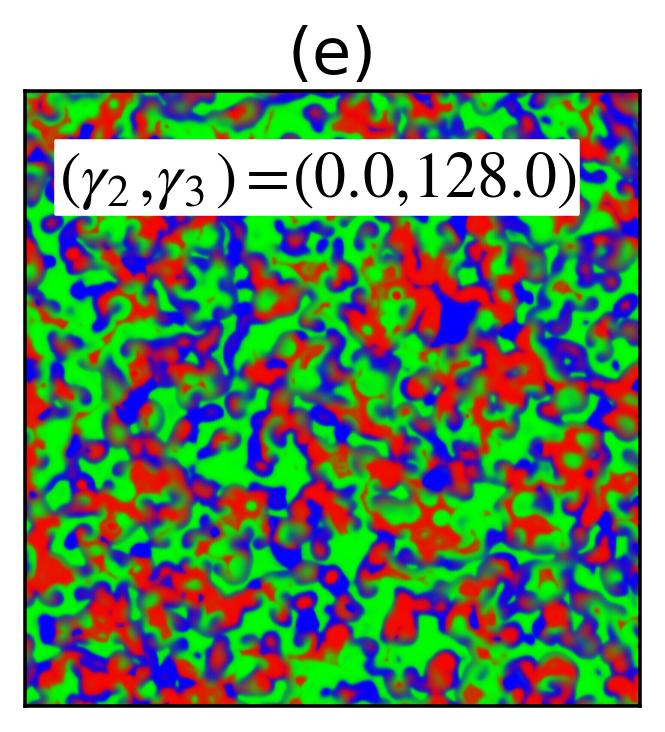}
  \includegraphics[width=0.156\textwidth]{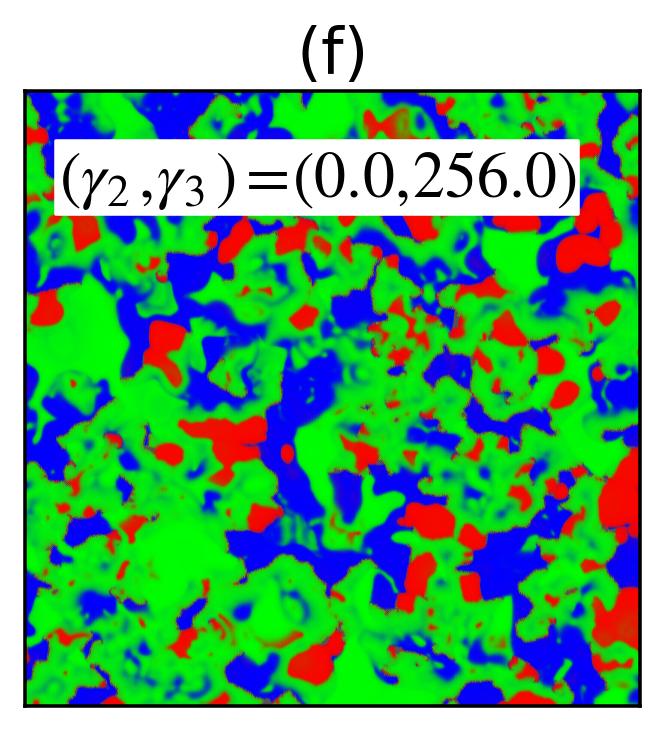}
  \caption{\footnotesize {\color{red} [color online]} Chaotic patterns induced by chase reactions.{\label{fig:nlchaos}}}
\end{figure}

Fig.~\ref{fig:nlchaos} shows six snapshots of $\phi$ corresponding to $N=2048\times 2048$, $\gamma_2=0$, $\gamma_3=64,128,256$, $\epsilon=9.70$ and $M=\infty$. We  obtained them by numerically integrating the SPDE with random initial conditions. To this aim, we used a semi-implicit Runge-Kutta scheme of order two, with time step $\rd t = 0.005$ and mobility reactions represented in configuration space (exponential time differencing is not possible in this case).

\noindent Plots (a)-(b)-(c) show $\phi$ at an early stage. They correspond respectively to $t=5,17,36$. As can be seen, strategy densities try to arrange initially into spatial patterns with size and shapes depending on $\gamma_3$. In particular, as $\gamma_3$ increases, patterns grow. Soon, chaos replaces all regular patterns. Plots (d)-(e)-(f) represent $\phi$ for $t=400$. No sign of spatial coherence is left. Patterns are chaotic for all values of $\gamma_3$. They keep evolving rapidly in time. Their size increases with $\gamma_3$ as expected, while their boundaries, as we observed in unreported snapshots, can be equally sharp or blurred.

\begin{figure}[!h]
  \centering
  \includegraphics[width=0.23\textwidth]{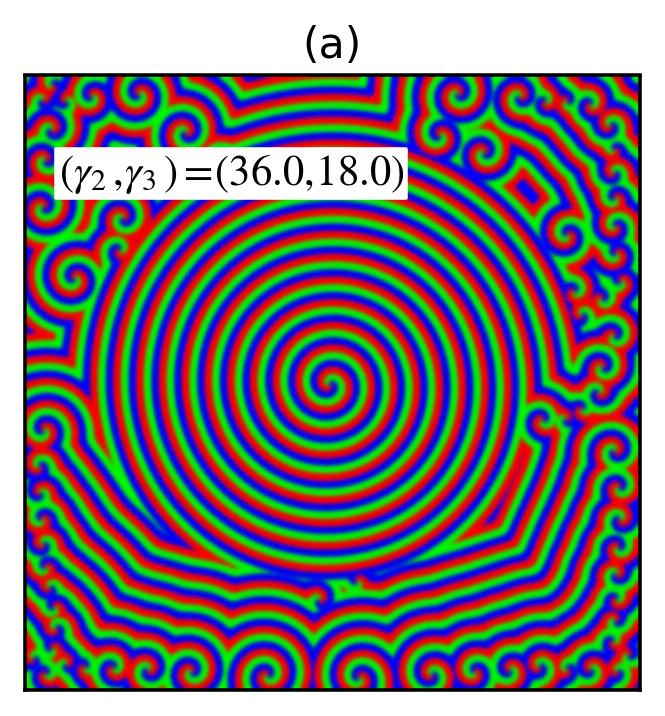}
  \includegraphics[width=0.23\textwidth]{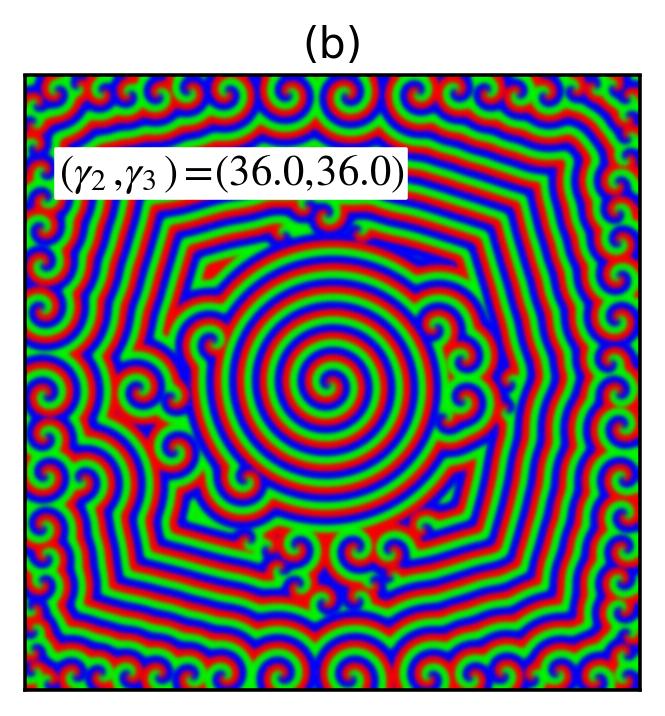}
  \\[1.0ex]
  \includegraphics[width=0.23\textwidth]{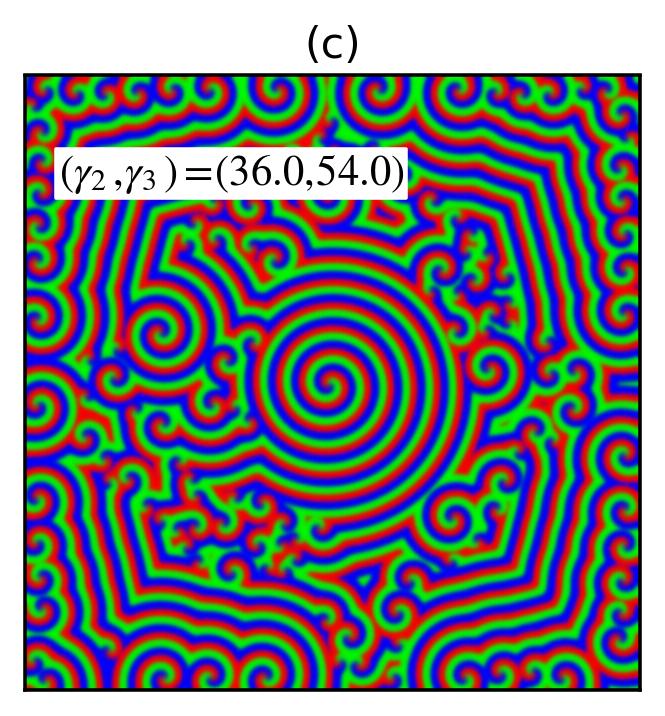}
  \includegraphics[width=0.23\textwidth]{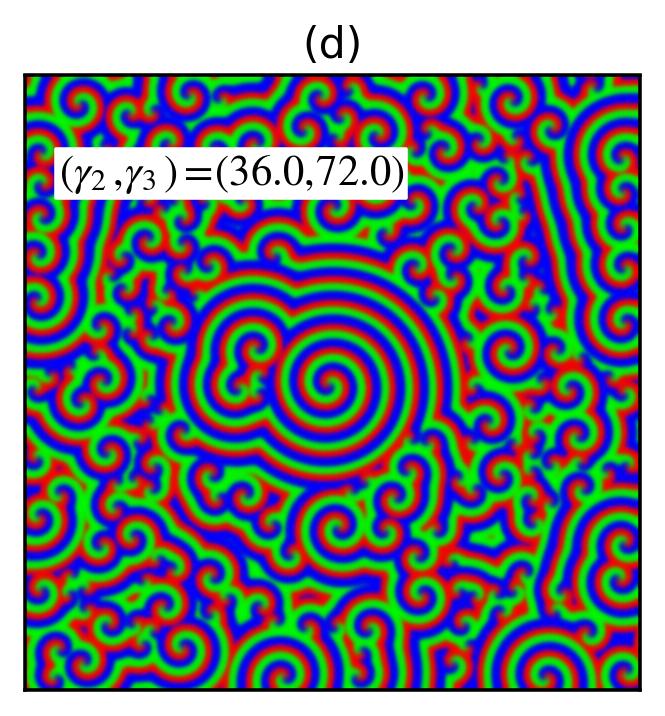}
  \caption{\footnotesize {\color{red} [color online]} Effects of chase reactions on the propagation of spiral waves.{\label{fig:nlwave}}}
\end{figure}

Fig.~\ref{fig:nlwave} shows four snapshots of $\phi$ corresponding to $N=3072\times 3072$, $\gamma_2=36$, $\gamma_3=18,36,54,72$, $\epsilon=9.70$ and $M=\infty$. We obtained them by numerically integrating the SPDE with initial conditions given by Eq.~(\ref{eq:initcond}). To this aim, we used the same integration scheme as explained above. We chose $\epsilon=9.70$ since we know from Fig.~\ref{fig:PDE}~(o) that the evolution of the strategy densities for $\gamma_3=0$ yields a fully extended and perfectly stable central spiral. Hence, this choice allows us to assess the effects of the chase operator in a controlled set up. 

Similar to stochastic noise, nonlinear diffusion is responsible for the far-field break-up of the central spiral. The break-up mechanism seems to be essentially the same: nonlinear disturbances propagate from the core outwards; they grow in size while propagating; spatial coherence breaks down as soon as disturbances overcome the carrier signal. However, a comparison with Fig.~\ref{fig:finiteM} highlights two differences regarding the breaking pattern:
\begin{itemize}[leftmargin=3.0mm]
\item[$\circ$]{stochastic noise deforms the profile of wavefronts near the break-up radius. This effect is absent in Fig.~\ref{fig:nlwave}, where all wavefronts up to the break-up radius are nearly perfect. If our interpretation is correct, the effects of nonlinear diffusion reveal suddenly, whereas stochastic noise is somewhat progressive;}
\item[$\circ$]{in Fig.~\ref{fig:finiteM}, spatial coherence is essentially lost beyond the break-up radius. Small spirals fill the environment without large emerging patterns. This behavior is at odds with Fig.~\ref{fig:nlwave}, where spatial coherence reappears beyond the break-up radius for $\gamma_3\lesssim\gamma_2$. External patterns consist of almost-linear wavefronts. Nearly oriented towards the main lattice directions, they form cusps along the main diagonals. Moreover, they resemble the boxy-shaped patterns of Fig.~\ref{fig:nlchaos}~(c). The effect of coherence recovery reduces progressively as $\gamma_3$ increases. Paradoxically, nonlinear diffusion has the power to rectify spiral waves.}
\end{itemize}
In this paper, we do not go beyond the above qualitative observations. Our goal was to show that mobility reactions, different from hopping and pair exchange, can have non-trivial and interesting consequences on the propagation of spiral waves. Reaction-diffusion equations with nonlinear diffusion mechanisms have attracted much attention in the past years, thanks to their potential applications to many natural phenomena. We refer the reader to the Refs.~\cite{Gilding1,Gilding2,Gilding3,Gilding4,Vazquez1} for in-depth studies. 

\section{Conclusions}

To summarize, we have investigated several aspects of a variant of the cyclic Lotka-Volterra model, featuring three-agent interactions. We aimed at studying cyclic dominance, mediated by cooperative predation, in a simple theoretical setting.

Nonlinear analysis of the underlying rate equations in a well-mixed environment has revealed the existence of degenerate Hopf bifurcations. They occur for specific values of the rate constants. More precisely, in our model, reactions involving two prey and one predator equilibrate the system, while reactions involving two predators and one prey polarize it. Bifurcations correspond to non-trivial equilibria between the rates of the former and the latter reactions. If equilibrating and polarizing reactions have homogeneous rates, respectively $\deq$ and  $\dpol$, the rate equations bifurcate for $\deq=\dpol$. This condition describes predators hunting in a group or alone with equal propensity. In a metapopulation model with patches hosting several agents, rotating spiral waves appear only for $\deq<\dpol$, i.e., when the propensity for cooperative predation is stronger than for individual hunting. 

Theoretical methodologies used in this paper have been developed elsewhere in the literature. We have just applied them to our model to compare features of the underlying dynamics with other existing models.

In particular, we have derived the magnitude of the stochastic noise at the bifurcation point for homogeneous rates (where the rate equations predict neutrally stable orbits), to make a comparison with the original cyclic Lotka-Volterra model~\cite{Reichenbach5}. Three-agent interactions are intrinsically noisier than two-agent ones since they involve more fluctuating degrees of freedom. Nevertheless, the extinction probability is uniformly lower in our model than in Ref.~\cite{Reichenbach5}. This apparent paradox has a simple solution. When three agents interact, strategies fluctuate longer around the reactive fixed point before one of them prevails on the others. Stochastic noise has no preferred direction. Hence, it acts as an equilibrating force. Group interactions help the system stay in equilibrium. Doing so, they promote species coexistence.

Similarly, we have studied the phase portrait for heterogeneous rates, to make a comparison with a model in which group interactions involve an agent and its four von-Neumann neighbors or its eight Moore neighbors~\cite{Szolnoki2}. Although we consider only three interacting agents, the phase portrait of our model shows a richer structure. As far as we understand, the reason for this result is that our reaction rates are fully independent. Besides, we have shown that spatial topology plays a critical role in shaping the phase portrait. Indeed, Hopf bifurcations disappear on a two-dimensional lattice (with one agent per lattice site) as a consequence of the locality of the interactions.

Then, we have studied the effects of individual mobility in a lattice metapopulation model, with patches hosting several agents. It turns out that our rotating spirals are qualitatively similar to those arising in the spatially structured version of the May-Leonard model~\cite{Reichenbach2}. Although our model differs from Ref.~\cite{Reichenbach2} in that we assume no empty sites, no birth and no selection-removal interactions, the observed similarity is not surprising, for three reasons. First of all, models of cyclic dominance without mutations usually undergo degenerate Hopf bifurcations. Secondly, it is usually possible to map a system of reaction-diffusion equations, with supercritical Hopf bifurcation, onto a complex Landau-Ginzburg equation~\cite{Reichenbach1,Frey1}. The map is possible provided diffusion is linear. Finally, the latter equation yields a good description, even when the bifurcation is degenerate. As a result, the regimes of species coexistence in our model and the spatially structured May-Leonard model fall in the same universality class. As such, they are in one-to-one correspondence. 

To conclude, we have shown that one can build nonlinear continuous mobility operators starting from three-agent chase reactions on a lattice. We have focused on one of several possible definitions. In particular, we have studied the effects of chase reactions on the propagation of spiral waves. Similar to nonlinear diffusion, arising in models where hopping and pair exchange occur with different rates, chase reactions produce far-field break-up of spiral waves. However, the breaking pattern is peculiar. Nonlinear diffusion could play a role in explaining structural patterns in growing bacterial colonies. Our study shows that group interactions provide viable mechanisms for both predation and dispersal. 

\section*{Acknowledgments}

F.P. acknowledges his little daughter Giulia for (re)inventing {\tt RPS} with three players and for suggesting it as a research topic. The computing resources used for our numerical simulations and the related technical support have been provided by the CRESCO/ENEA\-GRID High Performance Computing infrastructure and its staff~\cite{Ponti}. CRESCO ({\color{red}C}omputational  {\color{red} RES}earch centre on {\color{red} CO}mplex systems) is funded by ENEA and by Italian and European research programmes.
\vskip 0.1cm
{\bf Author contribution statement} F.P. performed theoretical calculations, numerical simulations, and data analysis with assistance from S.F. and S.T. All authors contributed to the discussion and the writing of the manuscript.

\bibliographystyle{hunsrt}
\bibliography{main}

\end{document}